\documentclass[reprint]{sigplanconf-eurosys}

\usepackage{fancyhdr}
\usepackage[normalem]{ulem}
\usepackage{hyperref}
\usepackage{epsfig}
\usepackage{graphicx}
\usepackage{epstopdf}
\usepackage{setspace} 
\usepackage{subfig}
\usepackage[ruled,vlined,linesnumbered]{algorithm2e}
\usepackage{url}
\usepackage{amsmath}
\usepackage{pifont}
\usepackage{fixltx2e}
\usepackage{anyfontsize}
\usepackage{kantlipsum}
\usepackage{multirow}
\usepackage[font=bf]{caption}
\usepackage{array}
\usepackage{flushend}
\usepackage{tikz}
\newcommand*\blkcircled[1]{\tikz[baseline=(char.base)]{
            \node[shape=circle,fill=black,font=\bfseries,inner sep=1pt] (char) {\textcolor{white}{#1}};}}
\newcommand*\redcircled[1]{\tikz[baseline=(char.base)]{
            \node[shape=circle,fill=red,font=\bfseries,inner sep=1pt] (char) {\textcolor{white}{#1}};}}
\newcommand*\orgcircled[1]{\tikz[baseline=(char.base)]{
            \node[shape=circle,fill=orange,font=\bfseries,inner sep=1pt] (char) {\textcolor{white}{#1}};}}
\newcommand*\bluecircled[1]{\tikz[baseline=(char.base)]{
            \node[shape=circle,fill=blue,font=\bfseries,inner sep=1pt] (char) {\textcolor{white}{#1}};}}
\newcommand*\greencircled[1]{\tikz[baseline=(char.base)]{
            \node[shape=circle,fill=green,font=\bfseries,inner sep=1pt] (char) {\textcolor{white}{#1}};}}

\newcommand\blfootnote[1]{%
  \begingroup
  \renewcommand\thefootnote{}\footnote{#1}%
  \addtocounter{footnote}{-1}%
  \endgroup
}

\begin{document}

 %8.5in,11in
\setlength{\pdfpageheight}{\paperheight}
\setlength{\pdfpagewidth}{\paperwidth}  
\titlebanner{banner above paper title}        % These are ignored unless
\preprintfooter{short description of paper}   % 'preprint' option specified.

%\makeatletter
%\def\@maketitle{\newpage
% \null
% \setbox\@acmtitlebox\vbox{%
%\baselineskip 20pt
%\vskip -2em                   % Vertical space above title.
%   \begin{center}
%    {\ttlfnt \@title\par}       % Title set in 18pt Helvetica (Arial) bold size.
%    \vskip -1.5em                % Vertical space after title.
%	%This should be the subtitle.
%{\subttlfnt \the\subtitletext\par}\vskip 1.25em%\fi
%    {\baselineskip 16pt\aufnt   % each author set in \12 pt Arial, in a
%     \lineskip .5em             % tabular environment
%     \begin{tabular}[t]{c}\@author
%     \end{tabular}\par}
%    \vskip 1.5em               % Vertical space after author.
%   \end{center}}
% \dimen0=\ht\@acmtitlebox
%	% \advance\dimen0 by -12.75pc\relax % comment by Marco Daniel
% \unvbox\@acmtitlebox
% \ifdim\dimen0<0.0pt\relax\vskip-\dimen0\fi}
%\makeatother

%%%%%%%%%%%---SETME-----%%%%%%%%%%%%%
%\title{FlashAbacus: A Storage-based Accelerator with Energy Efficient Multi-Kernel Execution} 
%\title{FlashAbacus: A Fully Self-Governing Accelerator For Embedded Systems}
%\title{FlashAbacus: A Fully Self-Governing Accelerator For Low-Power Systems}
%\title{FlashAbacus: A Storage-Integrated Hardware Accelerator For Fully Self-Governing Low-Power Systems}
\title{FlashAbacus: A Self-Governing Flash-Based Accelerator for Low-Power Systems}
%\title{FlashAbacus: Virtualized Data Processing Near Storage}

\authorinfo{\large{Jie Zhang and Myoungsoo Jung}

%\vspace{5pt}
\emph{Computer Architecture and Memory Systems Laboratory}

%\vspace{5pt}
\normalsize{School of Integrated Technology,Yonsei University,}

%\normalsize{Yonsei University,}

\textsf{http://camelab.org}
}

%\author{Jie Zhang and Myoungsoo Jung}
%%\authornote{Dr.~Trovato insisted his name be first.}
%%\orcid{1234-5678-9012}
%\affiliation{%
%  \institution{\emph{Computer Architecture and Memory Systems Laboratory}}
%  \department{\normalsize{School of Integrated Technology, Yonsei University,}}
%  %\city{\normalsize{}}
%  \state{\textsf{http://camelab.org}}
%}

%------------------------------------------------------------------------------
%                  Useful commands and acronymns used in the paper.
%------------------------------------------------------------------------------
% Commands
\newcommand{\mycomment}[1]{}
\newcommand{\ignore}[1]{}
%\ifpdf
%\renewcommand{\url}[1]{\texttt{\small{#1}}}
%\else
%\newcommand{\url}[1]{\texttt{\small{#1}}}
%\fi
\newcommand{\code}[1]{\texttt{#1}}
\newcommand{\codesm}[1]{\texttt{\small #1}}
\newcommand{\sm}[1]{{\small #1}}
\newcommand{\mycaption}[2]{\textbf{\textsf{\caption{#1}#2}}}

\newcommand{\hlc}[2][yellow]{ {\sethlcolor{#1} \hl{#2}} }

\newcommand{\todo}[1]{\color{white}\textbf{\hlc[black]{TODO: [#1]}}\color{black}\xspace}
\newcommand{\fixme}[1]{\color{red}\textbf{\hl{FIXME: [#1]}}\color{black}\xspace}
\newcommand{\rph}[1]{\color{white}\textbf{\hlc[red]{REPHRASE: [#1]}}\color{black}\xspace}
\newcommand{\remark}[1]{\color{blue}\textbf{\hlc[green]{REMARK: [#1]}}\color{black}\xspace}
\newcommand{\answer}[1]{\color{white}\textbf{\hlc[blue]{ANSWER: [#1]}}\color{black}\xspace}
\newcommand{\pointer}[1]{\color{white}\textbf{\hlc[red]{POINTER: [#1 is working here]}}\color{black}\xspace}
\newcommand{\newedit}[1]{#1\xspace}
% Acronymns
\newcommand{\apriori}{\textit{a priori}}
\newcommand{\adhoc}{ad hoc}
\newcommand{\etal}{\textit{et al.}}
\newcommand{\eg}{\textit{e.g.}}
\newcommand{\ie}{\textit{i.e.}}
\newcommand{\mybull}{\noindent $\bullet$~}
\newcommand{\ms}{$ms$~}
\newcommand{\us}{${\mu}s$~}

\newcommand{\mystar}{$\bigstar$}

% Section names, figure names and algorithm names.
\newcommand{\figref}[1]{Figure~\ref{#1}}
\newcommand{\sectref}[1]{Section~\ref{#1}}
\newcommand{\sectrefs}[2]{Sections~\ref{#1}-\ref{#2}}
\newcommand{\aref}[1]{Algorithm~\ref{#1}}
\newcommand{\sect}[1]{\section{#1}}
\newcommand{\subsect}[1]{\subsection{#1}}
\newcommand{\subsubsect}[1]{\subsubsection{#1}}
\newcommand{\mysect}[1]{\subsect{#1}}

\newtheorem{ourtask}{Task}

%don't want date printed
\date{}
%%%%%%%%%%%%%%%%%%%%%%%%%%%%%%%%%%%%

\maketitle

\setstretch{0.995}

\subsection*{Abstract}
Energy efficiency and computing flexibility are some of the primary design constraints of heterogeneous computing. In this paper, we present FlashAbacus, a data-processing accelerator that self-governs heterogeneous kernel executions and data storage accesses by integrating many flash modules in lightweight multiprocessors. The proposed accelerator can simultaneously process data from different applications with diverse types of operational functions, and it allows multiple kernels to directly access flash without the assistance of a host-level file system or an I/O runtime library. We prototype FlashAbacus on a multicore-based PCIe platform that connects to FPGA-based flash controllers with a 20 nm node process. The evaluation results show that FlashAbacus can improve the bandwidth of data processing by 127\%, while reducing energy consumption by 78.4\%, as compared to a conventional method of heterogeneous computing. \blfootnote{This paper is accepted by and will be published at 2018 EuroSys. This document is presented to ensure timely dissemination of scholarly and technical work.
}

%For this tight integration of storage and processors, we introduce multi-kernel execution that can perform flexible data processing near flash. This execution model can simultaneously process data from different applications with different types of operational functions, which can maximize the utilization of internal multiprocessors. In addition, we propose Flashvisor that allows multiple kernels to directly access flash without an assistance of a host-level file system or an I/O runtime library. 
%We prototype FlashAbacus on a multicore-based PCIe platform that connects to FPGA-based flash controllers at 20 nm node process. The evaluation results show that FlashAbacus can improve the bandwidth of data processing by 127\%, while reducing energy consumption by 78.4\%, compared to a conventional method of heterogeneous computing.

%We and converted assorted data-intensive applications into our new NDP framework. The evaluation results show that our FlashPhi can improve the performance by 7.8x and 75\%, while reducing energy consumption by 88\% and 57\%, compared to a CPU-driven and a GPU-based data processing approach, respectively.

\section{Introduction}
\label{sec:introduction}
% Data
Over the past few years, heterogeneous computing has undergone significant performance improvements for a broad range of data processing applications. This was made possible by incorporating many dissimilar coprocessors, such as general-purpose graphics processing units (GPGPUs) \cite{PASCAL} and many integrated cores (MICs) \cite{chrysos2014intel}. These accelerators with many cores can process programs offloaded from a host by employing hundreds and thousands of hardware threads, in turn yielding performance than that of CPUs many orders of magnitude \cite{manavski2007cuda, arora2012redefining, govindaraju2008high}. 

While these accelerators are widely used in diverse heterogeneous computing domains, their power requirements render hardware acceleration difficult to achieve in low power systems such as mobile devices, automated driving systems, and embedded designs.
For example, most modern GPGPUs and MICs consume as high as 180 watts and 300 watts per device \cite{pascalnvidia, chrysos2014intel}, respectively, whereas the power constraints of most embedded systems are approximately a few watts \cite{borriello2000embedded}. 
To satisfy this low power requirement, ASICs and FPGAs can be used for energy efficient heterogeneous computing. However, ASIC- or FPGA-based accelerators are narrowly optimized for specific computing applications. In addition, their system usability and accessibility are often limited due to static logic and inflexible programming tasks. Alternatively, embedded multicore processors such as the general purpose digital signal processor (GPDSP), embedded GPU and mobile multiprocessor can be employed for parallel data processing in diverse low power systems \cite{hegde2016caffepresso,jung2017nearzero}.

\ignore{
$$$$$$$$$$ can co go to related work  $$$$$$$$$$$$$
To satisfy this low power requirement, application-specific integrated circuits (ASICs) or field-programmable gate arrays (FPGAs) can be used for energy efficient heterogenous computing. However, ASIC-based or FPGA-based accelerators are narrowly optimized for specific computing applications. In addition, their system usability and accessability are often limited due to static logic and inflexible programming tasks. Alternatively, lightweight multicore processors (LWPs) such as general purpose digital signal processor (GPDSP) or mobile multiprocessor can be employed for parallel data processing in diverse low power systems \cite{levy2009embedded, azevedo2009parallel, ferri2010energy}. 
}

Although the design of low-power accelerators with embedded processors opens the door to a new class of heterogeneous computing, their applications cannot be sufficiently well tuned to enjoy the full benefits of the data-processing acceleration; this is because of two root causes: i) low processor utilization and ii) file-associated storage accesses. 
Owing to data transfers and dependencies in heterogeneous computing, some pieces of code execution are serialized, in turn limiting the processing capability of the low-power accelerators. 
%Single instruction multiple data (SIMD) is one of the most popular parallelism models to take advantage of multicore processors, but we observe that resources of low-power accelerators are significantly underutilized with SIMD processing. 
%but we observe that CPU utilization of SIMD-based data processing is lower than around 67\% under the execution of diverse throughput computing applications \cite{}. 
%The reason behind this poor resource utilization is data transfers and dependencies, which make some pieces of code execution serialized. 
In addition, because, to the best of our knowledge, all types of accelerators have to access external storage when their internal memory cannot accommodate the target data, they waste tremendous energy for storage accesses and data transfers. External storage accesses not only introduce multiple memory copy operations in moving the data across physical and logical boundaries in heterogeneous computing, but also impose serial computations for marshalling the data objects that exist across three different types of components (e.g., storage, accelerators, and CPUs). Specifically, 49\% of the total execution time and 85\% of the total system energy for heterogeneous computing are consumed for only data transfers between a low-power accelerator and storage.

\begin{figure*}
	\centering	
	\includegraphics[width=1\linewidth]{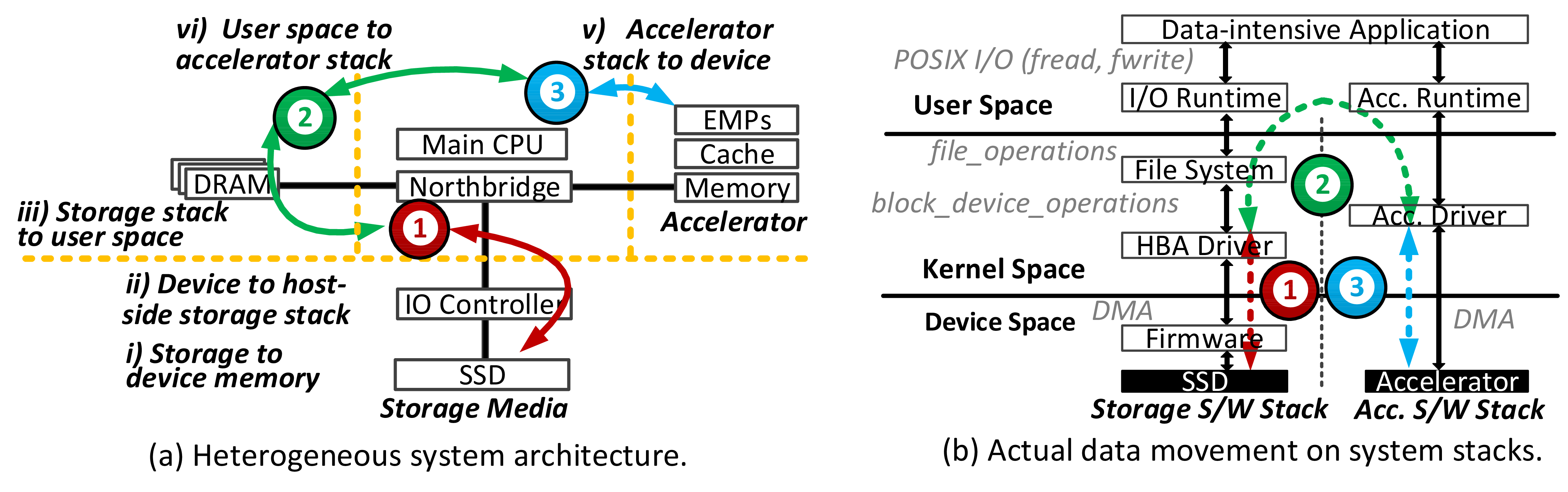}
	%\vspace{-5pt}
	\caption{Datapath analysis in terms of H/W and S/W.}
	%\vspace{-5pt}
	\label{fig:PCIe-com}
\end{figure*} 

In this paper, we propose \emph{FlashAbacus}, a data processing accelerator that self-governs heterogeneous computing and data storage by integrating many flash modules in lightweight multiprocessors to resemble a single low-power data processing unit. 
FlashAbacus employs tens of low-power functional units rather than hundreds and thousands of hardware threads that most manycore accelerators employ. %, which are used for single instruction multiple thread (SIMT) execution model in most manycore-based accelerators. 
Even though the computing capacity of the small number of functional units is not as high as that of the manycore accelerators, FlashAbacus can process data near flash with different types of applications or kernels that contain multiple functions or both. 
%FlashAbacus executes multiple kernels near flash in parallel rather than executing them as a serialized order. 
Such multi-kernel execution can maximize the resource utilization by dynamically scheduling multiple applications across all the internal processors. The kernel scheduler can also reorder the executions in an out of order manner by recognizing the code blocks that have no data dependency across different kernels of the target data-intensive application. Therefore, FlashAbacus offers high bandwidth with a relatively small number of processors, thereby maximizing energy efficiency.

A challenge in enabling such multi-kernel execution near flash is that the flash media integrated into the accelerator are not practically working memories; they operate in a manner similar to mass storage.
That is, the cores of accelerator cannot execute any type of kernels through typical load and store instructions. Therefore, the accelerator needs to employ an OS, but the OS will make the accelerator bulkier and will worsen the performance characteristics.
%it is required for the accelerator to employ OS. However, OS modules can make the accelerator much heavier and performance/energy characteristics even worse.  
To address these challenges, we introduce Flashvisor that allows the cores of accelerator to access flash directly without any modification of the instruction set architecture or assistance of the host-side storage stack. Flashvisor virtualizes storage resources by mapping the data section of each kernel to physical flash memory. In addition, it protects flash from the simultaneous accesses of multi-kernel execution by employing a range lock, which requires a few memory resources. We also separate the flash management tasks from their address translations \cite{kim2002space,gupta2009dftl,chung2009survey,park2008reconfigurable}, and allocate a different processor to take over the flash management tasks. This separation makes the flash management mechanisms (e.g., garbage collection) invisible to the execution of multiple kernels.  

%catches the all the memory request missed from each cores L2 cache and performs the address translation and memory protection between a network buffer and the underlying flash media. 

%Unlike the power-hungry accelerators that conventionally employ hundreds or thousand cores, our FlashAbacus is designed towards energy efficient data processing near flash with a relatively small number of cores. To alleviate a potential performance drop due to the limited resources, we design FlashAbacus for heterogeneous NDP-kernel execution, which can download different types of NDP applications and execute them in parallel, thereby maximizing user experience with better performance. 
%In our FlashAbacus, this scheduling can be either static or dynamic and can reorder the executions in an out-of-order fashion, which can satisfy diverse user demands. 

%Unlike other accelerators and in-storage processing approaches, our FlashAbacus can dynamically accommodate different types of NDP applications and process data near hundreds of flash, which in turn removes unnecessary data movements and exhibits high bandwidth with low power consumption. As our FlashAbacus can support a highly flexible NDP model and easy to program compute kernels, it neither requires an assistance of host-side applications nor needs to define the firmware-level primitives and programming interfaces at design time.

We built a prototype FlashAbacus on a low-power multicore \cite{TI6678} based PCIe platform that connects FPGA-based flash controllers with a 20 nm node process \cite{virtex}. To evaluate the effectiveness of FlashAbacus prototype, we converted diverse computing applications \cite{pouchet2012polybench} into our new execution framework. Our evaluation results show that FlashAbacus can offer 127\% higher bandwidth and consume 78\% less energy than a conventional hardware acceleration approach under data-intensive workloads.

%\section{Preamble to Heterogeneous Computing}
\section{Preliminaries}
\label{sec:background}
In this section, \newedit{we analyze the overheads of moving data between an accelerator and storage, which are often observed in a conventional heterogeneous computing system. This section then explains the baseline architecture of our FlashAbacus.}

%\begin{figure*}
%\centering
%\hspace{-5pt}
%\subfloat[High-level view.]{\label{fig:hw_org}\rotatebox{0}{\includegraphics[width=0.5\linewidth]{figs/hw_org}}}
%\hspace{3pt}
%\subfloat[Bridge internals.]{\label{fig:bridge}\rotatebox{0}{\includegraphics[width=0.19\linewidth]{figs/bridge}}}
%\hspace{3pt}
%\subfloat[Network buffer system internals.]{\label{fig:NBS}\rotatebox{0}{\includegraphics[width=0.29\linewidth]{figs/NBS}}}
%\hspace{-5pt}
%\vspace{-10pt}
%\caption{\label{fig:overview}The internal architecture of FlashPhi. }
%\end{figure*}

\begin{figure*}
\centering
\subfloat[System overview.]{\label{fig:overview}\rotatebox{0}{\includegraphics[width=0.33\linewidth]{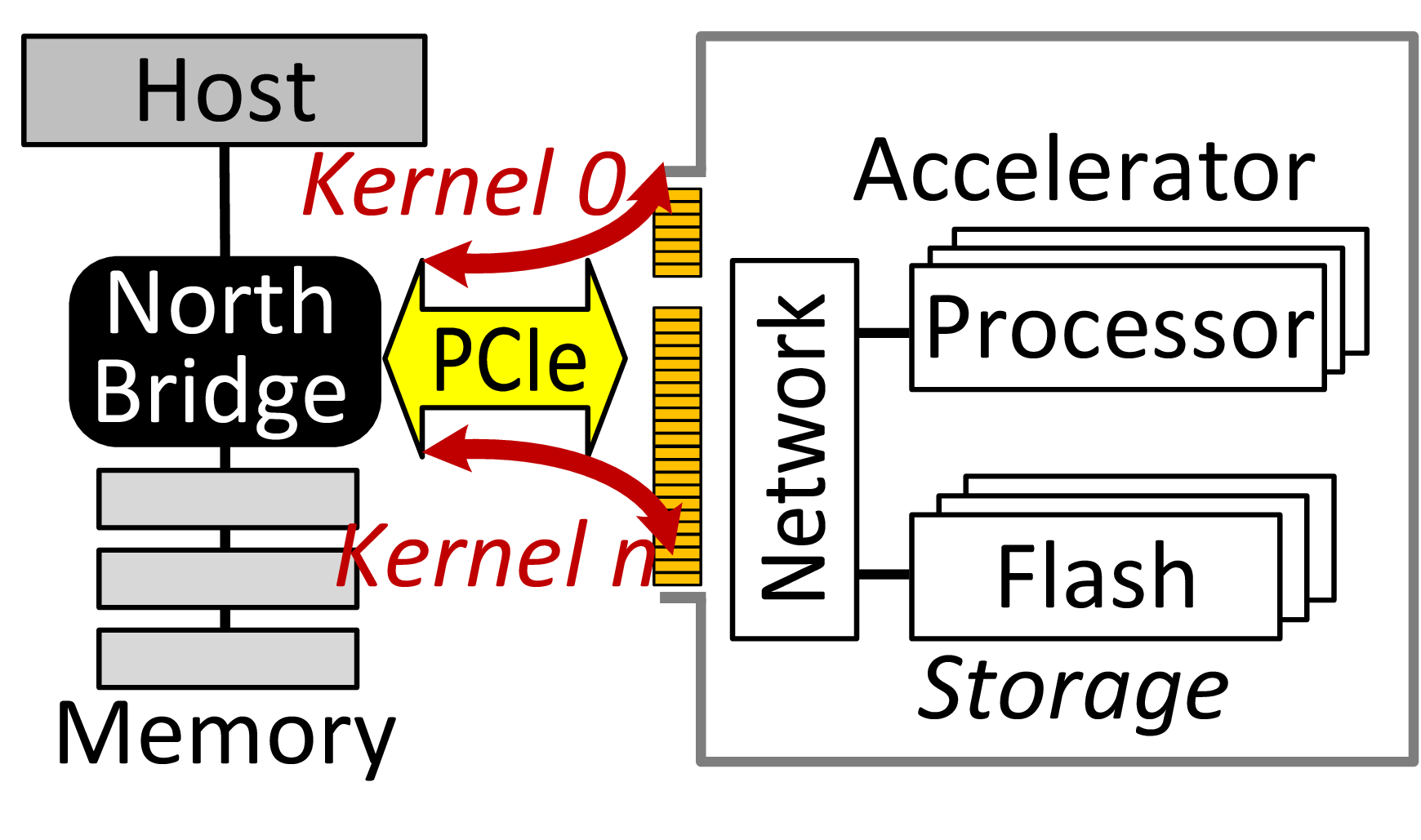}}}
\hspace{3pt}
\subfloat[FlashAbacus architecture.]{\label{fig:hw_org}\rotatebox{0}{\includegraphics[width=0.65\linewidth]{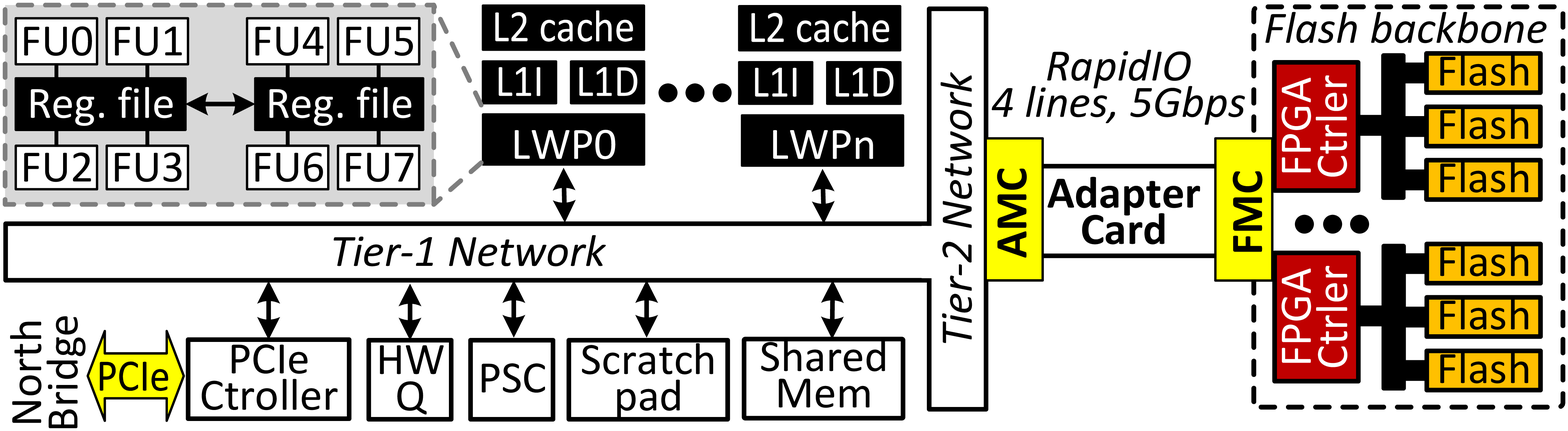}}}
%\vspace{-5pt}
\caption{\label{fig:overalloverview}System overview and internal architecture of the proposed FlashAbacus. }
%\vspace{-5pt}
\end{figure*}

\subsection{Physical and Logical Data Paths}
\label{sec:datapath}
\noindent \textbf{Hardware.} Figure \ref{fig:PCIe-com}(a) shows the physical datapath with a CPU, a low-power accelerator, and an SSD in conventional heterogeneous computing. In cases where the accelerator needs to process a large amount of data, the CPU generates I/O requests and issues them to the underlying SSD through I/O controllers such as AHCI \cite{jung2016exploring,AHCI}. An SSD controller then transfers the data from flash to its internal DRAM, and the host controller again moves such data from the internal DRAM to the host-side DRAM through a storage interface \cite{jung2013challenges, workgroup2001sata, budruk2004pci} (\redcircled{\small{1}}). During this time, the data can be reconstructed and set in order as a form of objects that the accelerator can recognize (\greencircled{\small{2}}). Finally, the CPU transfers the data from the host-side DRAM to the internal DRAM of the accelerator through the PCIe interface again (\bluecircled{\small{3}}). Note that, at this juncture, all kernel executions of the accelerator are still stalled because the input data are in being transferring and are not ready to be processed. %CPU the data to process are in transfer.  
Once the data are successfully downloaded, the embedded multicore processors (EMPs) of the accelerator can start processing the data, and the results will be delivered to the SSD in an inverse order of the input data loading procedure (\bluecircled{\small{3}}  $\rightarrow$  \greencircled{\small{2}}  $\rightarrow$  \redcircled{\small{1}}). 
The movement of data across different physical interface boundaries imposes the restriction of long latency before the accelerator begins to actually process data and leads to a waste of energy, resulting from the creation of redundant memory copies. In addition, the physical datapath deteriorates the degree of parallelism for kernel executions. For example, a single application task has to be split into multiple kernels due to capacity limit of the internal DRAM of the accelerator, in turn requiring file-associated storage accesses and thereby serializing the execution.

%Even though the overheads of CPU involved in moving data between the accelerator and the SSD may be reduced by employing an integrated network hardware  \cite{Mellanox, DSC -- micro2015 Jangwoo},  such memory copies over physical boundary are avoidable since the accelerator and the SSD operate in a different device domain. 

%More seriously, it may be required for the accelerator to split the target data processing task into multiple ones and execute them in a serial order by considering the size of its internal DRAM. 

%When a data-complex application needs to offload a function containing a body of loop from the GPU, 
\noindent \textbf{Software.} Another critical bottleneck for heterogeneous computing is the discrete software stacks, each of them exists for the accelerator and the SSD, respectively. 
%All host platforms need a group of programs that work in tandem to realize the underlying devices as an accelerator or storage. 
As shown in Figure \ref{fig:PCIe-com}(b), the host needs to employ a device driver and a runtime library for the accelerator, while it requires a storage stack that consists of flash firmware, a host block adapter (HBA), a file system, and an I/O runtime to employ the underlying device as storage. 
%it is required for storage stack to configure a flash firmware, a host block adapter (HBA), a file system, and an I/O runtime. 
The runtime libraries for the accelerator and the SSD offer different sets of interfaces, which allow a user application to service files or offload data processing, appropriately. In contrast, the accelerator driver and HBA driver are involved in transferring data between the device-side DRAM and host-side DRAM.  Therefore, the user application first needs to request data to the underlying SSD through the I/O runtime library (\redcircled{\small{1}}), and then it must write data to the accelerator through the accelerator runtime library (\bluecircled{\small{3}}). This activity causes multiple data copies within the host-side DRAM. Furthermore, when the file system and accelerator driver receive data from the application, all the data from user buffers must be copied to OS-kernel buffers, which again creates extra memory copies within the host-side DRAM (\greencircled{\small{2}}). 
This problem arises because OS-kernel modules cannot directly access the user memory space, as there is no guarantee that the current OS-kernel module is executing in the process that the I/O request was initiated.  
In addition to these unnecessary data copies (within the host-side DRAM), the discrete software stacks also increase data moving latency and consume energy because they enforce many user/privilege mode switches between their runtime libraries and OS-kernel drivers.

%Therefore, This data movement between two software stacks introduce another type of serious data movement overheads. Specifically, all the data accesses for the SSD and accelerator should be done by privilege mode while moving the data on runtime libraries are performed at user mode. Thus, these discrete software stacks cause great overheads imposed by not only context/mode switchings but also memory copies within even host-side DRAM to handle them between user and privilege mode threads. 

\subsection{Baseline Architecture}
\label{sec:prototype}
The datapath analysis in the previous section shows that the redundant memory copies across different devices are unnecessary if we can integrate the SSD directly into the accelerator, as shown in Figure \ref{fig:overview}. The accelerator can also execute a series of tasks without an interruption if there is an enough memory space that can accommodate input data. Motivated by these observations, we built an FPGA-based flash backbone with a 20 nm processor node \cite{virtex} and tightly integrated it into a commercially-available embedded SoC platform \cite{TI6678}, as shown in Figure \ref{fig:hw_org}. %The detailed discussion regarding the selection of this platform is described at Section \ref{}. 

%Figure \ref{fig:hw_org} shows the baseline architecture of FlashAbacus. 

%Specifically, our hardware platform consists of computing cores, flash controllers and buffer memory controllers, which are built on Texas Instrument TMS320C66X, Xilinx Spartan-6 FPGA and Altera Aria-10 FPGA chips, respectively. 

\noindent \textbf{Multicore.} To perform energy-efficient and low-power data processing near flash, our hardware platform employs multiple lightweight processors (\textit{LWPs}) \cite{TI6678}, which are built on a VLIW architecture. The VLIW-based LWPs require neither out-of-order scheduling \cite{lam1988software,fisher1984vliw} nor a runtime dependency check because these dynamics are shifted to compilers \cite{ellis1985bulldog,lee2000compiler}. Each LWP has eight functional units (FUs) that consist of two multiplication FUs, four general purpose processing FUs, and two load/store FUs; thus, we can reduce the hardware complexity of the accelerator, while simultaneously satisfying the diverse demands of low-power data processing applications. 
In this design, LWPs are all connected over a high crossbar network, and they can communicate with each other over message queue interfaces that we implemented by collaborating with a hardware queue attached to the network \cite{navigator}.

\noindent \textbf{Memory system.}
Our hardware platform employs two different memory systems over the network that connects all LWPs: i) \emph{DDR3L} and ii) \emph{scratchpad}. While DDR3L consists of a large low-power DRAM \cite{parameterddr3l}, the scratchpad is composed of eight high-speed SRAM banks \cite{CMOSSRAM}. \newedit{In our platform, DDR3L is used for mapping the data sections of each kernel to flash memory thereby hiding the long latency imposed by flash accesses. DDR3L is also capable of aggregating multiple I/O requests that head to the underlying storage modules, and feasible to buffer the majority of flash writes, which can take over the roles of the traditional SSD internal cache \cite{wei2014cbm}.} In contrast, the scratchpad serves all administrative I/O requests by virtualizing the flash and the entries queued by the communication interfaces as fast as an L2 cache. Note that all LWPs share a single memory address space, but have their own private L1 and L2 caches. 
%This is similar to SMP \cite{lovett1988symmetry}, but the memory address space is mapped to the underlying storage. 

%It should be note that the flash backbone opens the channel through tier-2 network while the backbone buffer exposes LPDDR3 and scratchpad to the tier-1 network
 
%In addition, this backend component of our hardware platform e. In our hardware platform, these two memory modules are managed by Altera Arria-10 FPGA \cite{}, which is referred to as \emph{backbone buffer}. While the LPDDR3 is used for flash managements and prefetching/buffering data to support latency hiding and actively aggregating I/O requests, scratchpad serves all administrative I/O requests on the processor network as fast as L2 cache. It should be note that the flash backbone opens the channel through tier-2 network while the backbone buffer exposes LPDDR3 and scratchpad to the tier-1 network. 

\begin{figure*}
\centering
\def\subfigcapskip{0pt}
\subfloat[Kernel execution.]{\label{fig:program-model}\rotatebox{0}{\includegraphics[width=0.19\linewidth]{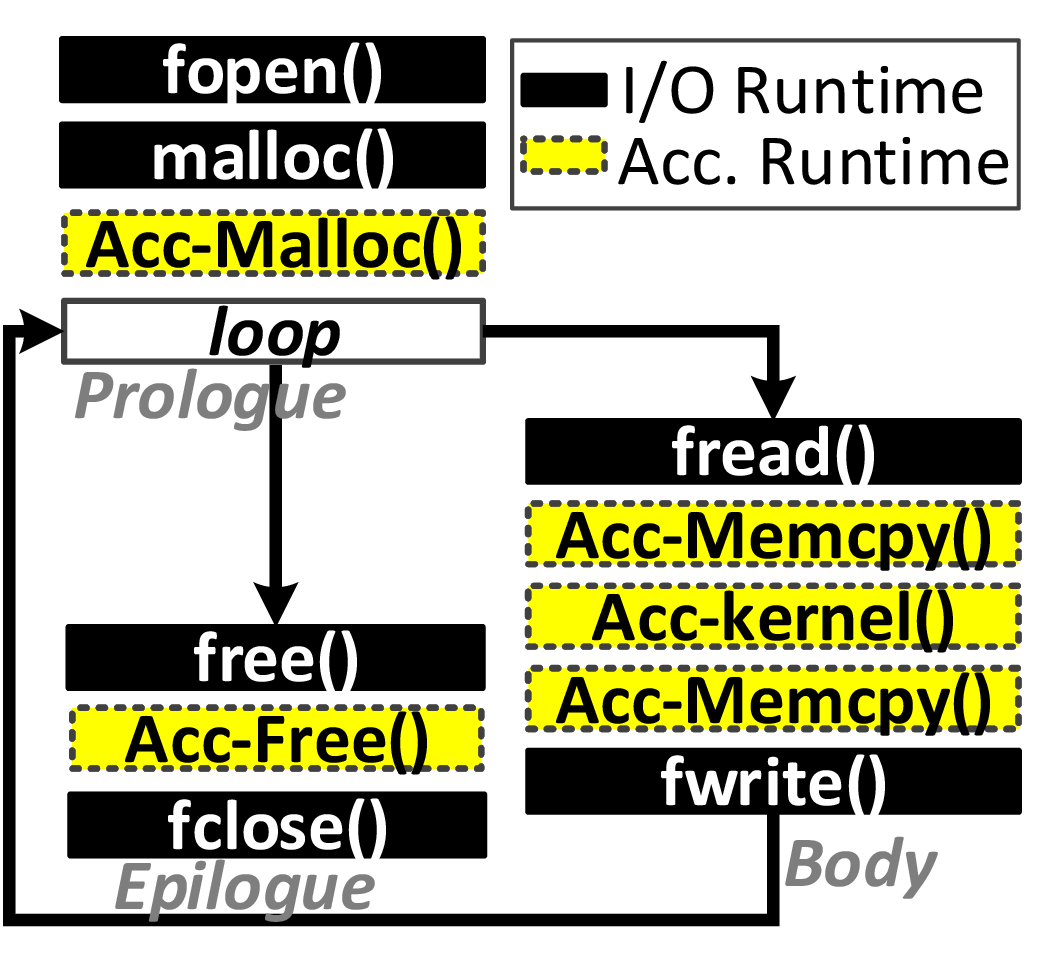}}}
\hspace{1pt}
\subfloat[Workload throughput.]{\label{fig:through-moti}\rotatebox{0}{\includegraphics[width=0.19\linewidth]{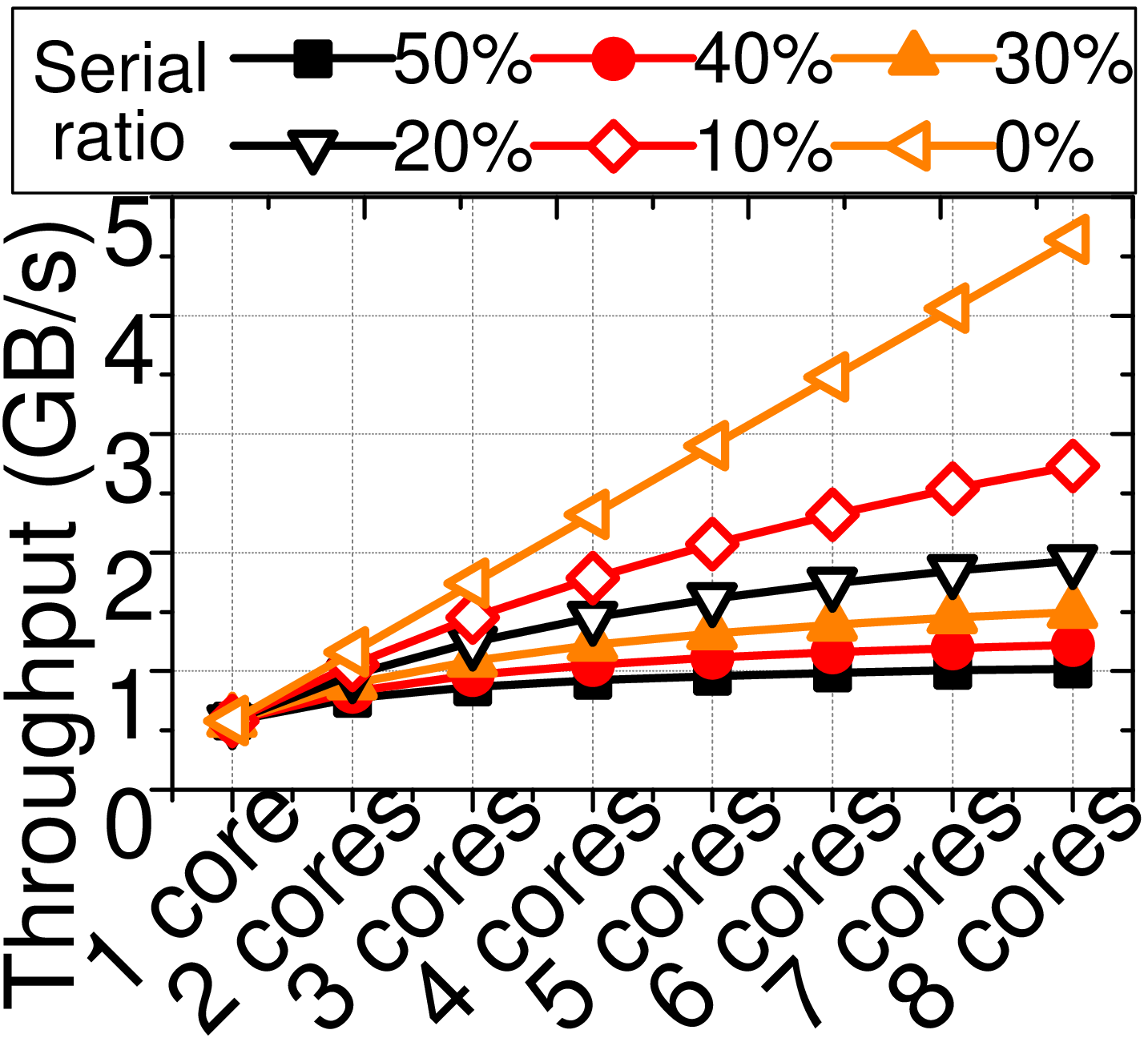}}} %
\hspace{1pt}
\subfloat[CPU utilization.]{\label{fig:util-moti}\rotatebox{0}{\includegraphics[width=0.19\linewidth]{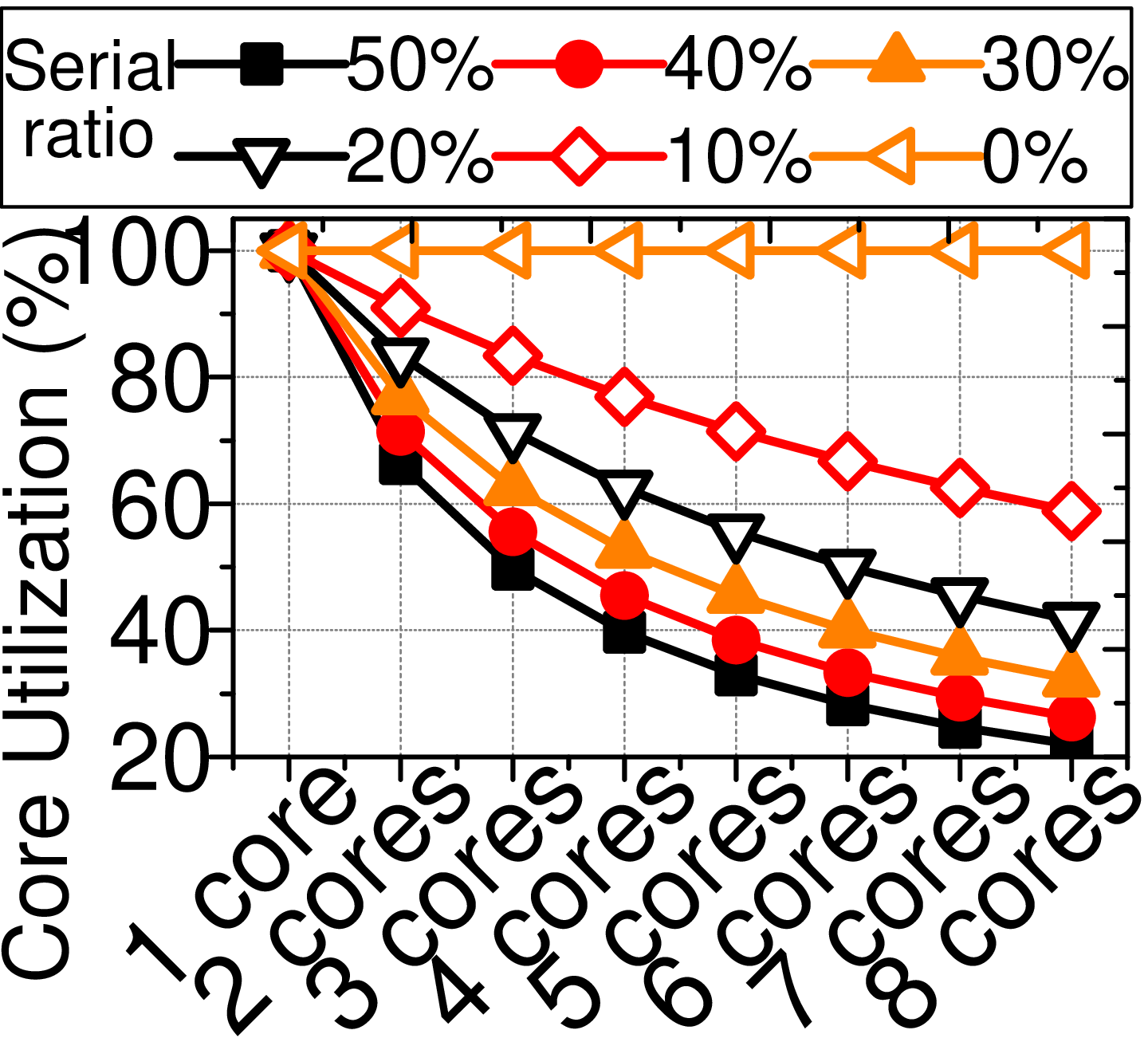}}} %
\hspace{1pt}
\subfloat[Latency breakdown.]{\label{fig:moti1}\rotatebox{0}{\includegraphics[width=0.19\linewidth]{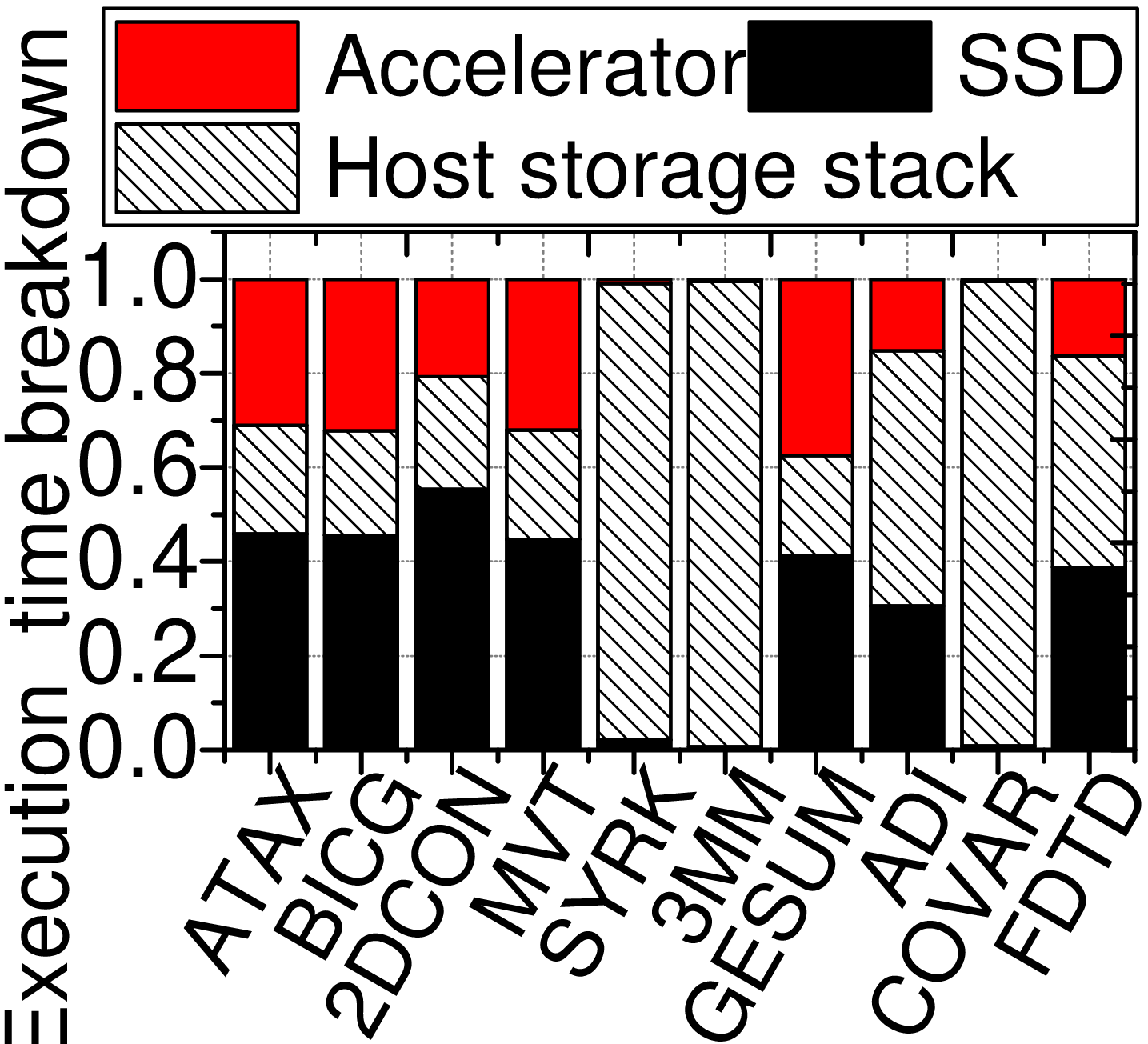}}} %
%\hspace{1pt}
\subfloat[Energy breakdown.]{\label{fig:moti2}\rotatebox{0}{\includegraphics[width=0.19\linewidth]{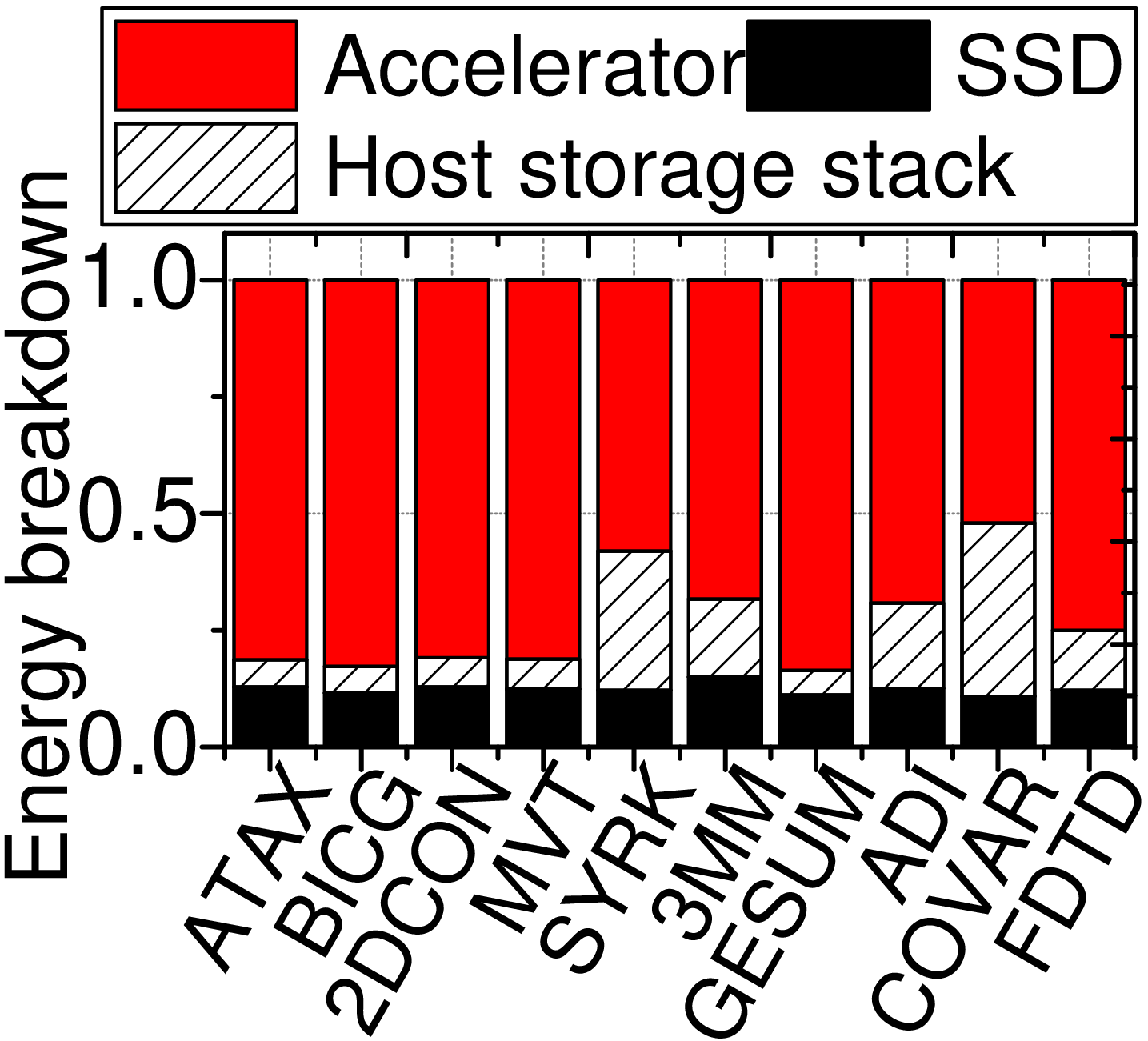}}} %

%\vspace{-5pt}
\caption{\label{fig:rdwr}Performance bottleneck analysis of low-power heterogeneous computing.}
\end{figure*}

\begin{table}
%\vspace{-5pt}
\centering
\resizebox{\linewidth}{!}{
\begin{tabular}{|c|c|c|c|c|}
\hline
{\bf Components}                                            & {\bf Specification}                                                     & {\bf \begin{tabular}[c]{@{}c@{}}Working\\   frequency\end{tabular} } & {\bf \begin{tabular}[c]{@{}c@{}}\newedit{Typical}\\   \newedit{power}\end{tabular}  } & {\bf \begin{tabular}[c]{@{}c@{}}Est.\\   B/W\end{tabular}} \\ 
\hline 

\hline
LWP                                                    & 8 processors                                                           & 1GHz & \newedit{0.8W/core}             & 16GB/s                                                                    \\ \hline
L1/L2 cache & 64KB/512KB & 500MHz                  & \newedit{N/A}             & 16GB/s                                                              \\ \hline
Scratchpad                                                         & 4MB                                                             & 500MHz                  & \newedit{N/A}             & 16GB/s                                                                     \\ \hline
Memory                                                        & DDR3L, 1GB                                                          & 800MHz                 & \newedit{0.7W}              & 6.4GB/s                                                                     \\ \hline
SSD                                                         & 16 dies, 32GB   & 200MHz                  & \newedit{11W}              & 3.2GB/s                                                                      \\ \hline
PCIe                                                        & v2.0, 2 lanes                                                    & 5GHz                    & \newedit{0.17W}               & 1GB/s                                                                      \\ \hline
Tier-1 crossbar & 256 lanes                                                               & 500MHz                  & \newedit{N/A}             & 16GB/s                                                                     \\ \hline
Tier-2 crossbar & 128 lanes                                                               & 333MHz                  & \newedit{N/A}             & 5.2GB/s                                                                      \\ \hline

\end{tabular}}
\vspace{5pt}
\caption{Hardware specification of our baseline.}
\vspace{5pt}
\label{tab:TI}
\end{table}

%%%%%%%%%%%%%%%%%%%%%%%%%%%%%%%%%
% should be revivied -- removed for SOSP
\noindent \textbf{Network organization.} Our hardware platform uses a partial crossbar switch \cite{sharma1996crossbar} that separates a large network into two sets of crossbar configuration: i) a streaming crossbar (\emph{tier-1}) and ii) multiple simplified-crossbars (\emph{tier-2}). The tier-1 crossbar is designed towards high performance (thereby integrating multiple LWPs with memory modules), whereas throughputs of the tier-2 network are sufficient for the performances of Advanced Mezzanine Card (AMC) and PCIe interfaces exhibit \cite{TIperformance}. 
%Specifically, tier-2 network limits the number of lanes and clock frequency of target devices to reduce the network's energy consumption and area overhead. In contrast, tier-1 network uses a point-to-point connection (thereby eliminating unnecessary bus arbitrations \cite{kang2000bus}) by employing the maximum number of lanes and highest clock frequency for each crossbar. 
These two crossbars are connected over multiple network switches \cite{teranet}. 

\ignore{
\noindent \textbf{Network organization.} Our hardware platform uses a partial crossbar switch \cite{sharma1996crossbar} that separates a large network into two sets of crossbar configuration: i) a streaming crossbar (\emph{tier-1}) and ii) multiple simplified-crossbars (\emph{tier-2}). The tier-1 crossbar is designed towards high performance (thereby integrating multiple LWPs with memory modules), whereas throughputs of tier-2 network are just enough to accept the performance that Advanced Mezzanine Card (AMC) and PCIe interfaces exhibit \cite{TIperformance}. Specifically, tier-2 network limits the number of lanes and clock frequency of target devices to reduce the network's energy consumption and area overhead. In contrast, tier-1 network uses a point-to-point connection (thereby eliminating unnecessary bus arbitrations \cite{kang2000bus}) by employing the maximum number of lanes and highest clock frequency for each crossbar. These two crossbars are connected together over multiple network switches \cite{TI6678}. 
}

\noindent \textbf{Flash backbone.} Tier-2 network's AMC is also connected to the FPGA Mezzanine Card (FMC) of the backend storage complex through four Serial RapidIO (SRIO) lanes (5Gbps/lane).  In this work, the backend storage complex is referred to as \emph{flash backbone}, which has four flash channels, each employing four flash packages over NV-DDR2 \cite{semiconductor2006open}. We introduce a FPGA-based flash controller for each channel, which converts the I/O requests from the processor network into the flash clock domain. To this end, our flash controller implements inbound and outbound ``tag'' queues, each of which is used for buffering the requests with minimum overheads. During the transition device domain, the controllers can handle flash transactions and transfer the corresponding data from the network to flash media through the SRIO lanes, which can minimize roles of flash firmware.

\newedit{It should be noted that the backend storage is a self-existence module, separated from the computation complex via FMC in our baseline architecture. Since flash backbone is designed as a separable component, ones can simply replace worn-out flash packages with new flash (if it needs).}

\noindent \textbf{Prototype specification.}
Eight LWPs of our FlashAbacus operate with a 1GHz clock and each LWP has its own private 64KB L1 cache and 512KB L2 cache. The size of the 8-bank scratchpad is 4MB; DDR3L also consists of eight banks with a total size of 1GB. On the other hand, the flash backbone connects 16 triple-level cell (TLC) flash packages \cite{TLC}, each of which has two flash dies therein (32GB). Each of the four flash packages is connected to one of the four NV-DDR2 channels that work on ONFi 3.0 \cite{semiconductor2006open}. The 8KB page read and write latency are around 81 us and 2.6 ms, respectively. Lastly, the flash controllers are built on the 2 million system logic cells of Virtex Ultrascale FGPA \cite{virtex}. The important characteristics of our prototype are presented in Table \ref{tab:TI}.

\section{High-level View of Self-Governing}
\label{sec:overview}
Figure \ref{fig:program-model} illustrates a kernel execution model for conventional hardware acceleration. In prologue, a data processing application needs to open a file and allocate memory resources for both an SSD and an accelerator.  Its body iterates the code segments that read a part of file, transfer it to the accelerator, execute a kernel, get results from the accelerator, and write them back to the SSD. Once the execution of body loop is completed, the application concludes by releasing all the file and memory resources. 
This model operates well with traditional manycore-based high-performance accelerators that have thousands of hardware threads. 
\newedit{However, in contrast to such traditional accelerators, the memory space of low-power accelerators is unfortunately limited and difficult to accommodate all data sets that an application requires to process. Thus, low-power accelerators demand more iterations of data transfers, which in turn can significantly increase I/O stack overheads.}
%However, it is not suitable for low-power accelerator design with fewer processing units but many irregular data accesses. As the file accesses of an application require going through different software runtimes (and stacks), significant overheads are introduced before the actual computation starts, thereby seriously degrading system performance. 
\newedit{Furthermore, a small memory size of the low-power accelerators can enforce a single data-processing task split into multiple functions and kernels, which can only be executed by the target accelerator in a serial order.} 

\begin{figure*}
\centering
%\vspace{-5pt}
\def\subfigcapskip{0pt}
\subfloat[Multiple kernels per app.]{\label{fig:MKPA}\rotatebox{0}{\includegraphics[width=0.4\linewidth]{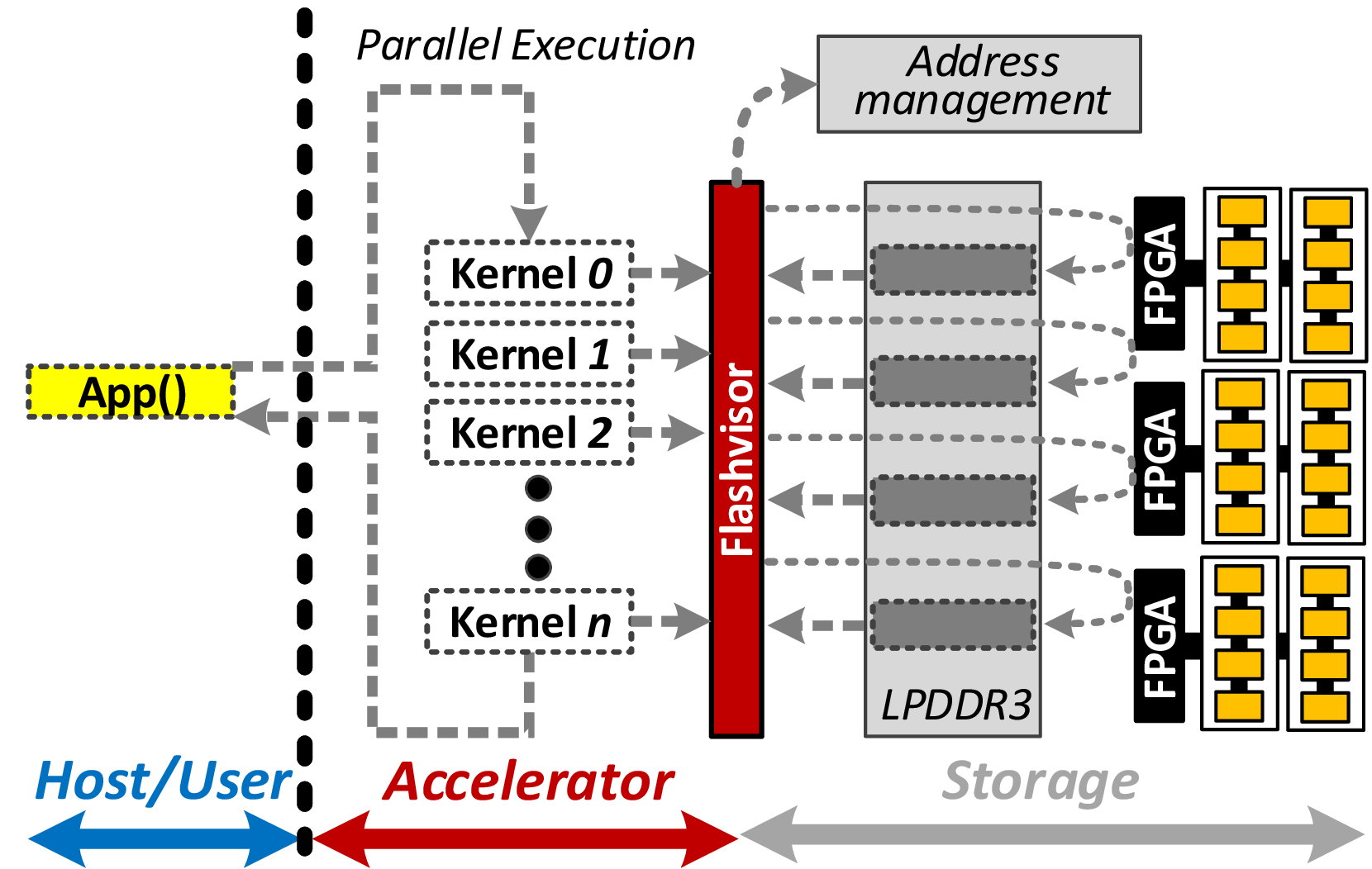}}}
\hspace{4pt}
\subfloat[Multiple applications and kernels.]{\label{fig:MAAK}\rotatebox{0}{\includegraphics[width=0.54\linewidth]{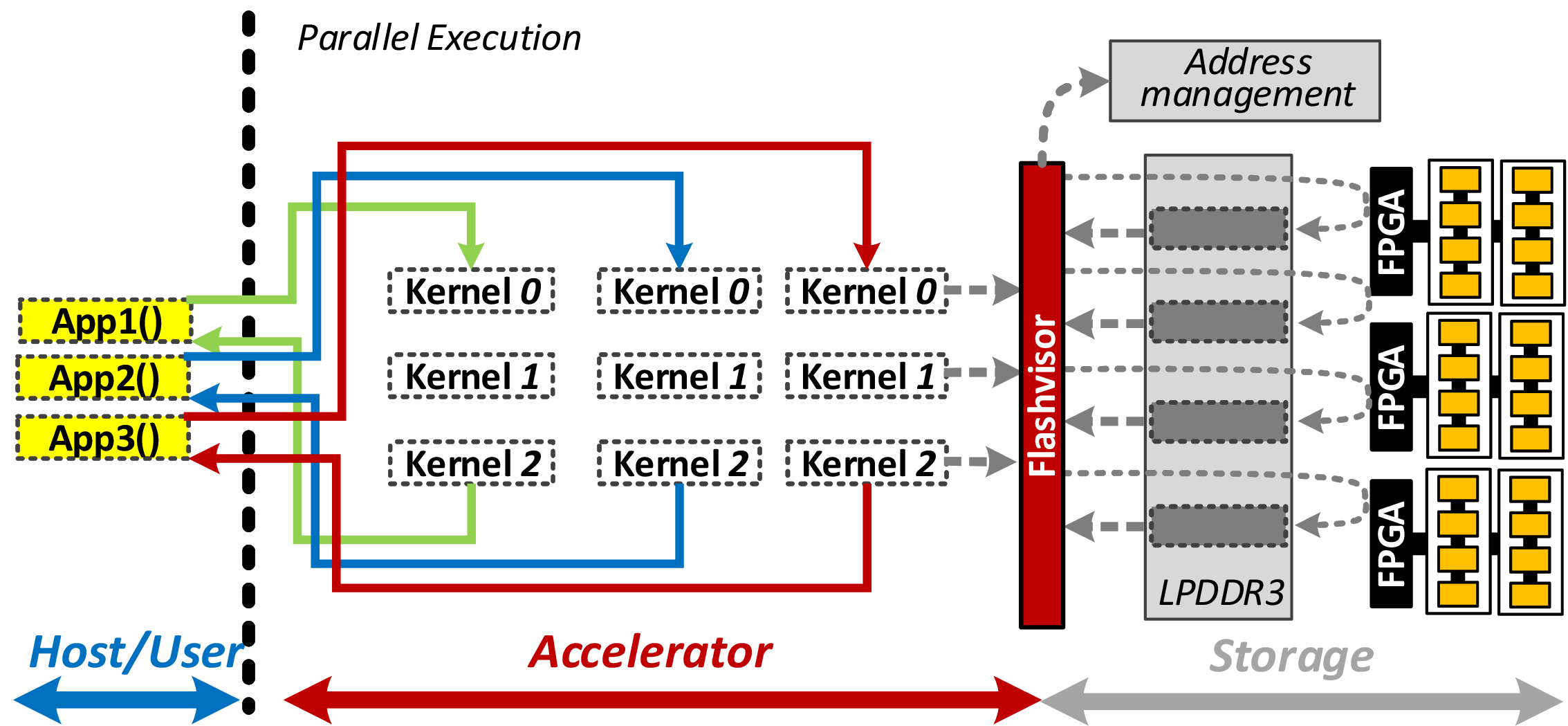}}}
%\vspace{-5pt}
\caption{\label{fig:}Multi-kernel execution. \vspace{-5pt}}
\end{figure*}

\subsection{Challenge Analysis}
For better insights on the aforementioned challenges, we built a heterogeneous system that employs the low-power accelerator described in Section \ref{sec:prototype} and an NVMe SSD \cite{NVMeSSD} as external storage instead of our flash backbone. 
%a low-power accelerator that employs our baseline architecture, exculding our flash backbone, and a high performance SSD \cite{NVMeSSD}. 

\noindent \textbf{Utilization.} Figures \ref{fig:through-moti} and \ref{fig:util-moti} show the results of our performance and utilization sensitivity studies in which the fraction of serial parts in kernel executions enforced by data transfers were varied. As the serial parts of a program increase, the throughput of data-processing significantly decreases (Figure \ref{fig:through-moti}). For example, if there exists a kernel whose fraction of serialized executions is 30\%, the performance degrades by 44\%, on average, compared to the ideal case (e.g., 0\%), thus the accelerator becomes non-scalable and the full benefits of heterogeneous computing are not achieved. Poor CPU utilization is the reason behind performance degradation (Figure \ref{fig:util-moti}). For the previous example, processor utilization is less than 46\%; even in cases where the serial parts take account for only 10\% of the total kernel executions, the utilization can be less than at most 59\%.   

\ignore{
-- in revision
We perform performance and utilization sensitivity studies by varying the fraction of serial parts in kernel executions enforced by data transfers. In this evaluation, we use OpenMP \cite{mitra2014implementation} to exploit SIMD model parallelism within the accelerator. 
As shown in Figure \ref{fig:through-moti}, as the one increases the serial parts of data processing bandwidth significantly degrade. If there exists a kernel whose faction of serialized executions is 30\%, the system performance degrades 44\%, on average, compared to the ideal case (e.g., 0\%). These kernel executions make the accelerator non-scalable and difficult to take the full benefits of heterogeneous computing. The reason behind this performance degradation is poor processor utilization as shown in Figure \ref{fig:util-moti}. In the previous performance example, the processor utilization is lower than 46\%, and even in cases where the serial parts of a program take account only 10\% of total kernel executions, the utilization can be lower than at most 59\%.  
}

\noindent \textbf{Storage accesses.} We also evaluate the heterogeneous system by performing a collection of PolyBench \cite{pouchet2012polybench} and decompose the execution time of each application into i) SSD latency to load/store input/output data, ii) CPU latency that host storage stack takes to transfer the data, and iii) accelerator latency to process the data\footnote{\newedit{In this analysis, our accelerator latency includes the execution time that overlaps with the corresponding data transfer (DMA) time; the details of evaluation method and workload characteristics are described in Section \ref{sec:result} and Table \ref{tbl:singlepro}, respectively}.}. Figure \ref{fig:moti1} shows that the data-intensive applications (e.g., \emph{ATAX, BICG, and MVT}) consume 77\% of the total execution time to transfer the data between the the accelerator and SSD.
%In Figure \ref{fig:moti2}, \newedit{the corresponding energy consumption behavior, which is profiled by using Intel VTune amplifier \cite{marowka2011performance} and our in-house power analyzer}.
Figure \ref{fig:moti2} illustrates that the storage stack accesses, including file system operations and I/O services, consume 85\% of the total energy in heterogeneous computing. 
Note that the data transfer overheads for computation-intensive applications (e.g., \emph{SYRK} and  \emph{3MM}) are not remarkable, but the corresponding energy consumed by the storage stack accounts for more than 77\% of the total energy consumed by the system, on average. \newedit{The detailed energy analysis of heterogeneous computing will be further discussed in Section \ref{sec:energy}.}

To address the above two challenges, FlashAbacus governs the internal hardware resources without an assistance of host or OS. We introduce a multi-kernel execution model and storage monitor to fully utilize LWPs and virtualize flash backbone into processors' memory address, respectively. 
%\newedit{To address the above two challenges, we propose FlashAbacus to govern the internal hardware resources without an assistance of host or OS. The main ideas of this paper can be described as follows: a multi-kernel execution model which can fully utilize LWPs and a storage monitor which can virtualize flash backbone into processors' memory address.}

%we introduce a parallel execution model to fully utilize LWPs and a technique to virtualize flash backbone into processors' memory address. 

%Motivated by these observations, our design goals of FlashAbacus are i) to remove the overheads imposed by multiple physical and logical boundaries between the SSD and the accelerator and ii) to fully parallelize the kernel executions with a small number of low-power processors. To this end, we introduce a different kernel execution model and simple flash management technique, which has no host-side software intervention. 

\subsection{Multi-Kernel Execution} 
%Figure \ref{} shows a high level view of FlashAbacus's programming and execution models. While a conventional kernel is a function that usually have a very simple iteration, our kernel can contain many functions and handle a deep depth of function call.  In addition, user can offload a set of kernels and execute them in parallel, which is referred to as \emph{multi-kernel} execution. 
Hand-threaded parallelism (using pthread or message passing interface) can offer fine control than the parallel execution, but for such parallelism, it requires to accommodate OS thread management, which is infeasible to our low-power accelerator, have to be accommodated. In this work, we allow all the LWPs to execute different types of kernels in parallel, and each kernel can contain various operational functions. This in turn enables users to offload diverse user applications and perform different types of data processing in concert; this is referred to as \emph{multi-kernel} execution. Figure \ref{fig:MKPA} shows an example of the multi-kernel execution model. While a conventional kernel is a function that usually has a very simple iteration, our kernel can contain many functions and handle a deep depth of function calls. In this execution model, a host can also offload multiple kernels, which are associated with different applications in our accelerator (Figure \ref{fig:MAAK}). While our multi-kernel execution is not as powerful as a thousand hardware thread executions that most manycore accelerators offer, it allows users to perform more flexible data processing near flash and opens up the opportunities to make data processing more energy efficient than in the conventional accelerators.

However, executing different kernels, each with many functions across multiple LWPs, can introduce other technical challenges such as load balancing and resource contention. 
%poor LWP utilization and unbalanced workloads. % resource utilization and load balancing.% on a wide spectrum of NDP workloads. 
To address these challenges, one can simply expose all internal LWP's resources to the host so that users can finely control everything on their own. Unfortunately, this design choice can lead to a serious security problem, as an unauthorized user can access the internal resources and put them to an improper use. This approach may also introduce another type of data movement overheads as frequent communications are required to use diverse FlashAbacus resources from outside.
Therefore, our accelerator internally governs all the kernels based on two different scheduling models: i) \emph{inter-kernel execution} and ii) \emph{intra-kernel execution}.
In general, in inter-kernel executions, each LWP is dedicated to execute a specific kernel that performs data processing from the beginning to the end as a single instruction stream. In contrast, the intra-kernel execution splits a kernel into multiple code blocks and concurrently executes them across multiple LWPs based on the input data layout. The scheduling details will be explained in Section \ref{sec:method}.

\subsection{Fusing Flash into a Multicore System}
%The reason why multi-kernel executions are feasible is that each LWP can execute a different instruction stream like SMP. 
%Even though each LWP of our hardware platform can execute a different instruction stream (kernel), it is infeasible to accommodate an operating system or a file system.  
The lack of file and runtime systems introduces several technical challenges to multi-kernel execution, including memory space management, I/O management, and resource protection. An easy-to-implement mechanism to address such issues is to read and write data on flash through a set of customized interfaces that the flash firmware may offer; this is the typically adopted mechanism in most active SSD approaches \cite{jun2015bluedbm, seshadri2014willow}. Unfortunately, this approach is inadequate for our low-power accelerator platform. Specifically, as the instruction streams (kernels) are independent of each other, they cannot dynamically be linked with flash firmware interfaces.
Furthermore, for the active SSD approaches, all existing user applications must be modified by considering the flash interfaces, leading to an inflexible execution model. 

Instead of allowing multiple kernels to access the flash firmware directly through a set of static firmware interfaces, we allocate an LWP to govern the memory space of the data section of each LWP by considering flash address spaces. This component, referred to as \emph{Flashvisor}, manages the logical and physical address spaces of the flash backbone by grouping multiple physical pages across different dies and channels, and it maps the logical addresses to the memory of the data section. Note that all the mapping information is stored in the scratchpad, while the data associated with each kernel's data sections are placed into DDR3L. In addition, Flashvisor isolates and protects the physical address space of flash backbone from the execution of multiple kernels.  
Whenever the kernel loaded to a specific LWP requires accessing its data section, it can inform Flashvisor about the logical address space where the target data exist by passing a message to Flashvisor. Flashvisor then checks a permission of such accesses and translates them to physical flash address. Lastly, Flashvisor issues the requests to the underling flash backbone, and the FPGA controllers bring the data to DDR3L.
\newedit{In this flash virtualization, most time-consuming tasks such as garbage collection or memory dump are periodically performed by a different LWP, which can address potential overheads brought by the flash management of Flashvisor (cf. Section \ref{sec:virtualization}). Note that, in contrast to a conventional virtual memory that requires paging over file system(s) and OS kernel memory module(s), our Flashvisor internally virtualizes the underlying flash backbone to offer a large-size of byte-addressable storage without any system software support.} The implementation details of flash virtualization will be explained in Section \ref{sec:virtualization}.

\ignore{
The master LWP employs \emph{flash diffusion machine} (FDM), which maintains all page table entries (PTEs) used for virtualizing the memory address space by combining DDRL3 of NBS and flash backbone addresses. This FDM also maintains the permission information to protect memory from multiple worker LWPs, which will be further described in Section \ref{}. 
In cases where FDMs need to bring the data from flash, the master LWP send the request the flash LWP that implements FTL. All the communications related to this flash management among multiple LWPs can be done through a message queue subsystem referred to as \emph{q-subsys}.

Even though the endpoint bridges of flash backbone can control low-level flash transactions, there are two more functionalities we need to incorporate: i) flash translation layer and ii) flash virtualization. While the firmware management is related to FTL and HIL implementations, flash virtualization is related to support the byte-addressability and memory protection on the flash backbone. In our architecture, all worker LWPs need to send memory request to the master LWP and ask permission to access flash. The master LWP employs \emph{flash diffusion machine} (FDM), which maintains all page table entries (PTEs) used for virtualizing the memory address space by combining DDRL3 of NBS and flash backbone addresses. This FDM also maintains the permission information to protect memory from multiple worker LWPs, which will be further described in Section \ref{}. In cases where FDMs need to bring the data from flash, the master LWP send the request the flash LWP that implements FTL. All the communications related to this flash management among multiple LWPs can be done through a message queue subsystem referred to as \emph{q-subsys}. The q-subsys implements a simple message buffer mechanism on the scratchpad in NBS and offers general queue interfaces such as \texttt{create()}, \texttt{open()}, \texttt{alloc\_msg()}, \texttt{put()}, \texttt{delete\_msg()}, and \texttt{delete()}, which creates a queue, opens a queue, allocates a message, sends a message to target, deletes a message and delete a queue, respectively. At the end of the q-subsys, flash LWP implements conventional flash translation layer. 
This FTL can see the entire flash modules and the corresponding physical address space in flash backbone, which allows flash LWP can efficiently parallelize data access and hide the underlying flash management complexity on behalf of HIL.   
Note that, as we separate the flash management from NDP-kernel execution, all SSD stack's functionalities will be not interfered by others. 
 }

\ignore{
\noindent \textbf{Accessibility.}
In our architecture, all internal components including each LWP cache, 
The memory address space of our proposed FlashAbacus is organized as \emph{uniformed memory access} (UMA) on the SMP-like architecture and multi-tiered network. As shown in Figure \ref{fig:UMA}, the UMA address layout allows all the DSP cores can access L2 caches of all DSP cores, SPM, and the network buffer memory.  
This UMA i) enables the master core to handle workers' L2 cache to manage NDP-kernel image, ii) retrieves message information form SPM without data transfer among the multiple DSP cores, and iii) allows each DSP worker to manage their own network buffer space.  
Even though this UMA is visible for all other DSP cores, it does not include the address space of storage backbone. This is because, unlike other memory components, flash memory accesses require address translation and other flash-specific managements. 
}

\section{Implementation Details}
\label{sec:method}
%In this section, we will introduce details of multi-kernel execution and scheduling procedure of FlashAbacus. We will then explain the flash virtualization, including memory protection, access control, and storage managements. 

%\subsection{Data Processing Controls}

%A data intensive application can offload its kernel(s) to the internal shared memory through PCIe interface as a form of executable object \cite{ti-coff}.  In our framework, each kernel can be compiled by either OpenCL \cite{} or a code generation tool \cite{}, and the compiled kernel is represented by a \emph{kernel description table}. The description table includes an executable that contains several sections like kernel code (\texttt{.text}), data section (\texttt{.ddr3\_arr}), heap (\texttt{.heap}), and stack (\texttt{.stack}), which is a variation of executable and linkable format (ELF) \cite{tool2001executable}. 

\noindent \textbf{Kernel.} The kernels are represented by an executable object \cite{COFF}, referred to as \emph{kernel description table}. The description table, which is a variation of the executable and linkable format (ELF) \cite{tool2001executable}, includes an executable that contains several types of section information such as the kernel code (\texttt{.text}), data section (\texttt{.ddr3\_arr}), heap (\texttt{.heap}), and stack (\texttt{.stack}). In our implementation, all the addresses of such sections point to the L2 cache of each LWP, except for the data section, which is managed by Flashvisor.

%\begin{figure}
%	\centering	
%	\includegraphics[width=1\linewidth]{./figs/intrakernel_execution.eps}
%	\vspace{-5pt}
%	\caption{Microblocks/screens in an NDP-kernel.}
%	\vspace{-5pt}
%	\label{fig:intrakernel_execution}
%\end{figure}

%\begin{figure*}
%	\centering	
%	\includegraphics[width=0.9\linewidth]{./figs/scheduling_examp2.eps}
%	\vspace{-7pt}
%	\caption{Examples of intra-kernel scheduling.}
%	\vspace{-5pt}
%	\label{fig:scheduling_examp2}
%\end{figure*}

\noindent \textbf{Offload.} A user application can have one or more kernels, which can be offloaded from a host to a designated memory space of DDR3L through PCIe. The host can write the kernel description table associated with the target kernel to a PCIe \emph{base address register} (BAR), which is mapped to DDR3L by the PCIe controller (cf. Figure \ref{fig:hw_org}).  
%At the beginning of host communication, PHY handles all PCIe timing requirements and passes the incoming request to the PCIe core. The PCIe core then parses/marshals data and forwards them to NBS or IPC-IR of the master LWP based on which BAR that the host indicates -- as shown in the figure, the addresses of NBS and IPC-IR are identically mapped to BAR1 and BAR2, respectively.
 
\noindent \textbf{Execution.}  After the completion of the kernel download(s), the host issues a PCIe interrupt to the PCIe controller, and then the controller internally forwards the interrupt to Flashvisor. Flashvisor puts the target LWP (which will execute the kernel) in sleep mode through power/sleep controller (PSC) and stores DDR3L address of such a downloaded kernel to a special register, called \textit{boot address register} of the target LWP. Flashvisor then writes an inter-process interrupt register of the target LWP, forcing this LWP to jump to the address written in the boot address register. Lastly, Flashvisor pulls the target LWP out of the sleep mode through PSC. Once this revocation process is completed, the target LWP begins to load and execute the specified kernel. Thus, Flashvisor can decide the order of kernel executions within an LWP across all LWPs.

\begin{figure}
	\centering	
	\includegraphics[width=1\linewidth]{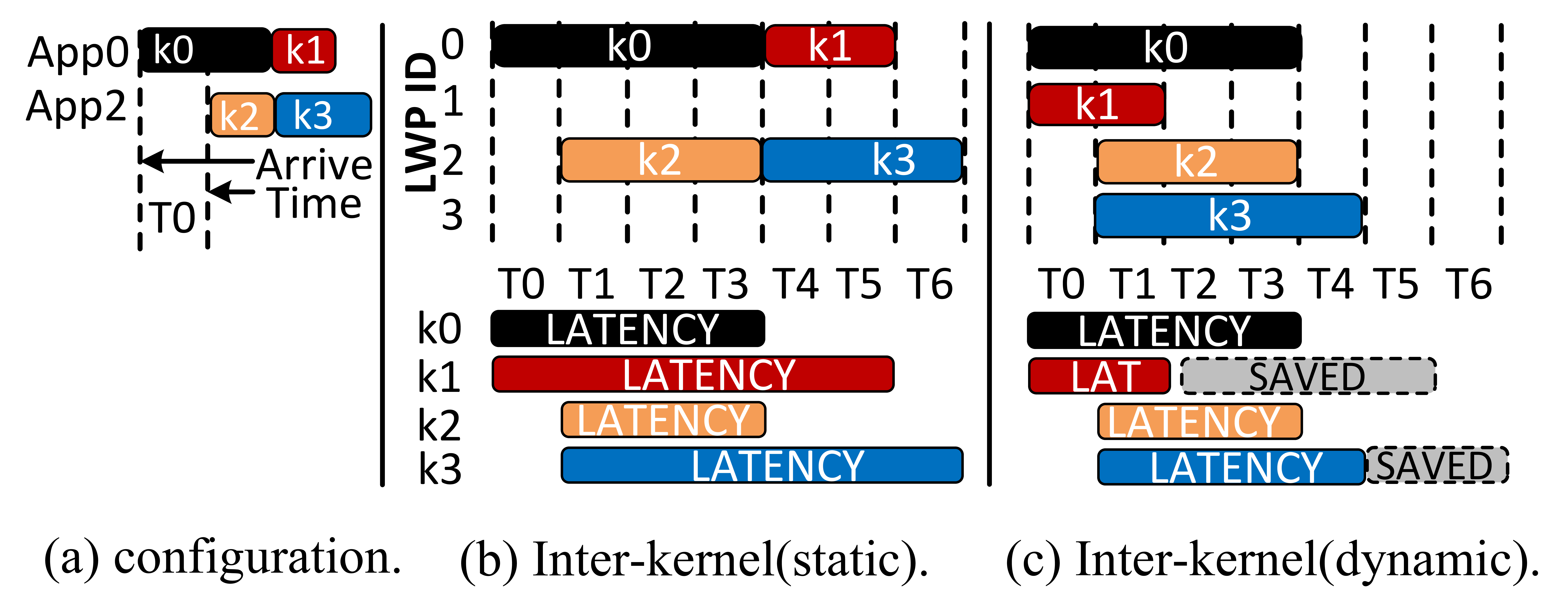}
	%\vspace{-15pt}
	\caption{Examples of inter-kernel scheduling.}
	%\vspace{-5pt}
	\label{fig:test}
\end{figure}

\begin{figure}
	\centering
	%\vspace{-10pt}	
	\includegraphics[width=1\linewidth]{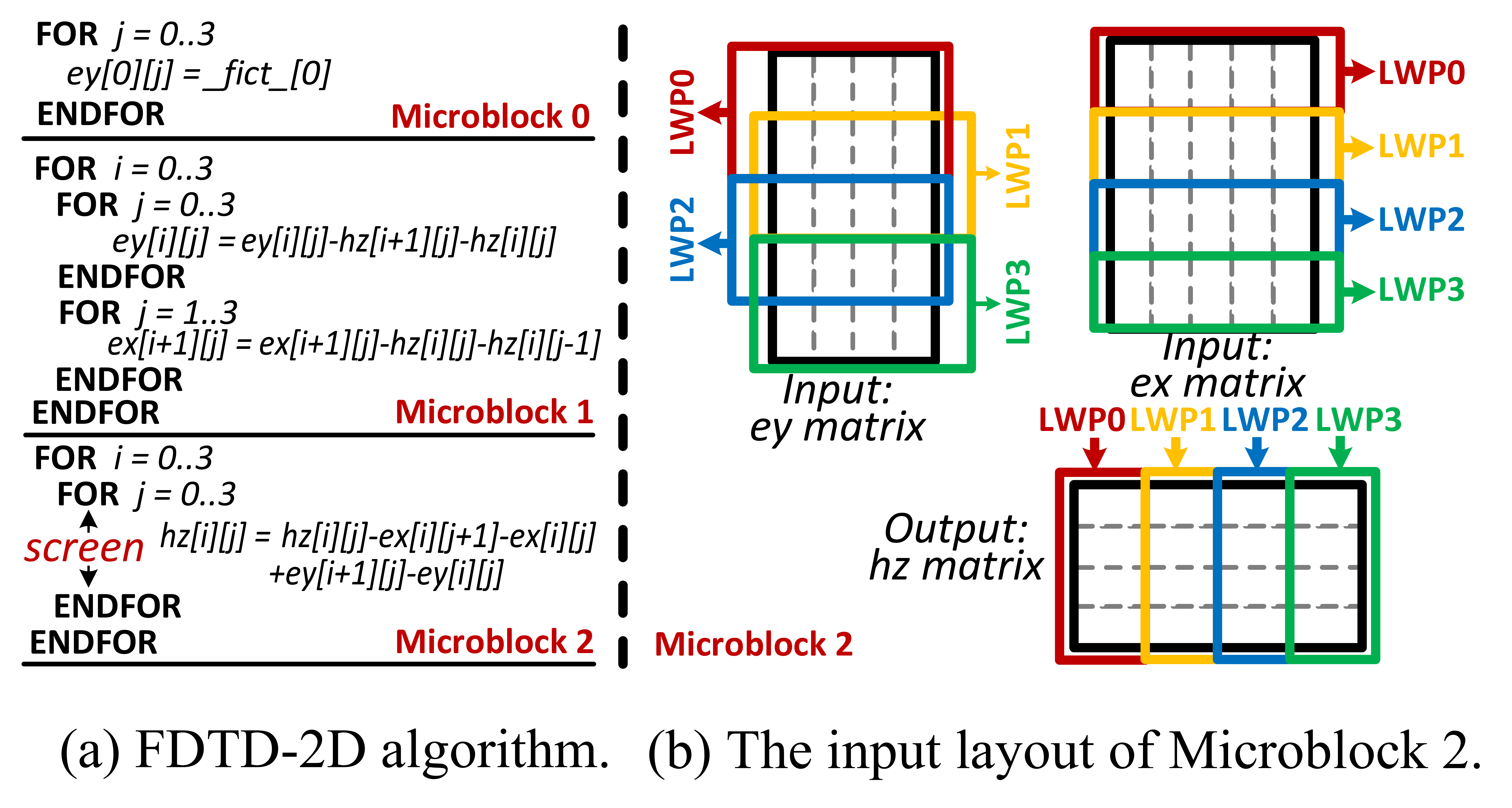}
	%\vspace{-15pt}
	\caption{Microblocks/screens in an NDP-kernel.}
	%\vspace{-5pt}
	\label{fig:microblock}
\end{figure}

%\subsection{Data Processing Controls}

\subsection{Inter-kernel Execution}

\noindent \textbf{Static inter-kernel scheduling.}
The simplest method to execute heterogeneous kernels across multiple LWPs is to allocate each incoming kernel statically to a specific LWP based on the corresponding application number. Figures \ref{fig:test}a shows that has two user applications, \texttt{App0} and \texttt{App2}, each of which contains two kernels, respectively (e.g., \texttt{k0/k1} for \texttt{App0} and \texttt{k2/k3} for \texttt{App2}). In this example, a static scheduler assigns all kernels associated with \texttt{App0} and \texttt{App2} to LWP0 and LWP2, respectively. Figure \ref{fig:test}b shows the timing diagram of each LWP execution (top) and the corresponding latency of each kernel (bottom). Once a host issues all the kernels in \texttt{App0} and \texttt{App2}, it does not require any further communication with the host until all computations are completed. Even though this static inter-kernel scheduling is easy to implement and manage in the multi-kernel execution model, such scheduling can unfortunately lead to a poor resource utilization due to the imbalance of kernel loads.  For example,  while LWP1 and LWP3 in idle, \texttt{k1} and \texttt{k3} should be suspended until the previously-issued \texttt{k0} and \texttt{k2} are completed.

\noindent \textbf{Dynamic inter-kernel scheduling.}
To address the poor utilization issue behind static scheduling, Flashvisor can dynamically allocate and distribute different kernels among LWPs. If a new application has arrived, this scheduler assigns the corresponding kernels to any available LWPs in a round robin fashion. As shown in Figure \ref{fig:test}c, \texttt{k1} and \texttt{k3} are allocated to LWP1 and LWP3, and they are executed in parallel with \texttt{k0} and \texttt{k2}. Therefore, the latency of \texttt{k1} and \texttt{k3} are reduced as compared to the case of the static scheduler, by 2 and 3 time units, respectively. 
Since each LWP informs the completion of kernel execution to Flashvisor through the hardware queue (cf. Figure \ref{fig:overalloverview}), Flashvisor can consecutively allocate next kernel to the target LWP. Therefore, this dynamic scheduler can fully utilize LWPs if it can secure sufficient kernel execution requests. However, there is still room to reduce the latency of each kernel while keeping all LWPs busy, if we can realize finer scheduling of kernels. 

%After that, whenever a worker LWP signals to the master LWP through IPC-IR (by completing an instance execution), the master LWP allocates the next available NDP-kernel instance to the work LWP back-to-back. As shown in Figure \ref{fig:scheduling_examp1}c, this dynamic scheduler is better than the static one since, in the same scenario, i1 and i3 can be served with i0 and i2 in parallel -- it saves two time slots for entire executions. 

%\begin{figure}
%	\centering	
%	\includegraphics[width=0.9\linewidth]{./figs/chain.eps}
%	\vspace{-5pt}
%	\caption{Data structure of multi-app execution chain.}
%	\vspace{-5pt}
%	\label{fig:chain}
%\end{figure}

%\begin{figure*}
%	\centering	
%	\includegraphics[width=1\linewidth]{./figs/test2.eps}
%	\vspace{-20pt}
%	\caption{Examples of intra-kernel scheduling and data structure of multi-app execution chain.}
%	\vspace{-5pt}
%	\label{fig:test2}
%\end{figure*}

\subsection{Intra-kernel Execution}

\noindent \textbf{Microblocks and screens.}
In FlashAbacus, a kernel is composed of multiple groups of code segments, wherein the execution of each depends on their input/output data. We refer to such groups as \emph{microblocks}. While the execution of different microblocks should be serialized, in several operations, different parts of the input vector can be processed in parallel. We call these operations (within a microblock) as \emph{screens}, which can be executed across different LWPs. 
For example, Figure \ref{fig:microblock}a shows the multiple microblocks observed in the kernel of (\emph{FDTD-2D}) in Yee's method \cite{pouchet2012polybench}. The goal of this kernel is to obtain the final output matrix, \texttt{hz}, by processing the input vector, \texttt{\_fict\_}.  Specifically, in microblock 0 (\texttt{m0}), this kernel first converts \texttt{\_fict\_} (1D array) to \texttt{ey} (2D array). The kernel then prepares new \texttt{ey} and \texttt{ex} vectors by calculating \texttt{ey}/\texttt{hz} and \texttt{ex}/\texttt{hz} differentials in microblock 1 (\texttt{m1}). These temporary vectors are used for getting the final output  \texttt{hz} at microblock 2 (\texttt{m2}). In \texttt{m2}, the execution codes per (inner loop) iteration generate one element of output vector, \texttt{hz}, at one time. Since there are no risks of write-after-write or read-after-write in \texttt{m2}, we can split the outer loop of \texttt{m2} into four screens and allocate them across different LWPs for parallel executions. Figure \ref{fig:microblock}b shows the input and output vector splitting and their mapping to the data sections of four LWPs to execute each screen.
%In contrast to m0, those two operations in m1 have no data dependency each other (within the microblock), such that the different loops (e.g., screens) can executed simultaneously. Similarly, in m2, while input matrix as shown in Figure \ref{fig:intrakernel_execution}b, we can split the outer loop of m2 into four screens and allocate them across four LWPs.

\begin{figure}
	\centering	
	\includegraphics[width=1\linewidth]{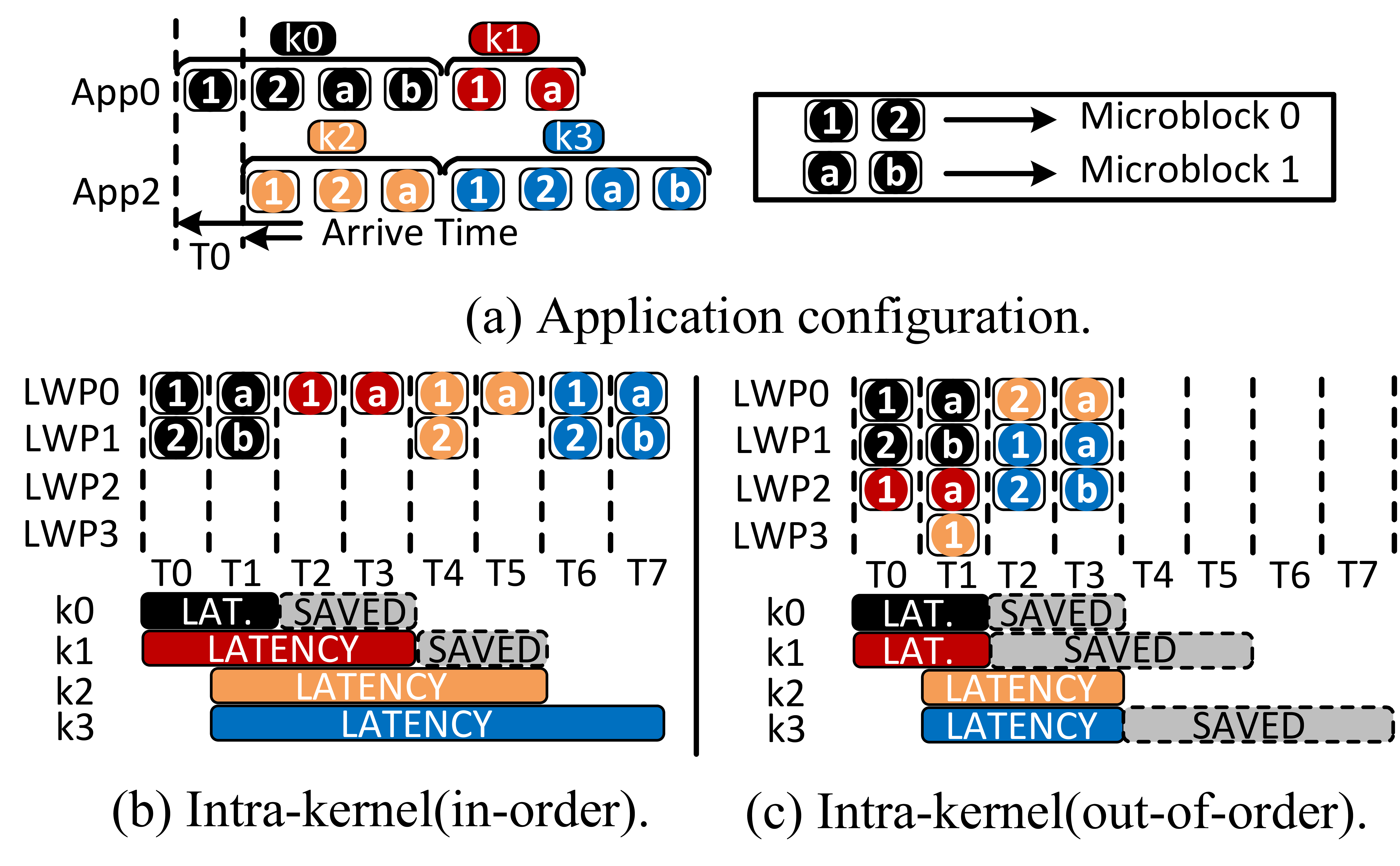}
	%\vspace{-15pt}
	\caption{Examples of intra-kernel scheduling.}
	%\vspace{-5pt}
	\label{fig:test2}
\end{figure}

\noindent \textbf{In-order intra-kernel scheduling.}
%The insight behind intra-kernel execution is that, while these microblocks should be executed in a certain order, within a microblock, there are several operations, called \emph{screens}, which can work in parallel on different parts of I/O matrices. 
This scheduler can simply assign various microblocks in a serial order, and simultaneously execute the numerous screens within a microblock by distributing them across multiple LWPs.
Figure \ref{fig:test2}b shows an example of an in-order inter-kernel scheduler with the same scenario as that explained for Figure \ref{fig:test}b. 
As shown in Figure \ref{fig:test2}a, \texttt{k0}'s \texttt{m0} contains two screens (\blkcircled{\small{1}} and \blkcircled{\small{2}}), which are concurrently executed by different LWPs (LWP0 and LWP1). This scheduling can reduce 50\% of \texttt{k0}'s latency compared to the static inter-kernel scheduler. Note that this scheduling method can shorten the individual latency for each kernel by incorporating parallelism at the screen-level, but it may increase the total execution time if the data size is not sufficiently large to partition as many screens as the number of available LWPs.

\begin{figure}
	\centering
	%\vspace{-10pt}	
	\includegraphics[width=1\linewidth]{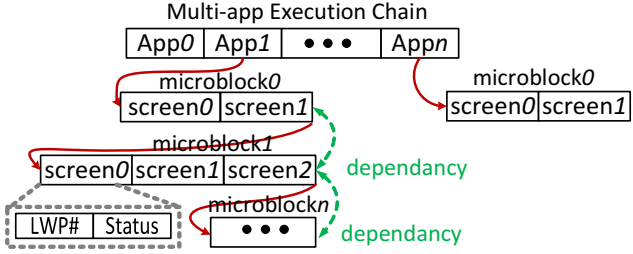}
	%\vspace{-15pt}
	\caption{Data structure of multi-app execution chain.}
	%\vspace{-5pt}
	\label{fig:chain}
\end{figure}

\begin{figure*}
\centering

\subfloat[Reads.]{\label{fig:reads}\rotatebox{0}{\includegraphics[width=0.45\linewidth]{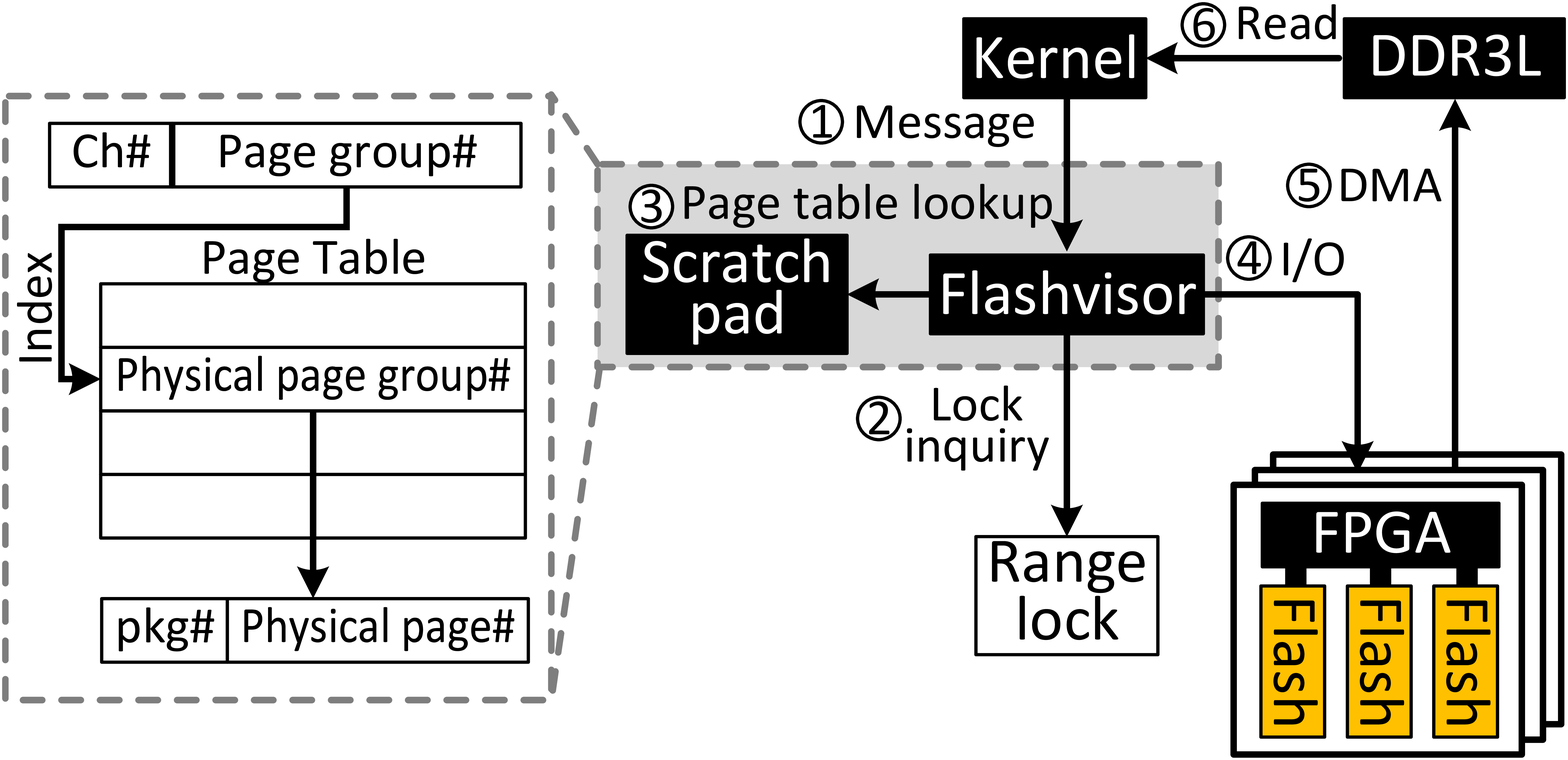}}}
\hspace{8pt}
\subfloat[Writes.]{\label{fig:writes}\rotatebox{0}{\includegraphics[width=0.46\linewidth]{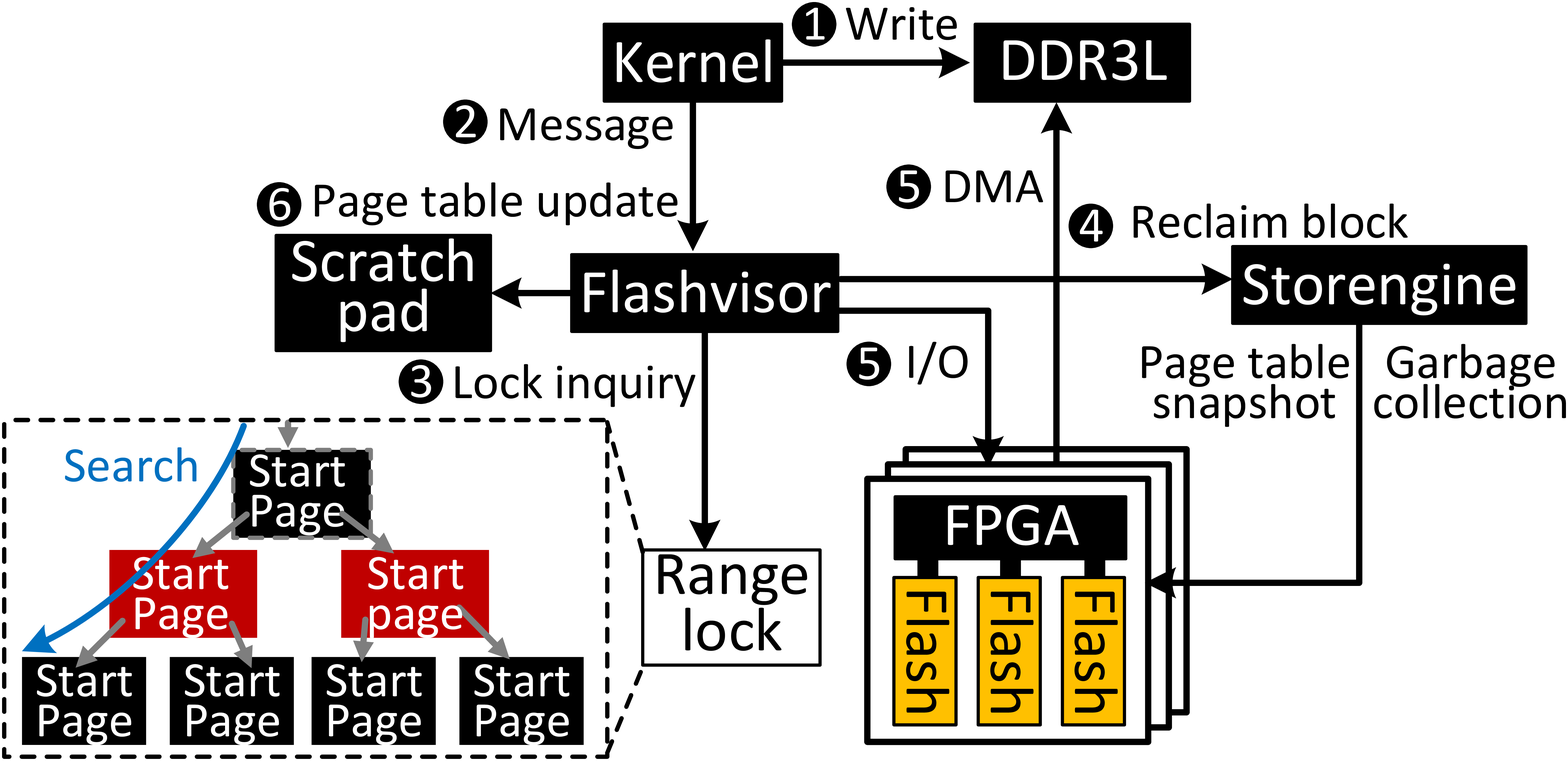}}}

%\vspace{-5pt}
\caption{\label{fig:rdwr}Examples of flash virtualization procedures in FlashAbacus. \vspace{-5pt}}
\end{figure*}

\noindent \textbf{Out-of-order intra-kernel Scheduling.}
This scheduler can perform an out of order execution of many screens associated with different microblocks and different kernels. The main insight behind this method is that, the data dependency only exists among the microblocks within an application's kernel. Thus, if there any available LWPs are observed, this scheduler borrows some screens from a different microblock, which exist across different kernel or application boundaries, and allocate the available LWPs to execute these screens. Similar to the dynamic inter-kernel scheduler, this method keeps all LWPs busy, which can maximize processor utilization, thereby achieving high throughput. Furthermore, the out-of-order execution of this scheduler can reduce the latency of each kernel. 
%This makes all LWPs as much as busy, which can reduce not only the latency of each microblock and but also improve the overall system performance. 
Figure \ref{fig:test2}c shows how the out-of-order scheduler can improve system performance, while maximizing processor utilization. 
As shown in the figure, this scheduler pulls the screen \redcircled{\small{1}} of \texttt{k1}'s microblock 0 (\texttt{m0}) from time unit (T1) and executes it at T0. This is because LWP2 and LWP3 are available even after executing all screens of \texttt{k0}'s microblock 0 (\texttt{m0}), \blkcircled{\small{1}} and \blkcircled{\small{2}}. With a same reason, the scheduler pulls the screen \redcircled{\small{a}} of \texttt{k1}'s microblock 2 (\texttt{m2}) from T3 and execute it at T1. Similarly, the screen \orgcircled{\small{1}} of \texttt{k2}'s microblock 1 (\texttt{m1}) can be executed at T1 instead of T2.  Thus, it can save 2, 4 and 4 time units for \texttt{k0}, \texttt{k1} and \texttt{k2}, respectively (versus the static inter-kernel scheduler).

Note that, in this example, no screen is scheduled before the completion of all the screens along with a previous microblock. In FlashAbacus, this rule is managed by \emph{multi-app execution chain}, which is a list that contains the data dependency information per application. Figure \ref{fig:chain} shows the structure of multi-app execution chain; the root contains multiple pointers, each indicating a list of nodes. Each node maintains a series of screen information per microblock such as LWP ID and status of the execution. Note that the order of such nodes indicates the data-dependency relationships among the microblocks.

%If there is a node that indicates that any LWP is in busy to handle a screen in the block, all the screens of the next one (in the same list) will be suspended. 
%To maintain an appropriate sequence of scheduling, our FlashAbacus employs \emph{NDP execution chain}, which is a list that contains the data dependency information per an NDP-kernel. Specifically, each node (per microblock) of the list contains the status of worker LWPs that execute the screens in the microblock, and the order of such list shows all the relationship of the data dependency. 

%If there is a node that indicates that any LWP is in busy to handle a screen in the block, all the screens of the next one (in the same list) will be suspended. %This execution chain allows our scheduler to freely execute the different screens without a complicated dependency checker at runtime.

\subsection{Flash Virtualization}
\label{sec:virtualization}
The kernels on all LWPs can map the memory regions of DDR3L pointed by their own data sections to the designated flash backbone addresses. As shown in Figure \ref{fig:rdwr}, individual kernel can declare such flash-mapped space for each data section (e.g., input vector on DDR3L) by passing a queue message to Flashvisor. That is, the queue message contains a request type (e.g., read or write), a pointer to the data section, and a word-based address of flash backbone. 
Flashvisor then calculates the page group address by dividing the input flash backbone address, with the number of channel. If the request type is a read, Flashvisor refers its page mapping table with the page group number, and retrieves the corresponding page table entry, which contains the address of physical page group number. It then divides the translated group number by the total page number of each flash package, which indicates the package index within in a channel, and the remainder of the division can be the target physical page number. Flashvisor creates a memory request targeting the underlying flash backbone, and then all the FPGA controllers take over such requests. \newedit{On the other hand, for a write request,} Flashvisor allocates a new page group number by simply increasing the page group number used in a previous write. In cases where there is no more available page group number, Flashvisor generates a request to reclaim a physical block.  Note that, since the time spent to lookup and update the mapping information should not be an overhead to virtualize the flash, the entire mapping table resides on scratchpad; to cover 32GB flash backbone with 64KB page group (4 channels * 2 planes per die * 8KB page), it only requires 2MB for address mapping. 
Considering other information that Flashvisor needs to maintain, a 4MB scratchpad is sufficient for covering the entire address space. 
Note that, since the page table entries associated with each block are also stored in the first two pages within the target physical block of flash backbone (practically used for metadata \cite{zhang2015opennvm}), the persistence of mapping information is guaranteed.

\noindent \textbf{Protection and access control.}
While any of the multiple kernels in FlashAbacus can access the flash backbone, there is no direct data path between the FGPA controllers and other LWPs that process the data near flash. 
That is, all requests related to flash backbone should be taken and controlled by Flashvisor. An easy way by which Flashvisor can protect the flash backbone is to add permission information and the owner's kernel number for each page to the page table entry. However, in contrast to the mapping table used for main memory virtualization, the entire mapping information of Flashvisor should be written in persistent storage and must be periodically updated considering flash I/O services such as garbage collection. Thus, adding such temporary information to the mapping table increases the complexity of our virtualization system, which can degrade overall system performance and shorten the life time of the underlying flash. 
Instead, Flashvisor uses a range lock for each flash-mapped data section. This lock mechanism blocks a request to map a kernel's data section to flash if its flash address range overlaps with that of another by considering the request type. For example, the new data section will be mapped to flash for reads, and such flash address is being used for writes by another kernel, Flashvisor will block the request. Similarly, a request to map for writes will be blocked if the address of target flash is overlapped with another data section, which is being used for reads.  Flashvisor implements this range lock by using red black tree structure \cite{tarjan1983updating}; the start page number of the data section mapping request is leveraged as a key, and each node is augmented with the last page number of the data section and mapping request type. % (e.g., read/write).

\begin{table*}
	\resizebox{\linewidth}{!}{
	%\begin{center}
		\begin{tabular}{| l | c | c | c | c | c | c |  c | c | c | c | c | c | c | c | c | c | c | c | c | c |} %{| m{1.6cm} | m{3.5cm} | m{0.6cm} | m{0.6cm} | m{0.4cm} | m{0.8cm} | m{0.4cm} |  m{0.03cm} | m{0.03cm} | m{0.03cm} | m{0.03cm} | m{0.03cm} | m{0.03cm} | m{0.03cm} | m{0.03cm} | m{0.03cm} | m{0.03cm} | m{0.03cm} | m{0.03cm} | m{0.03cm} | m{0.03cm} |}
		\hline
Name & Description & MBLKs & Serial & Input &  LD/ST & B/KI  & \multicolumn{13}{c}{Heterogeneous workloads}&  \\\cline{8-21}
&  &  & MBLK &(MB) &  ratio(\%) &  & 1 & 2 & 3 & 4 & 5 & 6 & 7 & 8 & 9 & 10 & 11 & 12 & 13 & 14\\
		\hline
		\hline
		ATAX & Matrix Transpose\&Multip. & 2 & 1 & 640  &  45.61 & 68.86  &\ding{108} & & & & \ding{108} & & & & & \ding{108} & \ding{108} & & &  \\\hline
		BICG & BiCG Sub Kernel & 2 & 1 & 640 &  46 & 72.3  &\ding{108} & & & & & \ding{108} & & & \ding{108} & & & & \ding{108} & \\\hline
		2DCON & 2-Dimension Convolution & 1 & 0 & 640 &  23.96 & 35.59  &\ding{108} & & & & \ding{108} & & & & & \ding{108} & & \ding{108} & \ding{108} &  \\\hline
		MVT & Matrix Vector Prod.\&Trans. & 1 & 0 & 640 &  45.1 & 72.05  &\ding{108} & & \ding{108} & & & \ding{108} & \ding{108} & & & \ding{108} & \ding{108} & \ding{108} & \ding{108} & \ding{108} \\\hline
		ADI & Direction Implicit solver & 3 & 1 & 1920 &  23.96 & 35.59 & & \ding{108} & \ding{108} & & & \ding{108} & \ding{108} & & \ding{108} & & \ding{108} & \ding{108} & \ding{108} & \ding{108} \\\hline
		FDTD & 2-D Finite Time Domain  & 3 & 1 & 1920  &  27.27  & 38.52   & & \ding{108} & & \ding{108} & & \ding{108} & \ding{108} & & & \ding{108} & \ding{108} & \ding{108} & & \ding{108} \\\hline
		GESUM & Scalar, Vector\&Multip. & 1 & 0 & 640  &  48.08  & 72.13   & & \ding{108} & \ding{108} & & \ding{108} & & & \ding{108} & & \ding{108} & \ding{108} & \ding{108} & & \ding{108} \\\hline
		SYRK & Symmetric rank-k op. & 1 & 0 & 1280 & 28.21  & 5.29   & \ding{108} & & \ding{108} & & \ding{108} & & & \ding{108} & \ding{108} & & & & & \\\hline
		3MM & 3-Matrix Multiplications & 3 & 1 & 2560 & 33.68  & 2.48  & \ding{108} & & \ding{108} & & & \ding{108} & \ding{108} & & & & & & &\\\hline
		COVAR & Covariance Computation & 3 & 1 & 640  &  34.33  & 2.86  & & \ding{108} & \ding{108} & \ding{108} & \ding{108} & & & \ding{108} & & & & & & \\\hline
		GEMM & Matrix-Multiply & 1 & 0 & 192 & 30.77  & 5.29  & & \ding{108} & & \ding{108} & \ding{108} & & & \ding{108} & \ding{108} & & & \ding{108} & \ding{108} & \ding{108} \\\hline
		2MM & 2-Matrix Multiplications & 2 & 1 & 2560 & 33.33  & 3.76   & & \ding{108} & & \ding{108} & & \ding{108} & \ding{108} & \ding{108} & \ding{108} & & & & \ding{108} & \\\hline
		SYR2K & Symmetric rank-k op. & 1 & 0 & 1280 & 30.19  & 1.85   & & & & \ding{108} & & & & \ding{108} & & \ding{108} & \ding{108} & & &\\\hline
		CORR & Correlation Computation & 4 & 1 & 640  &  33.04  & 2.79  & & & & \ding{108} & & & \ding{108} & & \ding{108} & & & & & \ding{108} \\\hline

		\end{tabular}}
	%\end{center}
	\vspace{5pt}
	\caption{Important characteristics of our workloads.}
	\vspace{5pt}
	\label{tbl:singlepro}
\end{table*}

\noindent \textbf{Storage management.}
Flashvisor can perform address translation similar to a log-structured pure page mapping technique \cite{kim2002space}, which is a key function of flash firmware \cite{agrawal2008design}. To manage the underlying flash appropriately, Flashvisor also performs metadata journaling \cite{agrawal2008design} and garbage collection (including wear-leveling) \cite{bux2010performance, chang2007efficient, shahidi2016exploring, jung2012taking}. However, such activities can be strong design constraints for SSDs as they should be resilient in all types of power failures, including sudden power outages. Thus, such metadata journaling is performed in the foreground and garbage collection are invoked on demand \cite{bux2010performance}. \newedit{In cases where an error is detected just before an uncorrectable error correction arises, Flashvisor excludes the corresponding block by remapping it with new one.} In addition, all the operations of these activities are executed in the lockstep with address translation. 

However, if these constraints can be relaxed to some extent, most overheads (that flash firmware bear brunt of) are removed from multi-kernel executions. Thus, we assign another LWP to perform this storage management rather than data processing. While Flashvisor is responsible for mainly address translation and multi-kernel execution scheduling, this LWP, referred to as \emph{Storengine}, periodically dumps the scratchpad information to the underlying flash as described in the previous subsection. In addition, Storengine reclaims the physical block from the beginning of flash address space to the end in background. Most garbage collection and wear-leveling algorithms that are employed for flash firmware select victim blocks by considering the number of valid pages and the number of block erases. These approaches increase the accuracy of garbage collection and wear-leveling, since they require extra information to set such parameters and search of the entire address translation information to identify a victim block. \newedit{Rather than wasting compute cycles to examine all the information in the page table, Storengine simply selects the target victim block from a used block pool in a round robin fashion and loads the corresponding page table entries from flash backbone. Storengine periodically migrates valid pages from the victim block to the new block, and return the victim to a free block in idle, which is a background process similar to preemtable firmware techniques \cite{jung2014hios, jung2012taking}.} Once the victim block is selected, the page mapping table associated with those two blocks is updated in both the scratchpad and flash. Note that all these activities of Storengine can be performed in parallel with the address translation of Flashvisor. Thus, locking the address ranges that Storengine generates for the snapshot of the scratchpad (journaling) or the block reclaim is necessary, but such activities overlapped with the kernel executions and address translations (are performed in the background).

\section{Evaluation}
\label{sec:result}
%\noindent \textbf{GPU-accelerated data processing.}
%We evaluate a NVIDIA GPU device \cite{GT610}, which has 48 CUDA cores and 1GB GDDR3 with 64-bit wide memory connector. All the 48 cores work at 850MHz, and their computation throughput is around 90 GFLOPS. The reason why we select this middle range GPU solution is that the the clock frequency for computation and memory configuration is similar to what DoubleFlash prototype employs (but the GPU has 8x more execution units than ours). This GPU uses PCIe 2.0 interface and consumes 29W at most. The system also employs the NVMe SSD (whose configurations are same with what CPU and DoubleFlash uses) over another PCIe slot. Overall, it offers 2.2GB/sec bandwidth under an ideal workload execution (sequential access). 

\begin{figure*}
\centering
%\vspace{-10pt}
\def\subfigcapskip{0pt}
\subfloat[Performance analysis for homogeneous workloads.]{\label{fig:exetime_singleapp_fig}\rotatebox{0}{\includegraphics[width=0.49\linewidth]{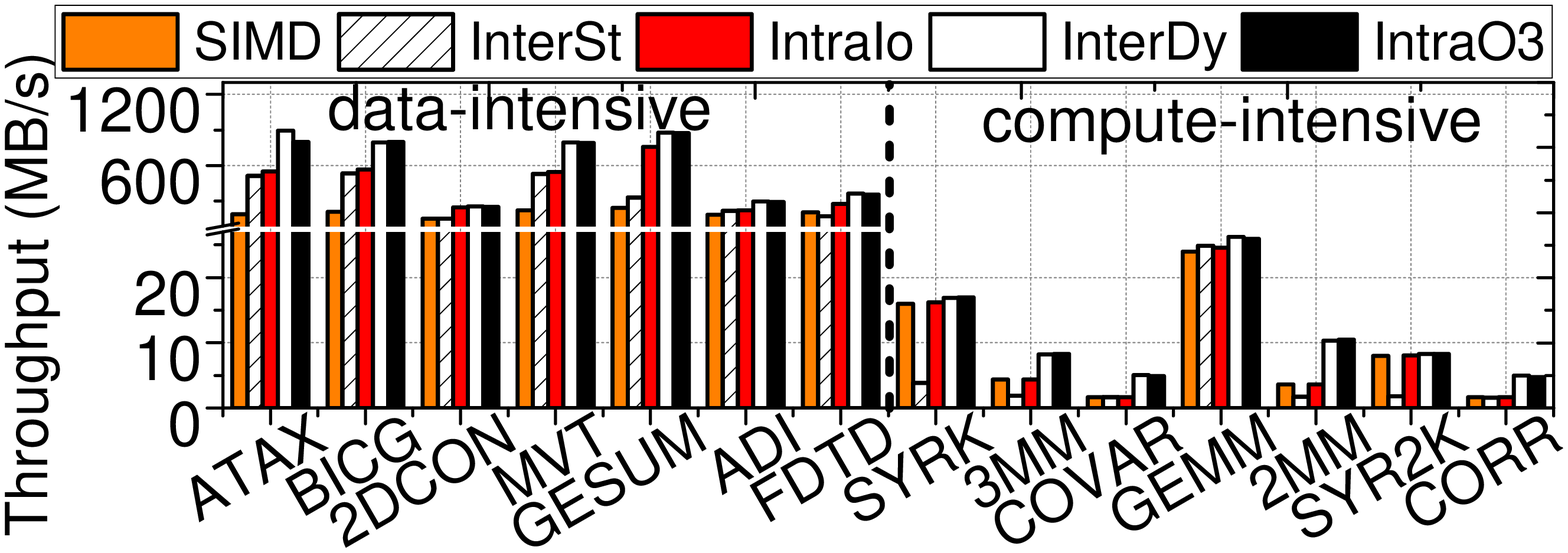}}}
\hspace{2pt}
\subfloat[Performance analysis for heterogeneous workloads.]{\label{fig:exetime_mix_fig}\rotatebox{0}{\includegraphics[width=0.49\linewidth]{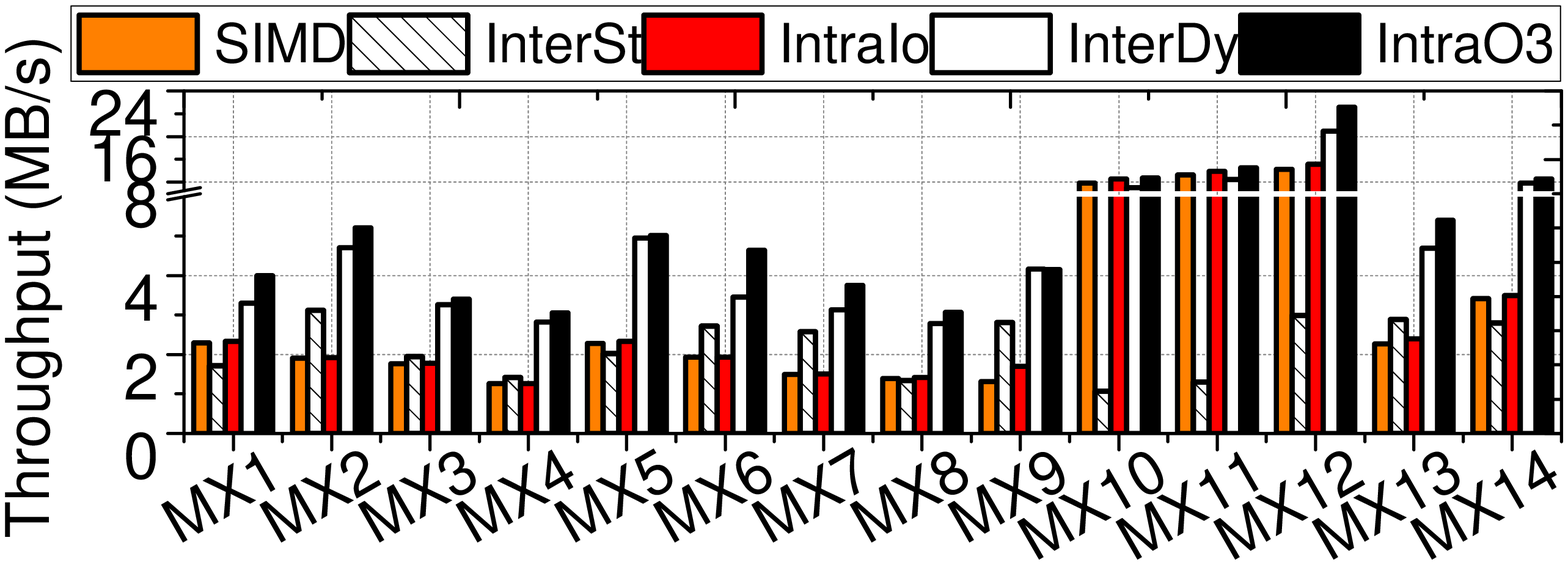}}}
%\vspace{-5pt}
\caption{Performance analysis (computation throughput normalized to \texttt{SIMD}).}
\label{fig:single-exetime-analysis}
%\vspace{-5pt}
\end{figure*}

%\noindent \textbf{Experiment setup.}
%We prototyped our proposed FlashAbacus on a real hardware which is described in Section \ref{sec:prototype}. Specifically, we allocate two LWPs as Flashvisor and Storengine, respectively, and assign six LWPs as worker LWPs. The Flashvisor, in this evaluation, employs four different kernel scheduling strategies: static inter-kernel scheduler, dynamic inter-kernel scheduler, in-order intra-kernel scheduler, and out-of-order intra-kernel scheduler. To compare the performance of our accelerators, we setup a heterogeneous computing system that employs accelerator and SSD as discrete PCIe devices attaching to a six-core Xeon CPU-based host \cite{E5Intel}. To better utilize the multiple LWPs within accelerator, host compiles the programs with parallel execution primitives \cite{mitra2014implementation}. Substantially, we configured five different data processing systems as follows:

%In this evaluation, 
\noindent \textbf{Accelerators.} We built five different heterogeneous computing options. \newedit{``\texttt{SIMD}'' employs a low-power accelerator that executes multiple data-processing instances using a single-instruction multiple-data (SIMD) model implemented in OpenMP \cite{mitra2014implementation}, and \texttt{SIMD} performs I/Os through a discrete high-performance SSD (i.e., Intel NVMe 750 \cite{NVMeSSD}).} On the other hand, ``\texttt{InterSt}'', ``\texttt{InterDy}'', ``\texttt{IntraIo}'', and ``\texttt{IntraO3}'' employ the static inter-kernel, dynamic inter-kernel, in-order intra-kernel, and out-of-order intra-kernel schedulers of FlashAbacus, respectively. In these accelerated systems, we configure the host with Xeon CPU \cite{E5Intel} and 32GB DDR4 DRAM \cite{standard2012ddr4}.

\noindent \textbf{Benchmarks.}
We also implemented 14 real benchmarks that stem from Polybench \cite{pouchet2012polybench} on all our accelerated systems\footnote{\newedit{In this implementation, we leverage a set of TI code generation tools \cite{CGT, MCSDK, MAD} to develop and compile the kernels of our FlashAbacus. Note that the memory protection and access control are regulated by the TI's tools, which may be limited for other types of acceleration hardware.}}. To evaluate the impact of serial instructions in the multi-core platform, we rephrase partial instructions of tested benchmarks to follow a manner of serial execution. The configuration details and descriptions for each application are explained in Table \ref{tbl:singlepro}. In this table, we explain how many microblocks exist in an application (denoted by MBLKs). Serial MBLK refers to the number of microblocks that have no screens; these microblocks are to be executed in serial. The table also provides several workload analyses, such as the input data size of an instance (Input), ratio of load/store instructions to the total number of instructions (LD/ST ratio), and computation complexity in terms of data volumes to process per thousand instructions (B/KI).
To evaluate the benefits of different accelerated systems, we also created 14 heterogeneous workloads by mixing six applications. All these configurations are presented at the right side of the table, where the symbol $\bullet$ indicates the correspondent applications, which are included for such heterogeneous workloads. 

\noindent \textbf{Profile methods.}
\newedit{To analyze the execution time breakdown of our accelerated systems, we instrument timestamp annotations in both host-side and accelerator-side application source codes by using a set of real-time clock registers and interfaces. The annotations for each scheduling point, data transfer time, and execution period enable us to capture the details of CPU active cycles and accelerator execution cycles at runtime. On the other hand, we use a blktrace analysis \cite{axboe2010blktrace} to measure device-level SSD performance by removing performance interference potentially brought by any modules in storage stack. We leverage Intel power gadget tool to collect the host CPU energy and host DRAM energy \cite{kim2014intel}. 
Lastly, we estimate the energy consumption of FlashAbacus hardware platform based on TI internal platform power calculator, and use our in-house power analyzer to monitor the dynamic power of SSD \cite{jung2016exploring, zhang2014power}.}

\begin{figure*}
\centering
%\vspace{-5pt}
\def\subfigcapskip{0pt}
\subfloat[The latency analysis for homogeneous workloads.]{\label{fig:lat3}\rotatebox{0}{\includegraphics[width=0.49\linewidth]{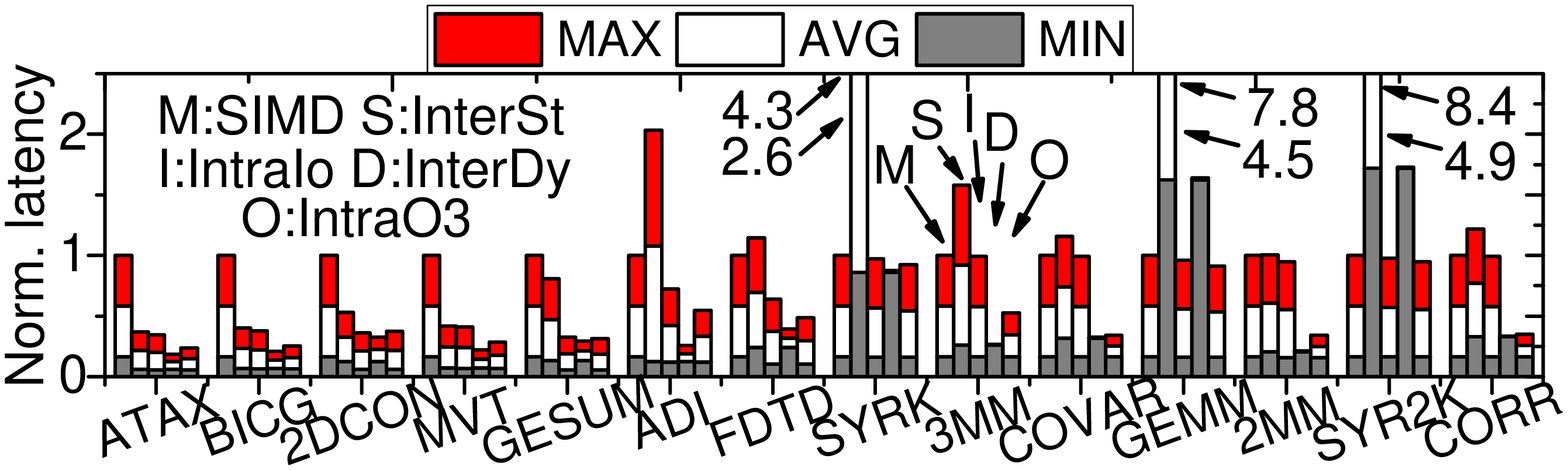}}}
\hspace{2pt}
\subfloat[The latency analysis for heterogeneous workloads.]{\label{fig:lat4}\rotatebox{0}{\includegraphics[width=0.49\linewidth]{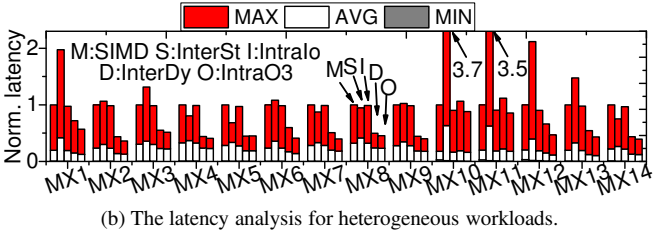}}}
%\vspace{-5pt}
\caption{Latency analysis (normalized to \texttt{SIMD}).}
\label{fig:latency-analysis}
%\vspace{-5pt}
\end{figure*}

\begin{figure}
\centering
%\vspace{-15pt}
\def\subfigcapskip{0pt}
\subfloat[Latency analysis (\textit{ATAX}).]{\label{fig:lat1}\rotatebox{0}{\includegraphics[width=0.49\linewidth]{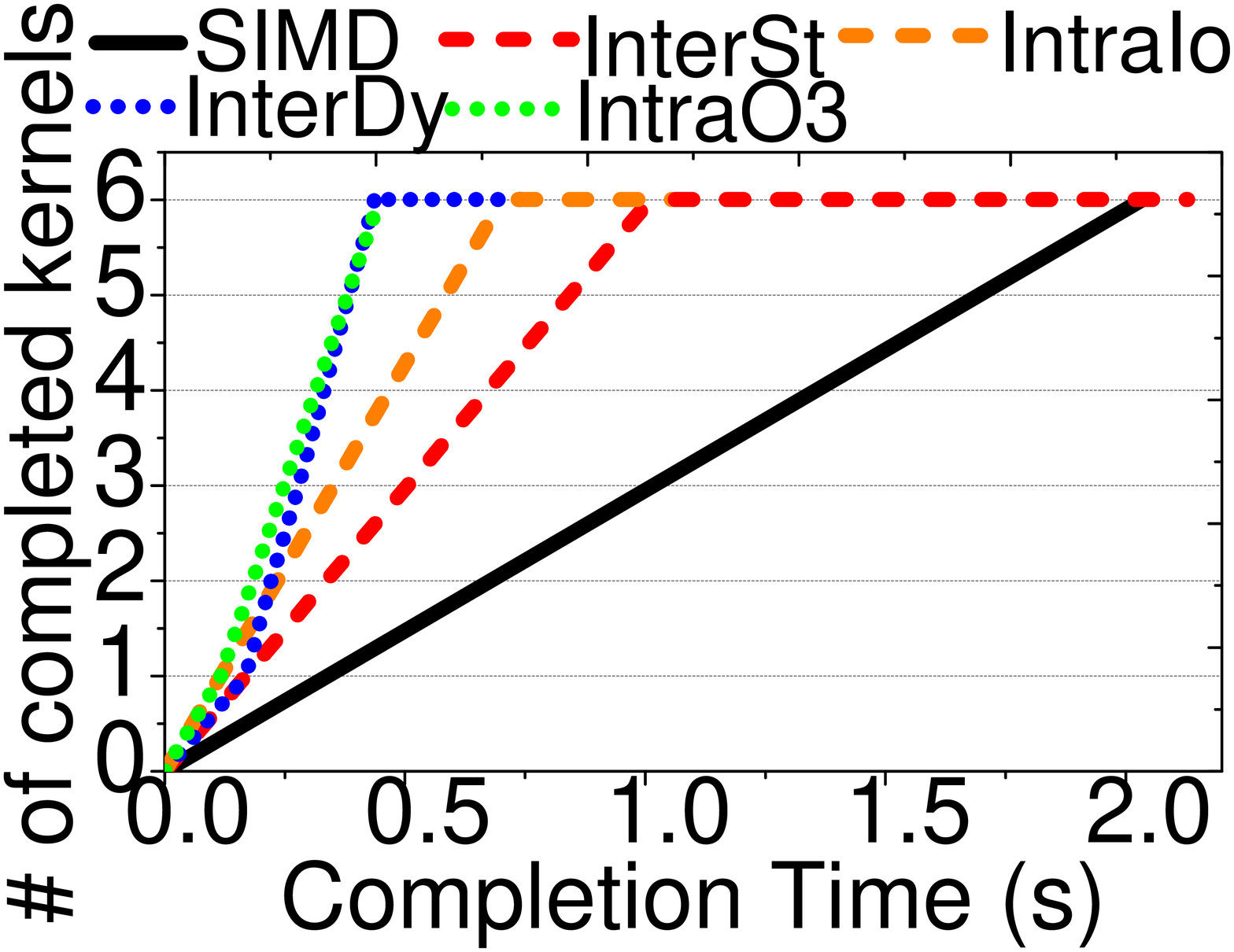}}}
\hspace{2pt}
\subfloat[Latency analysis (\textit{MX1}).]{\label{fig:lat2}\rotatebox{0}{\includegraphics[width=0.49\linewidth]{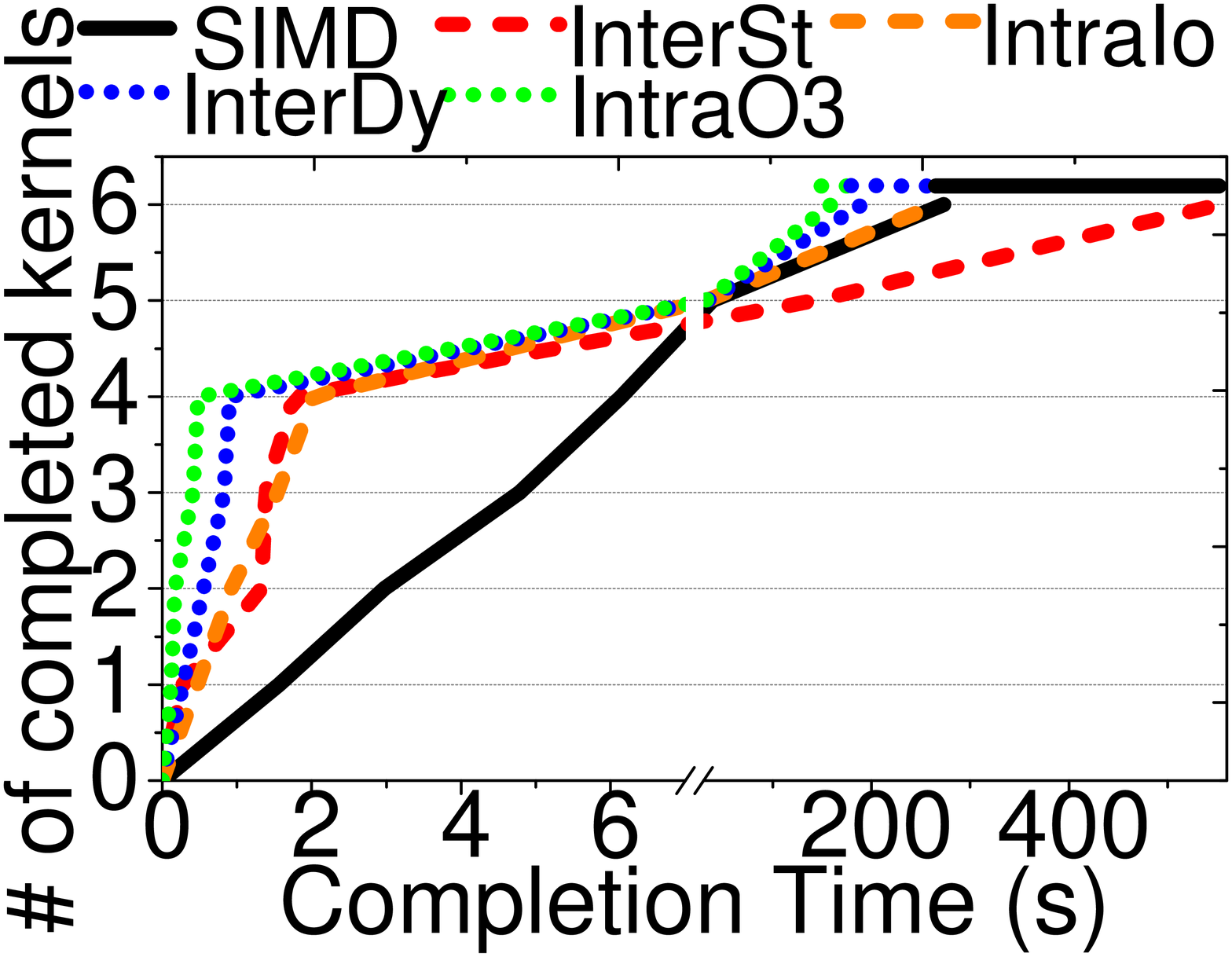}}}
%\vspace{-5pt}
\caption{CDF analysis for workload latency.}
\label{fig:lat-analysis}
%\vspace{-5pt}
\end{figure}

\begin{figure*}
\centering
%\vspace{-10pt}
\def\subfigcapskip{0pt}
\subfloat[Energy breakdown for homogeneous workloads.]{\label{fig:energybreak_singleapp_fi}\rotatebox{0}{\includegraphics[width=0.49\linewidth]{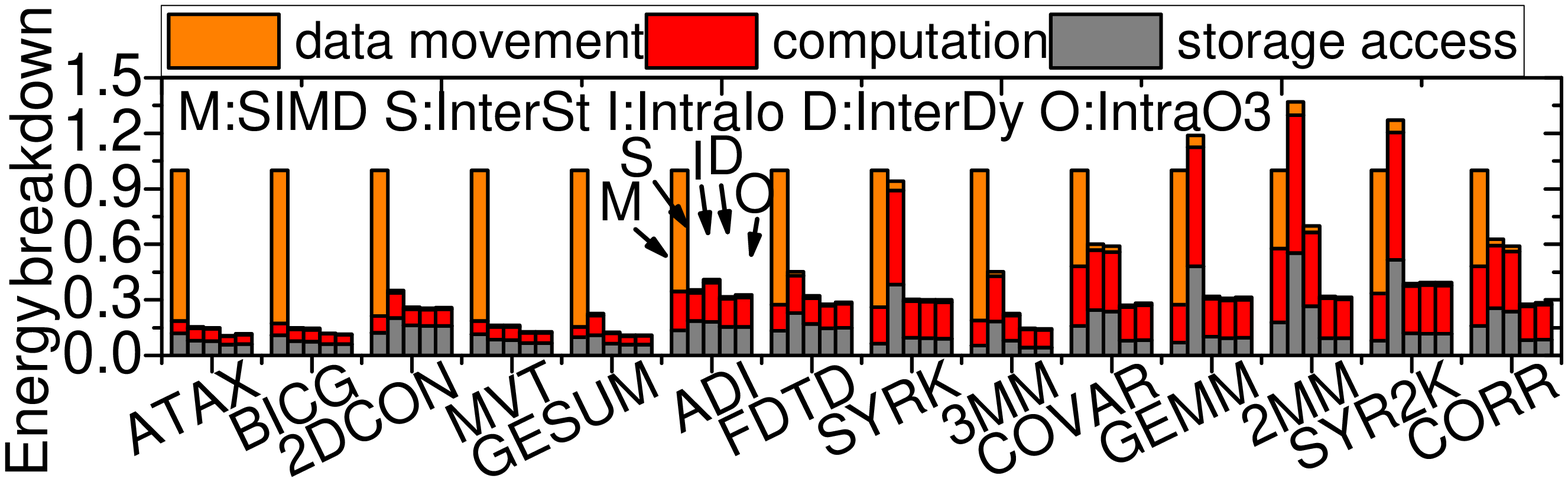}}}
\hspace{2pt}
\subfloat[Energy breakdown for heterogeneous workloads.]{\label{fig:energybreak_mix_fig}\rotatebox{0}{\includegraphics[width=0.49\linewidth]{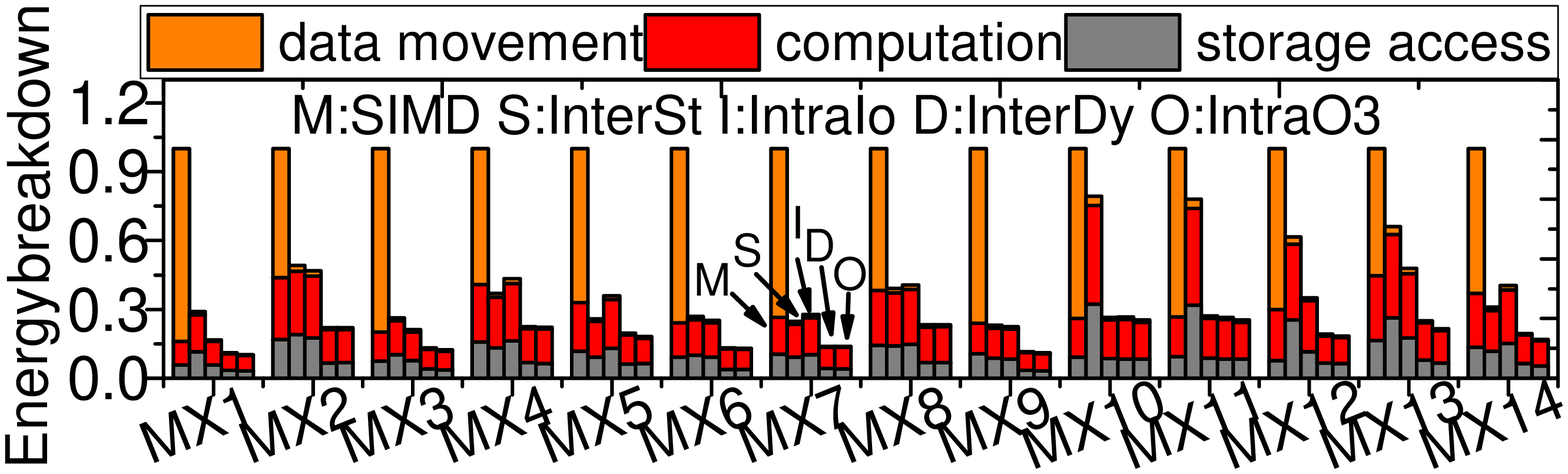}}}
%\vspace{5pt}
\caption{Analysis for energy decomposition (all results are normalized to \texttt{SIMD}).}
\label{fig:energy-analysis}
%\vspace{5pt}
\end{figure*}

\begin{figure*}
\centering
%\vspace{-15pt}
\def\subfigcapskip{0pt}
\subfloat[The ratio of LWP utilization for homogeneous workloads.]{\label{fig:single_util}\rotatebox{0}{\includegraphics[width=0.49\linewidth]{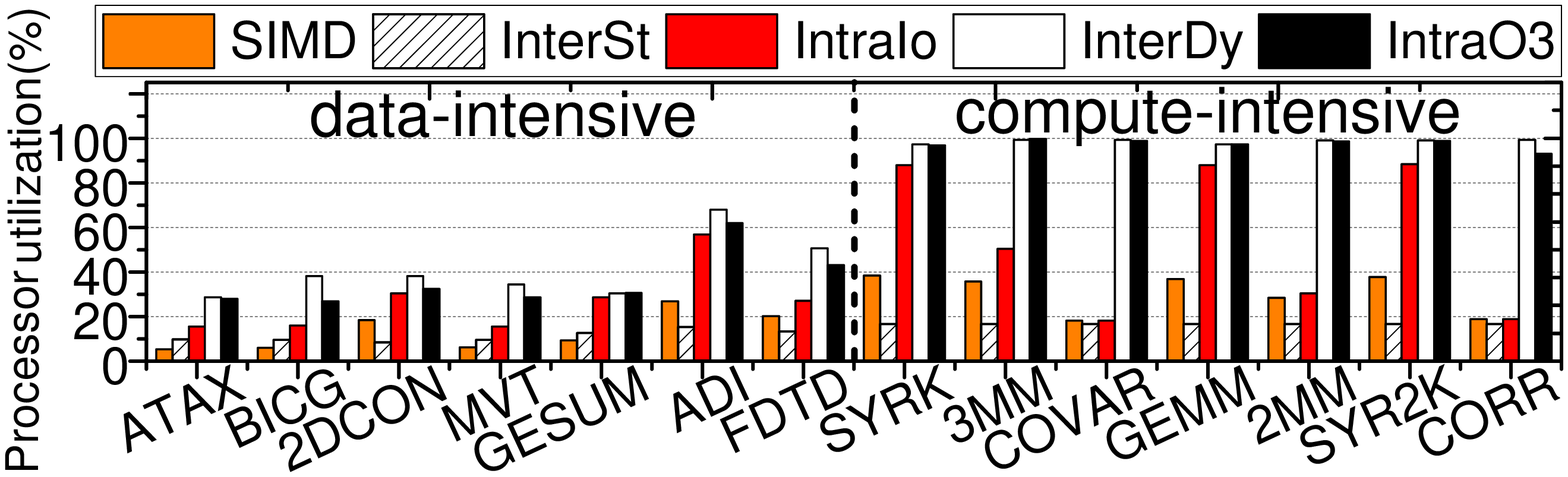}}}
\hspace{2pt}
\subfloat[The ratio of LWP utilization for heterogeneous workloads.]{\label{fig:mixutilfig}\rotatebox{0}{\includegraphics[width=0.49\linewidth]{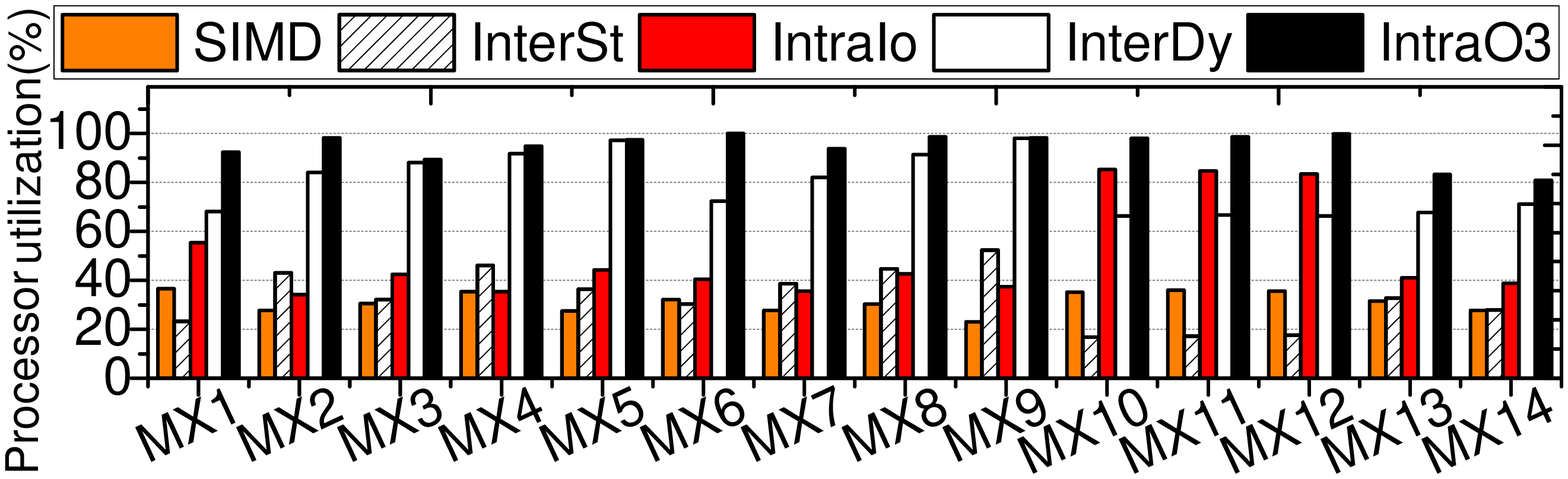}}}
%\vspace{5pt}
\caption{Analysis of processor utilization.}
\label{fig:util-analysis}
%\vspace{5pt}
\end{figure*}

\subsection{Data Processing Throughput}
\newedit{Figure \ref{fig:single-exetime-analysis} demonstrates the overall throughput of all hardware accelerators that we implemented by executing both homogeneous workloads and heterogeneous workloads. Generally speaking, the proposed FlashAbacus approach (i.e., \texttt{IntraO3}) outperforms the conventional accelerator approach \texttt{SIMD} by 127\%, on average, across all workloads tested.}

\noindent \textbf{Homogeneous workloads.} 
Figure \ref{fig:exetime_singleapp_fig} shows the overall throughput evaluated from homogeneous workloads. % by executing 14 individual workloads. 
In this evaluation, we generate 6 instances from each kernel. Based on the computation complexity (B/KI), we categorize all workloads into two groups: i) data-intensive and ii) computing-intensive. 
As shown in the figure, all our FlashAbacus approaches outperform \texttt{SIMD} for data-intensive workloads by 144\%, on average. This is because, although all the accelerators have equal same computing powers, \texttt{SIMD} cannot process data before the data are brought by the host system. However, \texttt{InterSt} has poorer performance, compared to \texttt{InterDy} and \texttt{IntraO3}, by 53\% and 52\%, respectively. The reason is that \texttt{InterSt} is limited to spreading homogeneous workloads across multiple workers, due to its static scheduler configuration. Unlike \texttt{InterSt}, \texttt{IntraIo} enjoys the benefits of parallelism, by partitioning kernels into many screens and assigning them to multiple LWPs. However, \texttt{IntraIo} cannot fully utilize multiple LWPs if there is a microblock that has no code segments and that cannot be concurrently executed (i.e., serial MBLK). Compared to \texttt{IntraIo}, \texttt{IntraO3} successfully overcomes the limits of serial MBLK by borrowing multiple screens from different microblocks, thereby improving the performance by 62\%, on average. For homogeneous workloads, \texttt{InterDy} achieves the best performance, because all the kernels are equally assigned to a single LWP and are simultaneously completed. While \texttt{IntraO3} can dynamically schedule microblocks to LWPs for a better load balance, IPC overheads (between Flashvisor and workers) and scheduling latency of out-of-order execution degrade \texttt{IntraO3}'s performance by 2\%, on average, compared to \texttt{InterDy}.

\noindent \textbf{Heterogeneous workloads.}
Figure \ref{fig:exetime_mix_fig} shows the overall system bandwidths of different accelerated systems under heterogeneous workloads. For each execution, we generate 24 instances, four from a single kernel. 
\texttt{SIMD} exhibits 3.5 MB/s system bandwidth across all heterogeneous workloads, on average, which is 98\% worse than that of data-intensive workloads. This is because, heterogeneous workloads contain computing-intensive kernels, which involve much longer latency in processing data than storage accesses. Although the overheads imposed by data movement are not significant in computing-intensive workloads, the limited scheduling flexibility of \texttt{SIMD} degrades performance by 42\% and 50\%, compared to \texttt{InterDy} and \texttt{IntraO3}, respectively. In contrast to the poor performance observed in homogeneous workloads, \texttt{InterSt} outperforms \texttt{IntraIo} by 34\% in workloads \textit{MX2,3,4,6,7,9} and \textit{13}. 
%\textit{MX2}, \textit{MX3}, \textit{MX4}, \textit{MX6}, \textit{MX7}, \textit{MX9}, and \textit{MX13}. 
This is because \texttt{SIMD} can statically map 6 different kernels to 6 workers. 
However, since different kernels have various data processing speeds, unbalanced loads across workers degrade performance by 51\%, compared to \texttt{IntraIo}, for workloads \textit{MX1,5,8,10,11} and \textit{12}. Unlike \texttt{InterSt}, \texttt{InterDy} monitors the status of workers and dynamically assigns a kernel to a free worker to achieve better load balance. Consequently, \texttt{InterDy} exhibits 177\% better performance than \texttt{InterSt}. Compared to the best performance reported by the previous homogeneous workload evaluations, performance degradation of \texttt{InterDy} in these workloads is caused by a stagger kernel, which exhibits much longer latency than other kernels. In contrast, \texttt{IntraO3} 
shortens the latency of the stagger kernel by executing it across multiple LWPs in parallel. As a result, \texttt{IntraO3} outperforms \texttt{InterDy} by 15\%.

\subsection{Execution Time Analysis}

\noindent \textbf{Homogeneous workloads.}
Figure \ref{fig:lat3} analyzes latency of the accelerated systems (with homogeneous workloads). 
For data-intensive workloads (i.e., \textit{ATAX}, \textit{BICG}, \textit{2DCONV}, and \textit{MVT}), average latency, maximum latency, and minimum latency of \texttt{SIMD} are 39\%, 87\%, and 113\% longer than those of FlashAbacus approaches, respectively. This is because, \texttt{SIMD} consumes extra latency to transfer data through different I/O interfaces and redundantly copy I/Os across different software stacks. \texttt{InterSt} and \texttt{InterDy} exhibit similar minimum latency. However, \texttt{InterDy} has 57\% and 68\% shorter average latency and maximum latency, respectively, compared to \texttt{InterSt}. This is because, \texttt{InterSt} can dynamically schedule multiple kernels across different workers in parallel. The minimum latency of inter-kernel schedulers (\texttt{InterSt} and \texttt{InterDy}) is 61\% longer than that of intra-kernel schedulers (\texttt{IntraIo} and \texttt{IntraO3}), because intra-kernel schedulers can shorten single kernel execution by leveraging multiple cores to execute microblocks in parallel. 

\noindent \textbf{Heterogeneous workloads.}
Figure \ref{fig:lat4} shows the same latency analysis, but with heterogeneous workloads. 
All heterogeneous workloads have more computing parts than the homogeneous ones, data movement overheads are hidden behind their computation somewhat. 
Consequently, \texttt{SIMD} exhibits kernel execution latency similar to that of \texttt{IntraIo}. 
Meanwhile, \texttt{InterSt} exhibits a shorter maximum latency for workloads \textit{MX4,5,8} and \textit{14}, compared to \texttt{IntraIo}. This is because, these workloads contain multiple serial microblocks that stagger kernel execution. Further \texttt{InterSt} has the longest average latency among all tested approaches due to inflexible scheduling strategy.
Since \texttt{IntraO3} can split kernels into multiple screens and achieve a higher level of parallelism with a finer-granule scheduling, it performs better than \texttt{InterDy} in terms of the average and maximum latency by 10\% and 19\%, respectively. 

\noindent \textbf{CDF analysis.} Figures \ref{fig:lat1} and \ref{fig:lat2} depict the execution details of workload \textit{ATAX} and \textit{MX1}, each representing homogeneous and heterogeneous workloads, respectively. As shown in Figure \ref{fig:lat1}, \texttt{InterDy} takes more time to complete the first kernel compared to \texttt{IntraIo} and \texttt{IntraO3}, because \texttt{InterDy} only allocates single LWP to process the first kernel. For all the six kernels, \texttt{InterO3} also completes the operation earlier than or about the same time as \texttt{InterDy}.
More benefits can be achieved by heterogeneous execution.  As shown in Figure \ref{fig:lat2}, for the first four data-intensive kernels, \texttt{SIMD} exhibits much longer latency compared to FlashAbacus, due to the system overheads. While \texttt{SIMD} outperforms \texttt{InterSt} for the last two computation-intensive kernels, \texttt{IntraO3} outperforms \texttt{SIMD} by 42\%, on average.

\subsection{Energy Evaluation}
\label{sec:energy}
In Figure \ref{fig:energy-analysis}, the energy of all five tested accelerated systems is decomposed into the data movement, computation and storage access parts. \newedit{Overall, \texttt{IntraO3} consumes average energy less than \texttt{SIMD} by 78.4\% for all the workloads that we tested.} 
Specifically, all FlashAbacus approaches exhibit better energy efficiency than \texttt{SIMD} for data-intensive workloads (cf. Figure \ref{fig:energybreak_singleapp_fi}). This is because, the host requires periodic transfer of data between the SSD and accelerator, consuming 71\% of the total energy of \texttt{SIMD}. Note that, \texttt{InterSt} consumes 28\% more energy than \texttt{SIMD} for computing-intensive workloads, \textit{GEMM}, \textit{2MM}, and \textit{SYR2K}. Even though \texttt{InterSt} cannot fully activate all workers due to its inflexibility, it must keep Flashvisor and Storengine always busy for their entire execution, leading to inefficient energy consumption behavior. 
However, \texttt{IntraIo}, \texttt{InterDy}, and \texttt{IntraO3} reduce the overall energy consumption by 47\%, 22\%, and 2\%, respectively, compared to \texttt{InterSt}. This is because \texttt{IntraO3} can exploit the massive parallelism (among all workers), saving energy, wasted by the idle workers (observed by \texttt{InterSt}).

\ignore{
%%%%%%%%%%%%%%%%%%%%%%%%%%%%%%%%%%%%%%%%%%%%%%%
$good for journal extension$
%%%%%%%%%%%%%%%%%%%%%%%%%%%%%%%%%%%%%%%%%%%%%%%
\begin{figure}
\centering

\subfloat[The execution time breakdown.]{\label{fig:datamove_brk}\rotatebox{0}{\includegraphics[width=0.48\linewidth]{figs/datamove_brk}}}
\hspace{2pt}
\subfloat[The energy breakdown.]{\label{fig:datamove_engbrk}\rotatebox{0}{\includegraphics[width=0.48\linewidth]{figs/datamove_engbrk}}}

%\vspace{-5pt}
\caption{\label{fig:datamove}Data movement analysis. }
%\vspace{-5pt}
\end{figure}

We also decompose the execution time and energy imposed by data movements across different data-intensive workloads. As shown in Figure \ref{fig:datamove_brk}, copying data between host DRAM and accelerator's DRAM (denoted as ``Acc. write" and ``Acc. read") takes 75\% of total transfer time. This long delay is introduced by limited PCIe bandwidth. Although data copy between user space and kernel space (``user-kernel copy") is much faster than ``Acc. write" and ``Acc. read", it still consumes 19\% of total time due to multiple context switches and CPU interventions. While the performance overhead of metadata operations in the host such as memory allocation (``CPU mem allocate") and memory free (``CPU mem free") are not significant, the metadata operations in accelerators (``Acc. mem allocate" and ``Acc. mem free") costs 6\% of total transfer time. %, due to the low performance of accelerator's core and iterative communication.
Figure \ref{fig:datamove_engbrk} show the energy breakdown of each hardware components during data movement. Since CPU is always involved in managing the data movement, it costs 70\% of overall energy consumption. Host-side DRAM also consumes 20\% of overall energy, as multiple reads and writes occur during transfers. However, the embedded DRAM of accelerator consumes energy less than 0.8\%. This is because, the embedded DRAM employed in accelerator has much smaller size (1GB) than host main memory and operates in lower frequency. Lastly, the SSD consumes only 9\% of total energy. 
}

\subsection{Processor Utilizations}
Figures \ref{fig:single_util} and \ref{fig:mixutilfig} show an analysis of LWP utilizations obtained by executing homogeneous and heterogeneous workloads. 
%Figure \ref{fig:util-analysis} shows the processor utilization to execute applications under various data processing approaches, which can explain how our \texttt{IntraO3} can achieve better performance and less energy consumption than other ones. 
In this evaluation, LWP utilization is calculated by dividing the LWP's actual execution time (average) by the latency to execute entire application(s) in each workload. 
For data-intensive ones, most of LWP executions are stalled due to the long latency of storage accesses, which make \texttt{SIMD}'s LWP utilizations lower than \texttt{InterO3} by 23\%, on average. 
On the other hand, \texttt{InterDy} keeps all processors busy by 98\% of the total execution time, which leads the highest LWP utilization among all accelerated systems we tested. \texttt{IntraO3} schedules microblocks in out-of-order manner and only a small piece of code segment is executed each time, which makes its LWP utilization slightly worse than \texttt{InterDy} ones (less than 1\%) for the execution of homogeneous kernels. In contrast, for heterogeneous kernel executions, \texttt{IntraO3} achieves processor utilization over 94\%, on average, 15\% better than that of \texttt{InterDy}. Since  LWP utilization is strongly connected to the degree of parallelism and computing bandwidth, we can conclude that \texttt{IntraO3} outperforms all the accelerated systems.

\begin{figure}
\centering
%\vspace{-10pt}
\def\subfigcapskip{0pt}
\subfloat[The function unit utilization of heterogeneous workloads.]{\label{fig:latTSA_MIX}\rotatebox{0} % \textit{ATAX}, \textit{BICG}, \textit{2DCON}, and \textit{MVT}
{\includegraphics[width=0.98\linewidth]{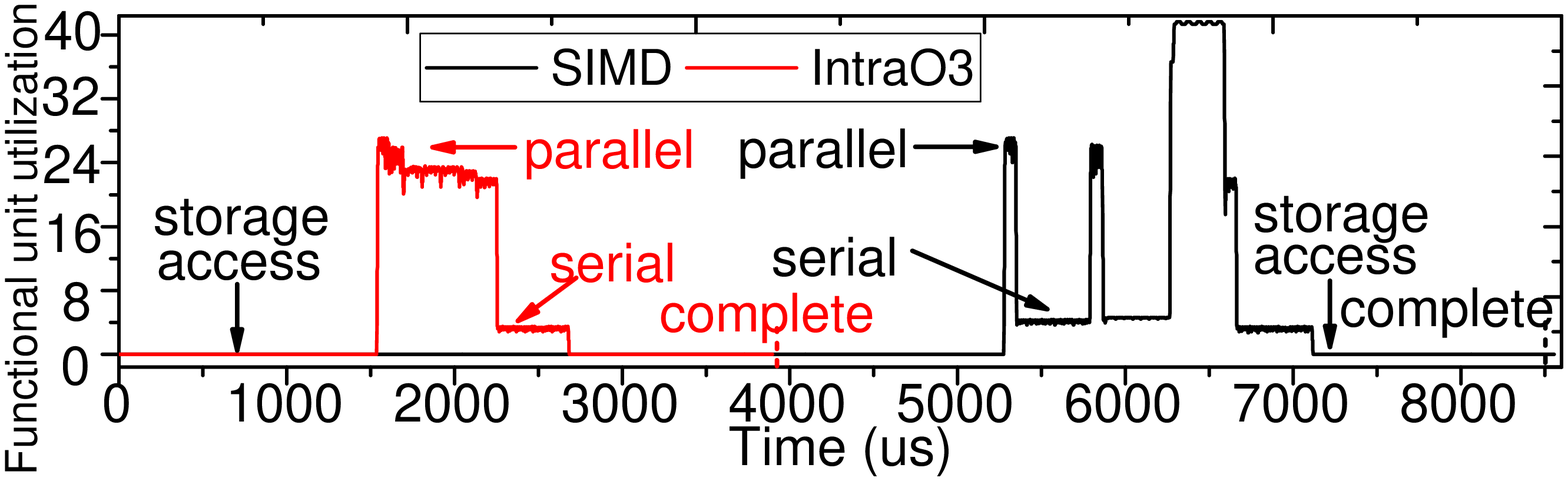}}}
%\vspace{-8pt}

\subfloat[Power analysis of heterogeneous workloads.]{\label{fig:latTSA_MIXeng}\rotatebox{0}{\includegraphics[width=0.98\linewidth]{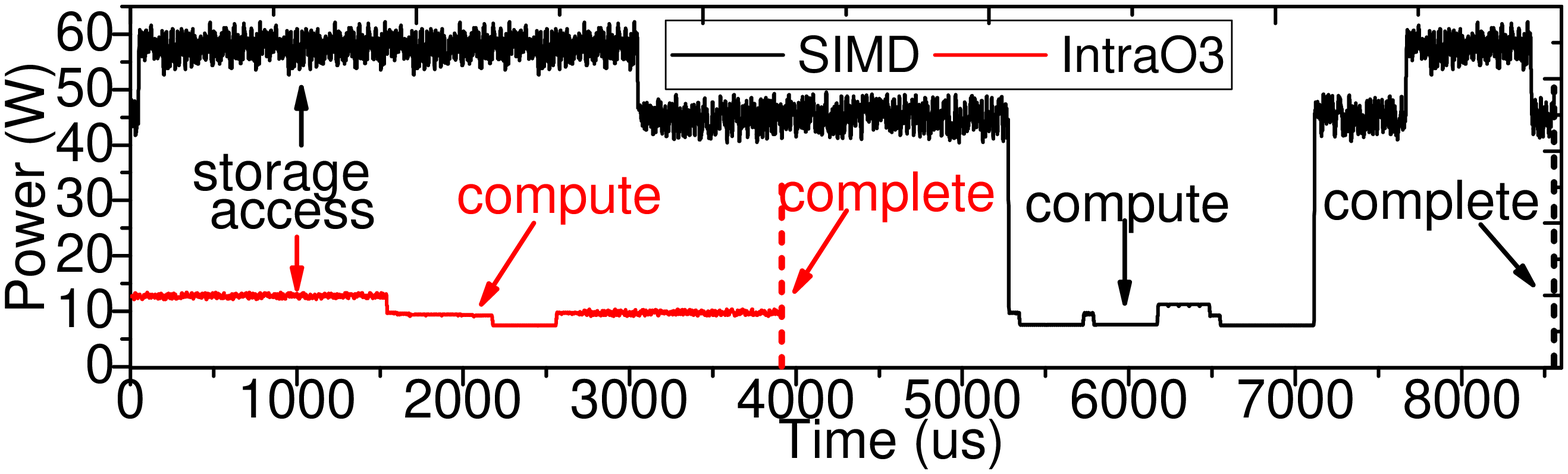}}}
%\vspace{-5pt}
\caption{Resource utilization and power analysis.}
\label{fig:latTSA}
%\vspace{-5pt}
\end{figure}

\begin{figure}
\centering
%\vspace{-10pt}
\def\subfigcapskip{0pt}
\subfloat[Performance analysis.]{\label{fig:perf1}\rotatebox{0} % \textit{ATAX}, \textit{BICG}, \textit{2DCON}, and \textit{MVT}
{\includegraphics[width=0.49\linewidth]{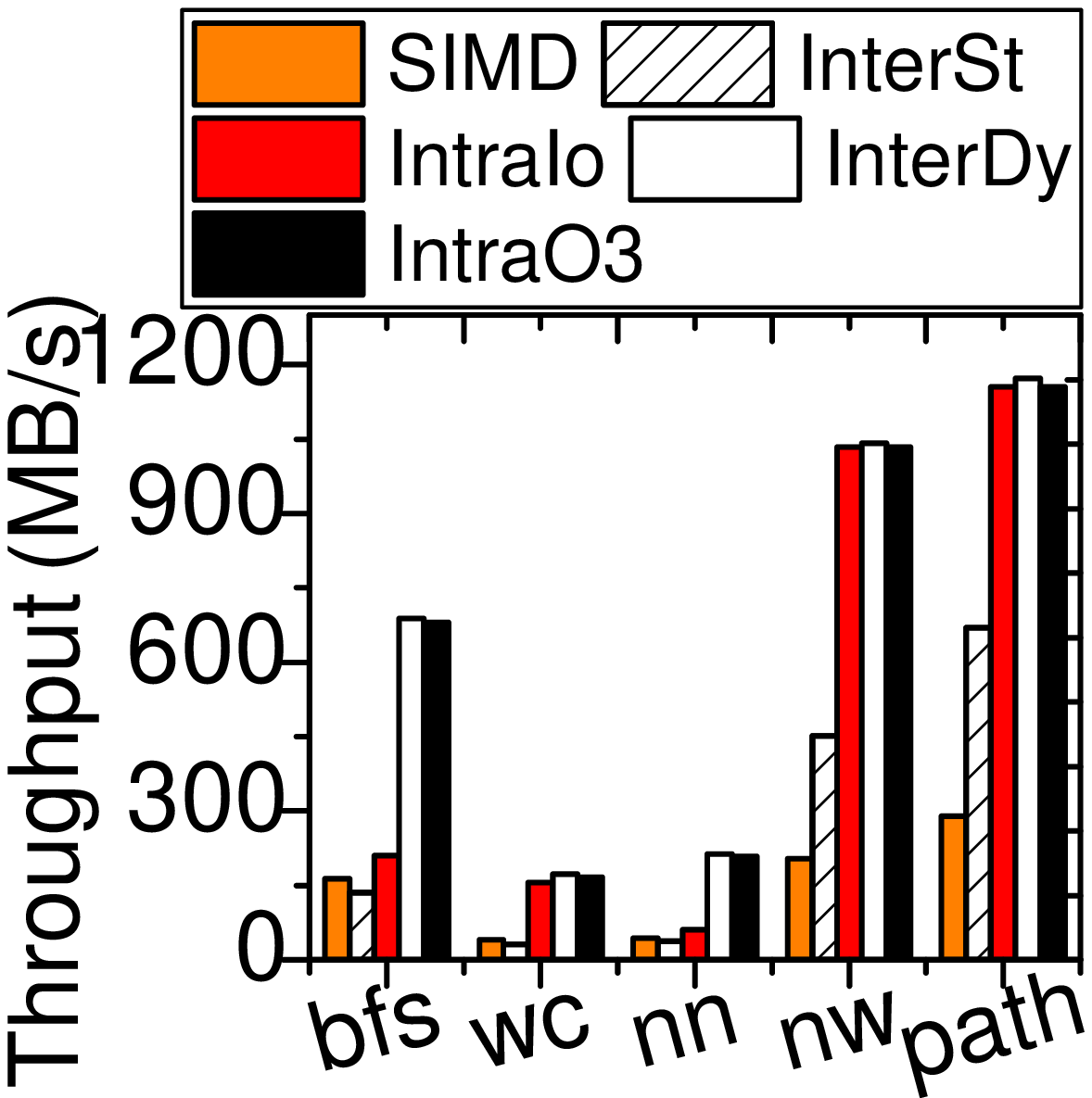}}}
\hspace{1pt}
\subfloat[Energy breakdown.]{\label{fig:energybrk1}\rotatebox{0}{\includegraphics[width=0.49\linewidth]{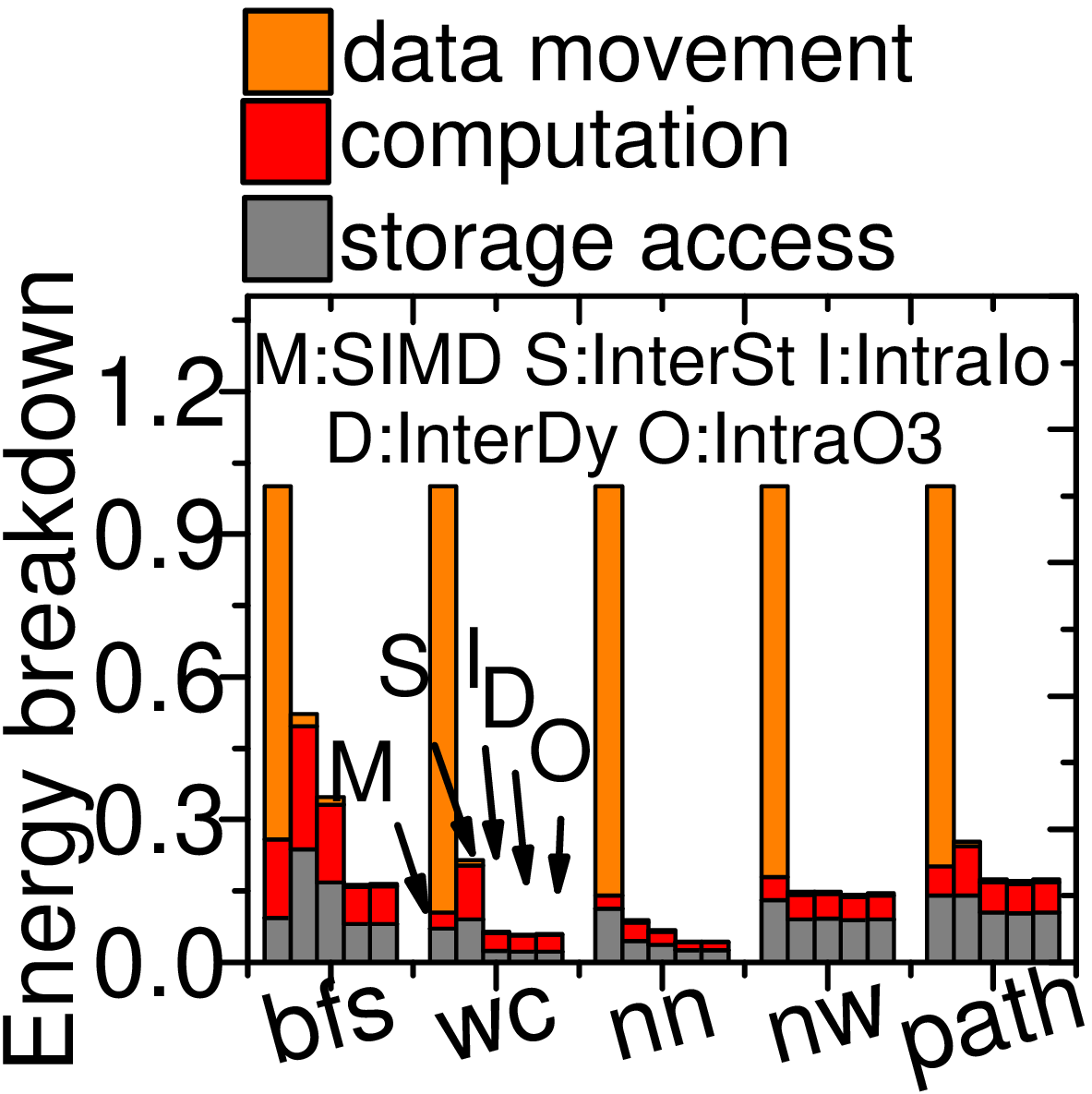}}}
%\vspace{-5pt}
\caption{Analysis of graph/big data applications.}
\label{fig:real_workload}
%\vspace{10pt}
\end{figure}

%\subsection{Effectiveness of FlashAbacus}
\subsection{Dynamics in Data Processing and Power}
Figures \ref{fig:latTSA_MIX} and \ref{fig:latTSA_MIXeng} illustrate the time series analysis results for utilizing functional units and power usage of workers, respectively. In this evaluation, we compare the results of \texttt{IntraO3} with those for \texttt{SIMD} to understand in detail the runtime behavior of the proposed system. As shown in Figure \ref{fig:latTSA_MIX}, \texttt{IntraO3} achieves shorter execution latency and better function unit utilization than \texttt{SIMD}. This is because, \texttt{IntraO3} can perform parallel execution of multiple serial microblocks in multiple cores, which increases the function unit utilization. In addition, due to the shorter storage access, \texttt{IntraO3} can complete the execution 3600 us earlier than \texttt{SIMD}. On the other hand, \texttt{IntraO3} exhibits much lower power consumption than \texttt{SIMD} during storage access (Figure \ref{fig:latTSA_MIXeng}). This is because \texttt{SIMD} requires the assistance of host CPU and main memory in transferring data between the SSD and accelerator, which in turn consumes 3.3x more power than \texttt{IntraO3}. Interestingly, the pure computation power of \texttt{IntraO3} is 21\% higher than \texttt{SIMD}, because \texttt{IntraO3} enables more active function units to perform data processing, thus taking more dynamic power. 

\subsection{Extended Evaluation on Real-world Applications} 
\noindent \textbf{Application selection.}
\newedit{In this section, we select five representative data-intensive workloads coming from graph benchmarks \cite{che2009rodinia} and bigdata benchmarks \cite{he2008mars} in order to better understand system behaviors with real applications: i) K-nearest neighbor (\textit{nn}), ii) graph traversal (\textit{bfs}), iii) DNA sequence searching (\textit{nw}), iv) grid traversal (\textit{path}) and v) mapreduce wordcount (\textit{wc})).}

\noindent \textbf{Performance.}
\newedit{Figure \ref{fig:perf1} illustrates data processing throughput of our FlashAbacus by executing the graph/bigdata benchmarks. One can observe from this figure that the average throughput of FlashAbacus approaches (i.e., \texttt{IntraIo}, \texttt{InterDy} and \texttt{IntraO3}) outperform \texttt{SIMD} by 2.1x, 3.4x and 3.4x, respectively, for all data-intensive workloads that we tested. Even though \texttt{SIMD} can fully utilize all LWPs, and it is expected to bring an excellent throughput for the workloads that have no serialized faction of microblocks (i.e., \textit{nw} and \textit{path}), \texttt{SIMD} unfortunately exhibits poor performance than other FlashAbacus approaches. This performance degradation is observed because LWPs are frequently stalled and bear brunt of comparably long latency imposed by moving the target data between its accelerator and external SSD over multiple interface and software intervention boundaries.  
Similarly, while \texttt{SIMD}'s computation capability is powerful much more than \texttt{InterSt}'s one, which requires to execute multiple kernels in a serial order, \texttt{InterSt}'s average throughput is 120\% and 131\% better than \texttt{SIMD}'s performance for \textit{nw} and \textit{path}, respectively. }

\noindent \textbf{Energy.}
\newedit{
We also decompose the total energy of the system, which is used for each data-intensive application, into i) data movement, ii) computation, and iii) storage access; each item represents the host energy consumed for transferring data between the accelerator and SSD, the actual energy used by accelerator to compute, and the energy involved in serving I/O requests, respectively. The results of this energy breakdown analysis is illustrated by Figure \ref{fig:energybrk1}. One can observe from this figure that \texttt{InterSt}, \texttt{IntraI0}, \texttt{InterDy} and \texttt{IntraO3} can save the total average energy for processing all data by 74\%, 83\%, 88\% and 88\%, compared to \texttt{SIMD}, respectively. 
One of the reasons behind of this energy saving is that the cost of data transfers account for 79\% of the total energy cost in \texttt{SIMD}, while FlashAbacus eliminates the energy, wasted in moving the data between the accelerator and SSD. We observe that the computation efficiency of \texttt{InterDy} and \texttt{IntraO3} also is better than that \texttt{SIMD} for even the workloads that have serial microblocks (i.e., \textit{bfs} and \textit{nn}). This is because OpenMP that \texttt{SIMD} use should serialize the executions of serial microblocks, whereas Flashvisor of the proposed FlashAbacus coordinates such serial microblocks from different kernels and schedule them across different LWPs in parallel.
}

%%%%%%%%%%%%%%%%%%%%%%%%%%%%%%%%%%%%%%%%%%%%%%%%%%%%%%%%%%%%%%%%%%%%%%%%%%%%%%%%%%%%%%%%%%%%%%
\ignore{

\begin{figure}
\centering
%\vspace{-10pt}
\def\subfigcapskip{0pt}
\subfloat[Data-intensive (\textit{ATAX}).]{\label{fig:cores_atax}\rotatebox{0}{\includegraphics[width=0.48\linewidth]{./figs/cores_atax}}}
\hspace{2pt}
\subfloat[computing-intensive (\textit{CORR}).]{\label{fig:cores_corr}\rotatebox{0}{\includegraphics[width=0.48\linewidth]{./figs/cores_corr}}}
\vspace{-8pt}
\caption{A sensitivity analysis (varying number of worker LWP).}
\label{fig:cores-analysis}
\vspace{-5pt}
\end{figure}

\subsection{Sensitive Analysis}
\noindent \textbf{Various computation power.}
Figures \ref{fig:cores_atax} and \ref{fig:cores_corr} show the throughput of different DoubleFlash approaches with various number of LWP workers in representative data-intensive workload \textit{ATAX} and computing-intensive workload \textit{CORR}. 
%Each workload contains 6 instances from the same application. 
Since \textit{ATAX} is a data-intensive workload, the extra data movement overhead of \texttt{GPU} results in its similar performance with \texttt{CPU}, which is shown in Figure \ref{fig:cores_atax}. Since \texttt{ACT}, \texttt{DFS}, and \texttt{DFI} is unable to fully utilize the computation parallelism of multiple LWPs, they experience limited performance improvement, when the number of worker LWPs increases.
%the throughput of \texttt{DFS} has no improvement similar to DFU with single worker LWP, while \texttt{DFI} only achieves limited improvement. That is because, \texttt{DFS} has fixed assignment of applications, which will not span the NDP-kernels across multiple workers. And the performance of \texttt{DFI} is throttled by the serial microblocks in ATAX. Although the number of workers have been increased, the serial microblocks have to be serialized to execute in single worker at one time. 
On the other hand, \texttt{DFO} and \texttt{DFD} can achieve throughput bandwidth proportional to the number of worker LWPs. Since \texttt{DFO} can assign the computation tasks to each worker LWP in a finer granularity, \texttt{DFO} can achieve better performance compared to DFD, when LWPs are not sufficient. As the number of workers reaches 6, the throughput bandwidth of \texttt{DFO} gets saturated, because the cumulative data access throughput (two times of throughput shown in Figure \ref{fig:cores_atax}) has reached the maximum bandwidth of underlying flash backbone (2GB/s). As for computing-intensive workload \textit{CORR}, which is shown in Figure \ref{fig:cores_corr}, the throughput of \texttt{GPU} is 4 times higher than \texttt{CPU}. 
%That is because, computation is the domain factor of execution time and \texttt{GPU} has much higher computation power compared to CPU-based platform. 
Compared to \texttt{GPU}, \texttt{DFO} and \texttt{DFD} have much lower throughput, when worker number is 1. However, when the number of LWPs is more than 4, \texttt{DFO} and \texttt{DFD} exceed the throughput of \texttt{GPU}. 
%But the throughput gets saturated, even if the number of workers increases. That is because, the memory accesses from multiple workers cause traffic congestion in network buffer system.
To sum up, the performance of our proposed \texttt{DFO} is scalable for multiple LWPs, and it also can provide better LWP utilization when the number of LWPs is few.
}
%%%%%%%%%%%%%%%%%%%%%%%%%%%%%%%%%%%%%%%%%%%%%%%%%%%%%%%%%%%%%%%%%%%%%%%%%%%%%%%%%%%%%%%%%%%%%%

\ignore{

%%%%%%%%%%%%%%%%%%%%%%%%%%%%%%%%%%%%%%%%%%%%%%%%%%%%%%%%%%%%%%%%%%%%%%%%%%%%%%%%%%%%%%%%%%
\subsection{Study of Data Movement Costs}
Obviously, the main benefit that our proposed DoubleFlash can deliver is the data movement cost reduction as it can remove the overheads to move data around the CPU and the storage. However, it is unclear where the the benefit comes from in real systems. In this section, we study such cost reduction benefits by decomposing the actual time that each component on storage stack needs to spend.

\noindent \textbf{Performance benefits.} 
Figure \ref{fig:data_transfer} nomailzes the data movement time that the host (GPSYS) requires in processing data to what our DoubleFlash requires. In addition, this figure also decomposes the actual latency for each component in the data movement path: i) PCIe interface overhead, ii) operating system (OS) overhead, and iii) flash access latency from the CPU viewpoint. 
As shown in the figure, overall, our proposed DoubleFlash removes the PCIe interface overheads, OS overheads and flash access overheads by 95\%, 93\% and 20\%, respectively. Especially, as target data resides in the underlying storage, our DoubleFlash is rarely involved in PCIe communication. It also has no or very few OS overheads imposed by user-and-kernel space data copies, file system operation latency, and the corresponding context switch activities since the master core, storage core, and DSP workers directly handle the data inside storage. In addition to these data movement cost reduction benefits, our DoubleFlash can also perform better I/O accesses being aware of the underlying storage backbone as there is no other 3rd party system kernels (e.g., buffer cache or block scheduler) that change the access pattern due to ignorance of such storage's physical layout.

\noindent \textbf{Energy benefit analysis.} Figure \ref{fig:data_transfer_eng} normalizes energy consumption for each hardware module on the storage stack to what our DoubleFlash consumes. While our DoubleFlash removes many PCIe interface overhead (in terms of latency), the energy benefit is not significant compared to others. This is because PCIe interface itself is already optimized into low power data movement. However, as our DoubleFlash can get rid of tremendous data movements in our model, it does not require copying and maintaining a larger-size data in DRAM. Therefore, our DoubleFlash shows 460\% energy reduction on DRAM module on the storage stack while the energy values required by accessing the underlying flash memory similar to each other. To sum up, our DoubleFlash reduces 250\% energy consumption that a conventional data processing approach requires, which makes our DoubleFlash more attractive to perform data processing near storage.

\subsection{Uniprocessor-based DoubleFlash}
\noindent \textbf{Overall performance.} We first evaluate uniprocessor-based NDP performance with other design optimizations and analyze the reason why we need multicore DSP processor. Specifically, we compare the performance of baseline NSSD-SGC with the cases where it has read-overlapping (NSSD-SGC-R), write-asynchronous (NSSD-SGC-W), and both techniques (NSSD-SGC-RW), and the results are shown in Figure \ref{fig:singlecore}. Even though NSSD-SGC-R, -W and -RW exhibit a better performance than the baseline by 4.6, 2.8, and 7.1 times under 2DCONV and 2MM, all these NDP burst and I/O burst overlapping schemes have less or no benefit for most workloads. 

\noindent \textbf{Execution time breakdown.} To better understand this overall performance behavior, we choose three workloads, 2DCONV, 2MM, and ATAX. We decompose the execution time and normalize them to GPSYS in Figure \ref{fig:Normalized_performance1}. 
%In this analysis, we choose three workloads, 2DCON, 2MM and ATAX, which show  
One can see from this figure that, the main contributor of 2DCONV is the flash access time, which takes into account 60\% of total execution time. Because of this, overlapping NDP burst and I/O burst can maximize the data processing bandwidth.
Unlike 2DCONV, the computation time of 2MM and ATAX accounts for more than 85\% of total execution time, whereas I/O time on flash memory only accounts for less than 10\%. As DSP computation time is the dominant factor to decide the overall NDP performance, all the NSSD-SGC options are not able to compete with GPSYS performance behavior.

\subsection{Multicore-based DoubleFlash}
Due to the limited computation capability, we show that uniprocessor-based DoubleFlash cannot improve performance even with NDP burst and I/O burst overlapping. To overcome this challenge, DoubleFlash also employ a scalable multicore DSP processor, which can accelerate the process within reasonable power constrains. We applied read-overlapping and write-asynchronous for all remaining parts as default. 

\noindent \textbf{Uni-program analysis.} Figure \ref{fig:Normalized_performance1} plots the evaluation results of multi-core system with uni-program workloads. As shown in the figure, while NSSD-SGC exhibits poor performance for most workloads we tested (except 2DCONV and 3DCONV), our DoubleFlash that employs inter-kernel execution (NSSD-MES and NSSD-MED) and intra-kernel execution (NSSD-MAI and NSSD-MAO), on average, improve the performance by 2.4 times and 2.9 times compared to GPSYS, respectively. Interestingly, while NSSD-MAI is worse than others, most multicore-based DoubleFlash platforms show similar behavior. This is because we evaluate single program, which exhibit a load balance characteristics to each DSP worker, but the performance of NSSD-MAI significantly degrade due to serial code fraction, which cannot fully utilize all DSP workers. 

\noindent \textbf{Multi-program analysis.} 
To have better comparison, we also analyze the multi-program workloads we tested, and the results are shown in Figure \ref{fig:Normalized_performance2}. One can observe from this figure that, for both inter- and intra-kernel execution, dynamic scheduling strategies (NSSD-MED and NSSD-MAO) exhibit better performance compared to the static methods (NSSD-MES and NSSD-MAI) by 2.1 times and 2.4 times, respectively. Meanwhile, NSSD-MED and NSSD-MAO perform better than GPSYS for all workload mixes. For MIX4, MIX10, and MIX13, their performance improvement can reach up to 3 times compared to that of GPSYS. Similar to the case of uni-program workloads, NSSD-SGC shows a inferior performance compared to GPSYS. Finally, NSSD-MES and NSSD-MAI, with their static scheduling strategy, have similar performance as GPSYS.    

%\fixme{The following sentences are needed to be completely revised.} 
%In workload ATAX, BICG, GESUMMV, and MVT, it shows highest computation throughput, which is xx MB/s. Because multiple DSP workers can significantly reduce the computation time. This computation time can overlap with regular NAND flash read access by using read overlapping and these workloads generate few output data which help them avoid NAND flash write overhead. As for 2DCONV, the throughput gets saturated when there is only single DSP worker. That's because, SSD read and write latency limits the data processing bandwidth. In this case, adding up more NAND flash resource can improve the performance.

%%%%%%%%%%%%%%%%%%%%%%%%%%%%%%%%%%%%%%%%%%%%%%%%%%%%%%%%%%%%%%%%%%%%%%%%%%%%%%%%%%%%%%%%%%%
\subsection{NDP Kernel Execution Strategy Study}
\noindent \textbf{Inter-kernel execution model.} %\fixme{The following sentences are needed to be completely revised.} 
The detailed analysis of our inter-kernel execution models with static and dynamic scheduling strategy on multi-program workloads are shown in Figure \ref{fig:cg_active} and \ref{fig:cg_various} as NSSD-MES and NSSD-MED, respectively. For workload MIX2 in Figure \ref{fig:cg_active}, NSSD-MES shows a unpredictable behavior, where worker 4 and worker 6 take 2X and 1.6X longer latency, compared to the uniform latency value of all workers in NSSD-MED. \ref{fig:cg_various} portrays latency analysis for all multi-program workloads. Here, NSSD-MES displays huge latency fluctuation between nearly 0\% to more than 400\%, compared to average latency. We believe this superior performance of NSSD-MED comes from the dynamic nature of its scheduling policy, as dynamic scheduling can carry out load-balancing in a more efficient way.

\noindent \textbf{Intra-kernel execution model.} 
Latency analysis of fine-granular intra-kernel execution model is shown in Figure \ref{fig:fg_active} and \ref{fig:fg_various}, where NSSD-MAI and NSSD-MAO represents in- and out-order scheduling policy. Figure \ref{fig:fg_active} shows serial code and parallel code distribution in different DSP workers. In NSSD-MAI, only DSP worker1 is allocated to execute serial code, while other workers are stalling. In contrast, NSSD-MAO assigns kernel blocks from other kernels to any DSP worker which is in idle. Thus, in figure \ref{fig:fg_active}, each worker in  NSSD-MAO has similar active time and worker 1 in NSSD-MAI experience much longer execution time. Figure \ref{fig:fg_various} shows the worker's execution time of NSSD-MAI in each multi-program workload. The NSSD-MAI execution time experience 425\% performance degradation in the worst case. We believe NSSD-MAO performs better for its dynamic scheduling.

Both NSSD-MED and NSSD-MAO strategies can balance DSP worker's load. But NSSD-MAO is compatible in more conditions and can provide better performance. Figure \ref{fig:CGFG} shows DSP worker's active time variance under different workloads and different number of commands from 6 to 576. A smaller variance value means each DSP worker has more similar execution time. From figure \ref{fig:CGFG}, when command number is small, it is limited to dynamically adjust each worker's work. NSSD-MAO assigns work to each worker based on fine-granular kernel blocks, which can tolerate such limitation.

%%%%%%%%%%%%%%%%%%%%%%%%%%%%%%%%%%%%%%%%%%%%%%%%%%%%%%%%%%%%%%%%%%%%%%%%%%%%%%%%%
\subsection{Energy Efficiency Analysis}
%\fixme{The following sentences are needed to be completely revised. -- need to touch energy consumption brake down as well as total energy graphs} 
DoubleFlash saves total energy by processing data inside SSD. Figure \ref{fig:tot_energy} shows the total energy consumption of each workload processed by GPSYS, NSSD-SGC, NSSD-MED, and NSSD-MAO. Multi-core DoubleFlash systems NSSD-MED and NSSD-MAO save 89\% and 90\% energy on average across all the workloads respectively. And NSSD-SGC also saves up to 65\% energy on average. Figure \ref{fig:energy-breakdown} shows the energy breakdown of GPSYS, NSSD-SGC, NSSD-MED, and NSSD-MAO separately. In GPSYS, CPU and DRAM consume more than 90\% energy. But in NSSD-SGC, NSSD-MED, and NSSD-MAO, the DSP workers as well as master core only consume 25\% and 45\% of total energy on average. Since only small-volume network buffer memory is used to accommodate program's input and output data, the total DRAM energy consumption is also few.

%%%%%%%%%%%%%%%%%%%%%%%%%%%%%%%%%%%%%%%%%%%%%%%%%%%%%%%%%%%%%%%%%%%%%%%%%%%%%%%%%

\subsection{Sensitive Tests}
\noindent \textbf{DSP workers.} DoubleFlash can improve the performance of a target program by employing multiple computation accelerators. Figure \ref{fig:various_core} demonstrates such performance improvement for increasing number of DSP workers based on NSSD-MAO system, where the speedup is normalized to GPSYS. Except for workload 2DCONV and 3DCONV, single-program workloads' performance (NSSD-MAO) increase linearly and exceeds that of CPU when the number of DSP workers is more than 4. 2DCONV and 3DCONV initially experience sharp improvement in performance, which then hits saturation because of limited SSD bandwidth.

\noindent \textbf{Storage backbone.} The number of NAND flash channels, dies, and planes utilized in a SSD configuration can significantly impact the SSD's internal read/write bandwidth. Figure \ref{fig:various_SSDconfig} portrays the comparative performance of five such configurations. By default, we use MLC NAND flash which takes 25us and 300us for flash cell read and write operations, respectively. Due to the limited bandwidth provided by the 1 PMM - 1 die SSD, each workload experiences significant performance degradation. Once we increase the number of PMMs and dies, our DoubleFlash (NSSD-MAO) achieves substantial performance improvement. In addition, we also test one configuration with TLC NAND flash, which takes 105us for a read and 1300us for a write. For that configuration, the system experiences some performance degradation because of TLC NAND flash's significant longer read and write latency.
}

\ignore{
%compared DF internally -- by showing each latency
\subsection{Execution Latency Analysis}

\noindent \textbf{Restrained execution.} 
%Figure \ref{} shows the execution time (normalized to CPU) of a single NDP-kernel instance for each workload. 
Compared to the throughput evaluation (Section 5.1), in this evaluation, we only execute a single NDP-kernel instance and measure the latency whose data is shown in Figure \ref{fig:latency_fig}. 
%In overall, GPU can deliver is only XX\% shorter latency, compared to CPU-based data processing approach. Even though \texttt{GPU} has more execution resources and can take advantage of massive threading, the amount of data that the single kernel instance needs to process is insufficient to create hundreds threads. 
%the latency overheads to access memory and data movement are relatively longer
%In contrast, even with single instance execution, our DoubleFlash (\texttt{DFO}) exhibit in overall XX\% better performance than CPU-based approach. 
As this evaluation executes just a single NDP-kernel instance, we group the five different DoubleFlash approaches into two categories: inter-kernel execution and intra-kernel execution. the inter-kernel execution includes \texttt{ACT}, \texttt{DFS}, and \texttt{DFD}, as the whole workload is only executed in single worker LWP in \texttt{ACT}, \texttt{DFS}, and \texttt{DFD} and they have same latency value. However, intra-kernel execution including \texttt{DFI} and \texttt{DFO}, is able to split a single instance into multiple NDP-microblocks and spread screens out across multiple LWPs. Thus, it improves performance by 3.4 times compared to inter-kernel execution. Note that NDP-kernel containing serial microblocks cannot fully utilize multiple worker LWPs. Thus, intra-kernel execution can reduce the execution latency by 38\% compared to inter-kernel execution in NDP-kernel with serial microblocks, while it only can reduce latency by 266\% in NDP-kernel without serial microblocks. To sum up, our proposed DFO provides euqal or shorter execution latency compared to other DoubleFlash approaches.

\ignore{
\noindent \textbf{Multi-instances.} 
Figures \ref{} and \ref{} show the latency of each platform with 24 workload executions. Unlike the restrained execution evaluation, GPU exhibit XX time better performance compared to CPU. This is because there many kernel resources that the underlying GPU can take massing thrashing. Specifically, while ATA and ADI has less benefit (as their computation fraction is lower than that of data managements), GPU accelerate performance by XX under SYRK and FLOYD workload execution. Our baseline DoubleFlash (\texttt{ACT}) degrades the latency of homogeneous and heterogeneous workloads compared to GPU by XX and XX\%, respectively. Even though \texttt{ACT} remove data movement and storage accesses compared to GPU by XX\% and XX\%, respectively, the low computation resource of LWP make performance worse. In contrast, our \texttt{DFD} and DFO shorten then latency of computation by employing the inter-kernel and intra-kernel executions. Specifically, our \texttt{DFD} and \texttt{DFO} shortens the latency of GPU-based acceleration approach by XX\% and XX\%, respectively.

Figures \ref{fig:energybreak_singleapp_fi} and \ref{fig:energybreak_mix_fig} show the decomposition of energy consumption for homogeneous and heterogeneous workloads. As shown in the figures, the fraction of data movement energy in GPU is 36\% higher than CPU under data-intensive workloads. That is because, GPU-based accelerator requires host supports to migrate data between GPU memory and SSD via PCIe, which causes a lot of host intervention. Even though the host is not involved in data processing, it takes a lot of host cycles to coordinate data access, which increases the total amount of energy. Unlike GPU, our DoubleFlash can significantly reduce the overall energy by eliminating the energy wasted by the host. As shown in the figure, the fraction of the data movement energy in our DoubleFlash is less than 3\%.  
Unlike data-intensive workloads, the fraction of data movement energy is around 2\% in computing-intensive and heterogeneous workloads, indicating the impact of data movement is negligible to the whole system. Since GPU power is much less than CPU and the heavy computation resources in GPU can largely shorten the execution time, GPU exhibits higher energy efficiency than CPU. Compared to the heavy computation components in GPU, our proposed \texttt{DFO} employs multiple light-weight processors and low-power DRAM, which consumes much less power than GPU. As \texttt{DFO}'s throughput is better than or equal to GPU, \texttt{DFO} offers higher energy efficiency than GPU. 

}
}

%\noindent \textbf{Heterogeneous multi-instances.}

\section{Discussion and Related Work}
\label{sec:relatedwork}
\noindent \textbf{Platform selection.}
\newedit{The multicore-based PCIe platform we select \cite{TI6678} integrates a terabit-bandwidth crossbar network for multi-core communication, which potentially make the platform a scale-out accelerator system (by adding up more LWPs into the network). In addition, considering a low power budget (5W $\sim$20W), the platform we implemented FlashAbacuse has been considered as one of the most energy efficient (and fast) accelerator in processing a massive set of data \cite{hegde2016caffepresso}. Specifically, it provides 358 Gops/s with 25.6 Gops/s/W power efficiency, while other low-power multiprocessors such as a many-core system \cite{EPIPHANY-III}, a mobile GPU \cite{nvidia2015nvidia, otterness2017evaluation}, a FPGA-based accelerator  \cite{xilinx-zc706} are available to offer 10.5, 256 and 11.5 Gops/s with 2.6, 25.6 and 0.3 Gops/s/W power efficiency, respectively.} 

\noindent \textbf{Limits of this study.}
\newedit{While the backend storage complex of FlashAbacus is designed as a self-existent module, we believe that it is difficult to replace with conventional off-the-shelf SSD solutions. This is because our solution builds flash firmware solution from scratch, which is especially customized for flash virtualization. We select a high-end FPGA for the flash transaction controls as research purpose, but the actual performance of flash management we believe cannot be equal to or better than microcoded ASIC-based flash controller due to the limit of low frequency of FPGA. Another limit behind this work is that we are unfortunately not able to implement FlashAbacus in different accelerators such as GPGPU due to access limits for such hardware platforms and a lack of supports for code generations. However, we believe that the proposed self-governing design (e.g., flash integration and multi-kernel execution) can be applied to any type of hardware accelerators if vendor supports or open their IPs, which will outperform many accelerators that require employing an external storage. }

%\noindent \textbf{FPGA implementation.} \newedit{Our FPGA-based NAND Flash controller controls multiple NAND Flash channels and seamlessly manages all flash transactions from each channel. In addition, it also implements the physical layer of RapidIO, which enables the communication with accelerator platform. This FPGA implementation is much more complex than the one proposed in OpenNVM \cite{zhang2015opennvm}. For research purpose, we select a high-end FPGA prototype which is flexible for future extension. However, it may not be superior to the FPGAs customized for off-the-shelf products, in terms of power efficiency and performance.}

\noindent \textbf{Hardware/software integration.}
\newedit{To address the overheads of data movements, several hardware approaches tried to directly reduce long datapath that sit between different computing devices. For example, \cite{shainer2011development, ahn2015dcs, foley2014nvlink} forward the target data directly among different PCIe devices without a CPU intervention. System-on-chip approachs that tightly integrate CPUs and accelerators over shared memory \cite{damaraju201222nm, wolf2008multiprocessor}, which can reduce the data transfers between. However, these hardware approaches cannot completely address the overheads imposed by discrete software stacks. Since SSDs are block devices, which can be shared by many other user applications, all the SSD accesses should be controlled by the storage stack appropriately. Otherwise, the system cannot guarantee the durability of storage, and simultaneous data access without an access control of storage stack can make data inconsistent, in turn leading to a system crash. In contrast, a few previous studies explored reducing the overheads of moving data between an accelerator and a storage by optimizing the software stack of the two devices \cite{zhang2015nvmmu, tseng2016morpheus, silberstein2013gpufs}. However, these studies cannot remove all limitations imposed by different physical interface boundaries, and unfortunately, their solutions highly depends on the target system's software and software environment. 
}

\noindent \textbf{In-storage processing.}
\newedit{Prior studies \cite{cho2013active, tiwari2013active, gu2016biscuit, acharya1998active, keeton1998case} proposed to eliminate the data movement overheads by leveraging the existing controller to execute a kernel in storage. However, these in-storage process approaches are limited to a single core execution and their computation flexibility is strictly limited by the APIs built at the design stage. \cite{jun2015bluedbm} integrates SSD controller and data processing engine within a single FPGA, which supports parallel data access and host task processing. While this FPGA-based approach is well optimized for a specific application, porting the current host programs to RTL design can bring huge burdens to programmers and still inflexible to adopt many different types of data-intensive applications at runtime. FlashAbacus governs the storage within an accelerator and directly executes multiple general applications across multiple light-weight processors without any limit imposed by the storage firmware or APIs.
}

\noindent \textbf{Open-channel or cross-layer optimization.}
\newedit{Many proposals also paid attention on cross-layer optimizations or open-channel SSD approaches that migrates firmware from the device to kernel or  partitions some functionalities of flash controller and implement it into OS \cite{bernhard2014cooperative, jung2012middleware, ouyang2014sdf, lee2016application, huang2015unified, jung2014triple}. While all these studies focused on managing the underlying flash memory and SSD efficiently, they have a lack of considerations on putting data processing and storage together in an hardware platform. In addition, since these proposals cannot be applied into an single hardware accelerator, which has no storage stack, but requires byte-addressability to access the underlying storage complex.
}

\ignore{
For example, \cite{zhang2015nvmmu} successfully eliminates redundant data copies in host memory and unnecessary software intervention by directly forwarding data requests between accelerator and storage. However, this approach requires the modification of host-side drivers or file systems, and still needs to bring data to host memory for one time. On the other hand, \cite{ahn2015dcs} proposes a FPGA-based data transfer engine to manage and monitor inter-device communications. While this hardware approach can avoid the host intervention, it still requires data to walk through the communication protocol as well as the physical links (i.e., PCIe).
Rather than moving a huge amount of data between an accelerator and a storage, our approach tightly integrates many flash modules within multi-core accelerators, which in turn drastically reduce system latency and energy overhead (only requires host to offload kernel executables to the accelerator).

\noindent \textbf{Near-data processing.} 
Prior studies \cite{cho2013active, tiwari2013active} proposed to eliminate the data movement overheads by leveraging the existing controller to execute kernel locally in SSDs. 
However, due to the power constraints and limited area space, the performance of SSD controller is unsatisfactory to process the computation tasks that host should perform. In addition, the SSD controller usually is fully utilized to support basic SSD operations such as I/O scheduling, ECC, and data compression, which in turn remains few space to process other tasks. 
\cite{gu2016biscuit} proposes to integrate dual ARM cores inside SSD. In this framework, data processing can be only done by a single processor, and it requires integrating customized interface flash firmware offers by modifying all data-intensive applications.
Other works such as \cite{cho2013xsd}, proposes to harness GPU inside SSD by leveraging GPU's massive computation power. While this approach can overcome the shortcoming of SSD controller, it suffers from GPU's huge power consumption. In addition, the SIMT execution model limits GPU from processing data-complex applications efficiently, which constrains its essence in practice.
\cite{jun2015bluedbm} integrates SSD controller and data processing engine within a single FPGA, which supports data access and host task processing in parallel. While this FPGA-based approach is well optimized for a specific application, porting current host programs to RTL design can bring huge burdens to programmers and still inflexible to adopt many different types of data-intensive applications.
Our FlashAbacus integrates NAND flash modules into low-power multiprocessors, in which each core independently processes either host tasks or SSD tasks. Specifically, we allocate one processor as flashvisor to schedule tasks and I/O access in harmony, and one processor to manage SSD-related background jobs such as garbage collection and page table snapshot. While single processor is not sufficient to process host tasks, we allocate multiple processors to process host tasks in parallel. Compared to GPU, our multiprocessors can efficiently process serialized code segments, while offering much more flexibility than FPGA.
Constrained by the limited computation power of SSD, \cite{tseng2016morpheus} and \cite{do2013query} explore the feasibility of preprocessing the application objects and performing relational analytic query processing locally in SSD. While offloading only simple program code segments can better utilize the cores inside SSD, it exposes the potential overhead of synchronization and communication between host and SSD compute cores.  
By leveraging convincing computation capability provided by multiprocessors, our approach offloads full programs to the SSD, which in turn can avoid CPU intervention and reduce the communication overhead between host and SSD. 
}

\section{Conclusion}
\label{sec:conclusion}
In this work, we combined flash with lightweight multi-core accelerators to carry out energy-efficient data processing. Our proposed accelerator can offload various kernels in an executable form and process parallel data. Our accelerator can improve performance by 127\%, while requiring 78\% less energy, compared to a multi-core accelerator that needs external storage accesses for data-processing. 

\section{Acknowledgement}
\label{sec:ack}
This research is mainly supported by NRF 2016R1C1B2015312 and MemRay grant (2015-11- 1731). This work is also supported in part by, DOE DEAC02-05CH 11231, IITP-2017-2017-0-01015, and NRF2015M3C4A7065645. The authors thank Gilles Muller for shepherding this paper. We also thank MemRay Corporation and Texas Instruments for their research sample donation and technical support. Myoungsoo jung is the corresponding author.

\bibliographystyle{abbrvnat}
\bibliography{references}

\end{document}